\documentclass[]{interact}
\usepackage{amsmath,amssymb}
\usepackage{amsthm}
\usepackage{epstopdf}
\usepackage[natbibapa,nodoi]{apacite}
\theoremstyle{plain}

\usepackage[colorlinks=true, citecolor=blue, linkcolor=blue, urlcolor=blue]{hyperref}
\usepackage{tikz}
\usetikzlibrary{calc}
\usepackage[font=scriptsize]{subcaption}

\usepackage{multirow}\usepackage{makecell}
\usepackage{url}
\usepackage{booktabs}

\usepackage{bm}

\newcommand{\mt}{\mathit{MT}}

\newcommand{\sr}{\mathit{SR}}

\newcommand{\margin}{\textsc{Margin}}
\newcommand{\mae}{\mathit{MAE}}
\newcommand{\rmse}{\mathit{RMSE}}
\newcommand{\mape}{\mathit{MAPE}}
\newcommand{\dedgex}{\mathit{D_{edge\_x}}}
\newcommand{\dedgey}{\mathit{D_{edge\_y}}}
\newcommand{\gammax}{\gamma_{1\_x}}
\newcommand{\gammay}{\gamma_{1\_y}}

\begin{document}

\articletype{Research Article. Word count: 9{,}787}

\title{Skewed Dual Normal Distribution Model: Predicting Touch Pointing Success Rates for Targets Near Screen Edges and Corners}

\author{
\name{Nobuhito Kasahara\textsuperscript{a}, Shota Yamanaka\textsuperscript{b,$\dagger{}$}, Homei Miyashita\textsuperscript{a}}
\affil{\textsuperscript{a} Meiji University, Nakano-ku, Tokyo, Japan; \textsuperscript{b}\thanks{${}^\mathrm{\dagger{}}$\ Corresponding author: S. Yamanaka. LY Corporation Research, LY Corporation, Kioi Tower, 102-8282, Tokyo Garden Terrace Kioicho, 1-3, Kioi-cho, Chiyoda-ku, Tokyo, Japan. Email: syamanak@lycorp.co.jp} LY Corporation, Chiyoda-ku, Tokyo, Japan}
}

\maketitle

\begin{abstract}
Typical success-rate prediction models for tapping exclude targets near screen edges.
However, design constraints often force such placements, and in scrollable user interfaces, any element can move close to the screen edges.
In this work, we model how target–edge distance affects touch pointing accuracy.
We propose the Skewed Dual Normal Distribution Model, which assumes the tap-coordinate distribution is skewed by a nearby edge.
The results showed that as targets approached the edge, the distribution's peak shifted toward the edge, and its tail extended away.
In contrast to prior reports, the success rate improved when the target touched the edge, suggesting a strategy of ``tapping the target together with the edge.''
Our model predicts success rates across a wide range of conditions, including edge‑adjacent targets.
Through three experiments of horizontal, vertical, and 2D pointing, we demonstrated the generalizability and utility of our proposed model.
\end{abstract}


\section{Introduction}
\subsection{Background}
Modeling human motor performance and refining models are central topics in HCI.
Research on Fitts' law \citep{fitts1954information} to model movement times (\(\mt\)s) for pointing is a typical example, as selecting icons or hyperlinks is one of the most frequently performed actions.
Another key usability indicator is the success rate (\(\sr\)), or conversely, the error rate: how accurately users perform pointing tasks.
As a prediction model, \cite{Bi2016DualGaussian} showed that touch pointing success can be estimated using the \emph{Dual Gaussian Distribution Model}, which accounts for ambiguity in finger tap-coordinates.
Building on this model, many refinements tailored to diverse task conditions have been proposed.

Most of these models focus on targets in the center of the screen.
Prior work has reported that in touch pointing, $\mt$ increases and $\sr$ tends to decrease as targets approach an edge.
Therefore, touch-based user interfaces (UIs) should avoid placing UI elements near the screen edge \citep{Avrahami2015EdgeTouch,Usuba2023EdgeTarget,Henze2011LargeExperiment}.
Nevertheless, placing targets near the edge or corner is common.
For example, a dense layout may necessitate such placement, or scrolling may move targets close to the edges and corners.
Existing models have a limitation in such cases: they estimate $\sr$s uniformly, regardless of target–edge distance.
Considering that all UI elements can potentially approach the screen edge or corner in scrollable UIs, existing models may only be applicable to the special case where targets are fixed at positions sufficiently far from all edges.

If we could accurately predict $\sr$s near the edge and corner, UI designers could utilize the entire screen more efficiently while maintaining user accuracy.
For example, combined with tools that compute $\sr$s from on-screen element size \citep{usuba24arxivTappy,Yamanaka24arXivFigmaTappy,LIFULL24tapAnalyzer}, designers could account for proximity to the edge and corner.
Another application is to enable more accurate selection by dynamically adjusting the UI layout based on target–edge distance after each scrolling operation.
Nonetheless, no model has been proposed that describes how $\sr$ changes as a function of target–edge distance.

\begin{figure}[t]
\centering
    \includegraphics[height = 6cm]{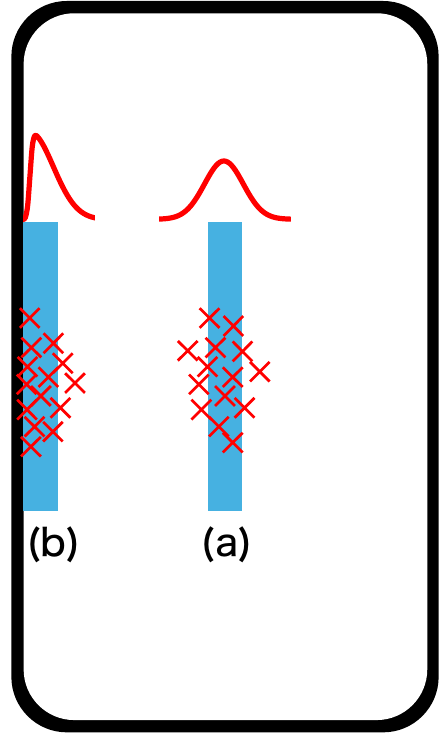}
\caption{The proposed Skewed Dual Normal Distribution Model assumes that the tap-coordinate distribution is skewed by the presence of a screen edge on one side of the target and uses the cumulative distribution function of the skew-normal distribution to estimate tap success rate. (a) When the target is sufficiently far from the screen edge, the tap-coordinate distribution is normal (Gaussian). (b) When the target is near the screen edge, the tap-coordinate distribution becomes skew-normal.}
\label{fig:Fig1}
\end{figure}

In light of this background, we model the relationship between target–edge distance and $\sr$.
Such interpretable mathematical models provide UI designers with quantitative and easy-to-understand design guidelines.
Our Skewed Dual Normal Distribution Model extends the Dual Gaussian Distribution Model proposed by \cite{Bi2016DualGaussian} (\autoref{fig:Fig1}). 
Unlike existing models that assume a Gaussian distribution for tap-coordinates, ours assumes that an edge near one side of the target skews the tap-point distribution and thus affects $\sr$.

To validate our proposed model, we conducted three touch-pointing experiments.
Experiment 1 was designed to verify two points: whether tap-coordinates follow a skewed normal distribution and whether the proposed model can appropriately capture this skewness.
To this end, we used 1D targets with horizontal constraints placed near the left edge of a smartphone.
Experiment 2 examined whether the model's characterization of skewness could be extended to the bottom edge (vertical axis).
Experiment 3 tested the model's applicability to 2D targets placed near the top-right corner.
Since the proposed model assumes a skew in the one-dimensional distribution of tap-coordinates, we first verified the model's validity with 1D pointing tasks.
Subsequently, we tested its applicability to 2D targets, which are more common in actual UIs, thereby confirming the generalizability of the model.
The results demonstrated that the proposed model can predict tap success rates with high accuracy, regardless of the coordinate axis (X or Y) or the target dimensionality (1D or 2D).

\subsection{Prior Publication Statement and Our Present Contributions}
This article is an extended version of our conference paper \citep{Kasahara2026chiSkew}.
The most substantial change in the present article is that we conducted Experiment 3 using 2D rectangular targets to further strengthen model validity.
We also reconstructed and enhanced the introduction, model derivation, and discussion.
Approximately 40\% of the present article (including \autoref{sec:exp3} onward and the parts mentioned above) is newly added for the present submission.
The applicability of our model was validated only for 1D targets with sufficient length (e.g., hyperlinks), which is a critical limitation for the purpose of UI design.
The current article resolves this issue, and readers can apply our model to more general conditions with 2D targets.

Our main contributions are as follows:
\begin{itemize}
  \item We theoretically derive an interpretable and highly accurate $\sr$ prediction model that covers edge-adjacent and corner-adjacent targets previously treated as exceptions in $\sr$ modeling.
  \item We experimentally show that our model accurately predicts $\sr$s for both vertical and horizontal 1D targets near the screen edges, as well as for 2D targets near the screen corners, thereby validating its usefulness and generalizability.
  \item We provide new insight into the user strategy of ``tapping together with the edge'' as a factor underlying the proposed model's predictive accuracy. This has deepened our understanding of the actual behavior that users perform in their smartphone operations.
\end{itemize}

\section{Related Work}
\subsection{Success Rate Estimation Models}
In HCI, $\sr$ prediction models for pointing have been widely studied.
\cite{Meyer1988Optimality} and \cite{Wobbrock2008_ERModel} proposed error-rate prediction models in cursor-based pointing.
In touch pointing, the Dual Gaussian Distribution Model is foundational \citep{Bi2016DualGaussian}.
It assumes that tap positions follow a Gaussian distribution characterized by a mixture of variance due to target size and finger-placement uncertainty.
The tap-coordinate variances along the \(x\)- and \(y\)-axes can be estimated as a function of target width \(W\) and height \(H\) by
\begin{equation}
\sigma_{x}^2=a_{x}W^2+b_x\ \ \mathrm{and}\ \ \sigma_{y}^2=a_{y}H^2+b_y,
\label{formula:Bi_sigma}
\end{equation}
where \(\sigma_x\) and \(\sigma_y\) are the standard deviations of tap-coordinates along \(x\) and \(y\), and $a_x, b_x, a_y,$ and $b_y$ are the regression constants.
For a 1D target whose width along the $x$-axis is $W$, the tap success rate ($\sr_x$) is defined as the probability that a tap falls within the interval $[-W/2, W/2]$ relative to the target center ($x=0$). 
Specifically, $\sr_x$ can be calculated using the cumulative distribution function (CDF) of the normal distribution, as
\begin{equation}
\begin{split}
\sr_x = P\left(-\frac{W}{2} \leq X \leq \frac{W}{2}\right) &= \int^{\frac{W}{2}}_{-\frac{W}{2}}p(x)dx \\&= \frac{1}{2}\left[\text{erf}\left(\frac{\frac{W}{2}-\mu_x}{\sigma_x\sqrt{2}}\right)-\text{erf}\left(\frac{-\frac{W}{2}-\mu_x}{\sigma_x\sqrt{2}}\right)\right].
\end{split}
\label{formula:DualGaussianX}
\end{equation}
where $p(x)$ is the probability density function of the normal distribution, $\mathrm{erf}(\cdot)$ is the Gaussian error function, and $\mu_x$ is the mean of the tap-coordinates along the $x$-axis.
$\sigma_x$ is derived from \autoref{formula:Bi_sigma}.
It is known that $\mu_x$ tends to be near the target center \citep{Henze2011LargeExperiment,Azenkot2012TouchBehavior,Bi2016DualGaussian}. 
Therefore, by substituting $\mu_x=0$ as a reasonable approximation, \autoref{formula:DualGaussianX} is simplified to
\begin{equation}
\begin{split}
\sr_x &= \frac{1}{2}\left[\text{erf}\left(\frac{\frac{W}{2}}{\sigma_x\sqrt{2}}\right)-\text{erf}\left(\frac{-\frac{W}{2}}{\sigma_x\sqrt{2}}\right)\right] = \text{erf}\left(\frac{W}{2\sqrt{2}\sigma_x}\right).
\end{split}
\label{formula:DualGaussianXeasy}
\end{equation}
A similar derivation applies to a 1D target whose height is constrained to $H$ along the $y$-axis. 
Thus, the success rate for 1D tapping is obtained by \autoref{formula:DualGaussian1D}:
\begin{equation}
\sr_x = \text{erf}\left(\frac{W}{2\sqrt{2}\sigma_x}\right)\mathrm{,}\ \ \sr_y = \text{erf}\left(\frac{H}{2\sqrt{2}\sigma_y}\right).
\label{formula:DualGaussian1D}
\end{equation}
Experimental results have demonstrated that this model predicts $\sr$ for 1D targets with high accuracy \citep{Bi2016DualGaussian,Yamanaka2024ISS}.
For a rectangular 2D target, the integration range can be independently defined as $-\frac{W}{2} < X < \frac{W}{2}$ and $-\frac{H}{2} < Y < \frac{H}{2}$.
Since the distributions along the $x$ and $y$ axes are known to be independent \citep{Bi2016DualGaussian,Yamanaka2020Rethinking}, the success rate $\sr_{rect}$ is
\begin{equation}
\begin{split}
\sr_{rect} = \sr_x \times \sr_y = \text{erf}\left(\frac{W}{2\sqrt{2}\sigma_x}\right) \times \text{erf}\left(\frac{H}{2\sqrt{2}\sigma_y}\right).
\end{split}
\label{formula:DualGaussianRectangle}
\end{equation}
This model has been experimentally shown to predict $\sr_{rect}$ accurately \citep{Usuba2022ER1Dto2D,Yamanaka23chiTuning}.

Furthermore, it has been demonstrated that modified models based on the Dual Gaussian Distribution Model are applicable to various scenarios.
Examples include 1D mouse-based pointing \citep{Yamanaka2021Crowdsource}, moving targets \citep{Huang2018Moving1D,Park2018,Lee18cue}, arbitrary target shapes \citep{Zhang2020ArbitraryShapedMovingTarget}, virtual reality \citep{Yu2019ERModelVR}, on-screen start tasks \citep{Yamanaka2020Rethinking}, different finger movement directions \citep{Ma2021RotationalDualGaussian}, displays with latency \citep{Yu2023TemporalERSpatialCorrespondence}, and adaptation to users and situations \citep{Zheng2021Scenario}.

\subsection{Pointing Near Screen Edges}
In indirect pointing (e.g., mouse), the screen edge is known as a ``magic pixel'' to afford fast and accurate selection \citep{tognazzini1999ask}.
This is because the edge stops the cursor, making the target size effectively infinite and preventing overshoot \citep{yamanaka2018sameEdge,Huot11torus}.
Experiments have shown that edge targets substantially reduce movement time compared to non-edge pointing \citep{Farris2001EdgeApplication,Dizmen2014EdgeNonEdge}, by as much as \(44\%\) \citep{Appert2008EdgePointing}.

With stylus-based direct pointing, similar benefits can be exploited. 
For devices where the screen is inset within the physical frame, a stylus can physically collide with the frame edge to prevent overshoot and improve time and accuracy \citep{Froehlich2007BarrierPointing,Wobbrock2003EdgeWrite}.

For finger-based touch pointing, edge targets have been suggested to degrade performance.
\cite{Avrahami2015EdgeTouch} and \cite{Usuba2023EdgeTarget} showed that smaller target–edge gaps increase $\mt$s.
A large-scale study (over 120 million taps) using an Android game app reported that targets near the edges are less accurate \citep{Henze2011LargeExperiment}.
When nearby distractors exist, $\mt$ and $\sr$ worsen, and tap distributions deviate from the normal distribution \citep{Yamanaka2018PenalDistractors,Yamanaka2018SurroundingDistractors,Yamanaka2019DistractorCrowdSource}.
Based on these findings, prior work has offered quantitative guidelines, e.g., distancing targets \(\geq\) 4 mm from edges/distractors \citep{Avrahami2015EdgeTouch,Usuba2023EdgeTarget,Yamanaka2018PenalDistractors}.
However, we found no studies that model how target–edge distance affects pointing accuracy.
Such a model would enable UI designs that also exploit edge-adjacent areas efficiently.

\section{Proposed Model: Skewed Dual Normal Distribution Model}
\subsection{Model Overview}
\begin{figure}[ht]
\centering
\begin{minipage}[b]{0.23\textwidth}
    \centering
    \includegraphics[height = 3.4cm]{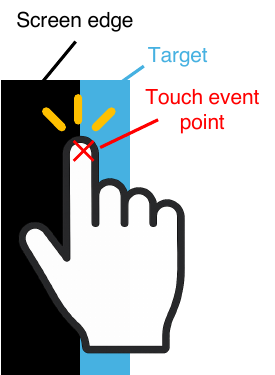}
    \subcaption{Touch event near edge}
    \label{fig:FingerEdge}
\end{minipage}
\begin{minipage}[b]{0.50\textwidth}
    \centering
    \includegraphics[height = 3.4cm]{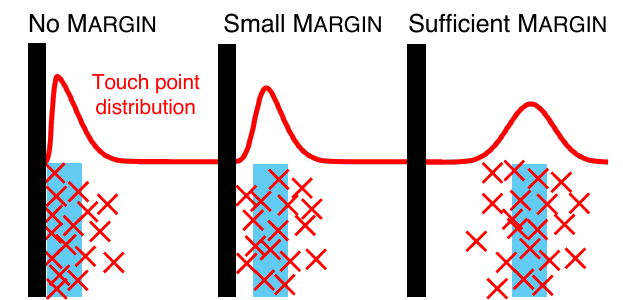}
    \subcaption{Distributional changes with edge distance}
    \label{fig:DistributionChange}
\end{minipage}
\begin{minipage}[b]{0.25\textwidth}
    \centering
    \includegraphics[height = 3.4cm]{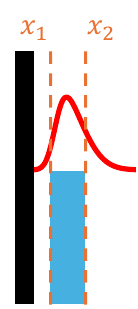}
    \subcaption{Computing \(\sr\) by CDF}
    \label{fig:CalculateSR by CDF}
\end{minipage}
\caption{(a) Our model assumes that a touch that would have occurred outside the screen edge is triggered inside the screen due to finger thickness. (b) Thus, as the target approaches the edge, the tap-coordinate distribution becomes more skewed. (c) We compute \(\sr\) using the skew-normal CDF.}
\label{fig:proposedModel}
\end{figure}

While the baseline Dual Gaussian Distribution Model \citep{Bi2016DualGaussian} accurately estimates $\sr$ for screen-center targets by assuming a normal tap distribution, as we later show, this assumption fails near edges where tap events cannot occur off-screen.
Similar to how distractors deviate tap points from normality \citep{Yamanaka2018PenalDistractors}, screen edges act as constraints that skew the distribution.

In contrast, our model assumes that tap-coordinates follow a skew-normal distribution, as touches intended for a smartphone bezel are triggered just inside the edge (\autoref{fig:FingerEdge}). 
Consequently, the distribution becomes increasingly skewed as the target approaches the edge (\autoref{fig:DistributionChange}).
By modeling this skewness, we can predict the distribution shape accurately and estimate $\sr$ using the skew-normal cumulative distribution function (CDF) (\autoref{fig:CalculateSR by CDF}).
Since the skew-normal distribution reverts to a normal distribution when the skewness parameter is zero, our model generalizes the Dual Gaussian Distribution Model for the entire screen.

\begin{figure}[ht]
\centering
\begin{minipage}[b]{0.58\textwidth}
    \centering
    \includegraphics[height = 3.0cm]{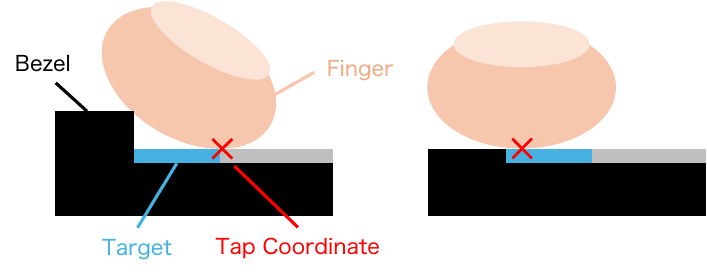}
    \subcaption{The difference between raised and flat bezels}
    \label{fig:FingerBezel}
\end{minipage}
\begin{minipage}[b]{0.40\textwidth}
    \centering
    \includegraphics[height = 3.0cm]{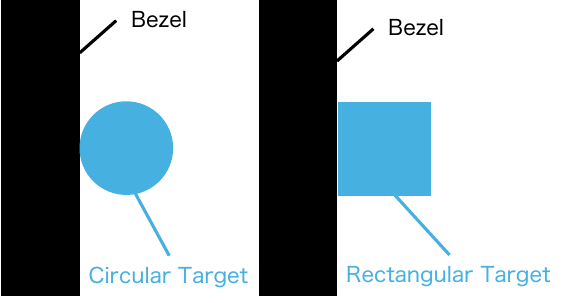}
    \subcaption{The difference due to target shape}
    \label{fig:CircleRectangle}
\end{minipage}
\caption{(a) On devices with a raised bezel, tapping the area immediately adjacent to the bezel is physically difficult. In contrast, a modern flat bezel allows users to contact the edge-adjacent area without physical interference. (b) For the circular targets used in prior studies, the target and screen edge contact only at a single point, making edge tapping an irrational strategy. Conversely, with rectangular targets, the target and bezel share a linear boundary. In this case, tapping near the edge becomes a rational strategy to minimize potential errors occurring toward the screen center.}
\label{fig:FingerBezelCircleRectangle}
\end{figure}

Our model assumes increased tap frequency near screen edges, seemingly contradicting previous findings that users avoid these areas \citep{Henze2011LargeExperiment,Avrahami2015EdgeTouch,Usuba2023EdgeTarget}.
However, in Avrahami's study, physically raised frames placed on tablets made edge tapping difficult (\autoref{fig:FingerBezel}).
Furthermore, their use of circular targets, which contact edges only at a single point, might have discouraged edge tapping (\autoref{fig:CircleRectangle}).
In comparison, our model focuses on flat bezels adopted by most modern smartphones in recent years and rectangular targets.
Since the screen edge and target meet along a line, tapping both simultaneously becomes a rational strategy to minimize errors toward the screen center.
Our experimental results confirmed higher $\sr$ for edge-adjacent targets, with qualitative feedback supporting this strategy.
Note, however, that the model may not apply to thick protective cases or raised bezels that hinder access to the edge-adjacent area, or when UI targets do not share a linear boundary with the edge.

\subsection{Probability of a Successful Tap}
For the rectangular targets addressed in this study, if the distributions along the $x$- and $y$-axes are independent, the overall success rate $\sr$ is the probability that the tap-coordinates fall within the target boundaries on both the $x$- and $y$-axes:
\begin{equation} 
\begin{split}
    \sr = \sr_x \times \sr_y.
\end{split} 
\label{formula:DualSkewRect} 
\end{equation}
Here, $\sr_x$ and $\sr_y$ represent the probabilities that the tap-coordinates fall within the target on the $x$- and $y$-axes, respectively.
When the tap-coordinate distribution on each axis follows a skew-normal distribution, the probability of the tap falling within the target can be calculated using the skew-normal CDF (\autoref{fig:CalculateSR by CDF}).
For a target size $S$ along a given axis $A$ (either $x$ or $y$), the success rate $\sr_{A}$ on that axis is
\begin{equation}
\sr_{A} = P\left(-\frac{S}{2} \leq A \leq \frac{S}{2}\right) = \int_{-S/2}^{S/2}f(a)da = \int_{-\infty}^{S/2}f(a)da - \int_{-\infty}^{-S/2}f(a)da.
\label{Formula:-S/2toS/2}
\end{equation}
Here, $\int_{-\infty}^{a}f(t)dt$ is the skew-normal CDF up to $a$, and $f(a)$ is the skew-normal probability density function (PDF) at $a$.
The PDF of a skew-normal distribution is determined using the normal PDF ($\phi(z)$) and CDF ($\Phi(z)$) as
\begin{equation}
\begin{gathered}
    f(a) = \frac{2}{\omega}\phi\left(\frac{a-\xi}{\omega}\right) \Phi\left(\alpha\frac{a-\xi}{\omega}\right),\\
    \phi\left(z\right) = \frac{1}{\sqrt{2\pi}}\exp\left(-\frac{z^2}{2}\right), \ \ \Phi\left(z\right) = \int_{-\infty}^z\phi(t)dt,
\end{gathered}
\label{formula:skewNormalPDF}
\end{equation}
where $\omega$ is the scale, $\xi$ is the location, and $\alpha$ is the shape parameter representing the direction and magnitude of the skew.
The CDF can be transformed using the normal CDF ($\Phi(z)$) and Owen's T-function ($T(h_T,a_T)$):
\begin{equation}
\begin{gathered}
    \int_{-\infty}^{a}f(t)dt = \Phi\left( \frac{a-\xi}{\omega}\right)-2T\left(\frac{a-\xi}{\omega}, \alpha\right)\\
    T(h_T, a_T) = \frac{1}{2\pi}\int_{0}^{a_T}\frac{\exp\left(-\frac{1}{2}h_T^2\left(1+t^2\right)\right)}{1+t^2}dt.
\end{gathered}
\label{Formula:SkewNormalCDF}
\end{equation}
The normal CDF ($\Phi(z)$) can be expressed using the Gaussian error function $\text{erf}(\cdot)$:
\begin{equation}
    \Phi\left(z\right) = \frac{1}{2}\left(1+\text{erf}\left(\frac{z}{\sqrt{2}}\right)\right).
\label{Formula:PhiERF}
\end{equation}
By transforming \autoref{Formula:-S/2toS/2} using the above definitions, the success rate $\sr_{A}$ for a single axis is derived as
\begin{equation}
\begin{split}
\sr_A = &\left(\frac{1}{2}\left(1+\text{erf}\left(\frac{\frac{S}{2}-\xi}{\sqrt{2}\omega}\right)\right)-2T\left(\frac{\frac{S}{2}-\xi}{\omega}, \alpha\right)\right) \\
&- \left(\frac{1}{2}\left(1+\text{erf}\left(\frac{-\frac{S}{2}-\xi}{\sqrt{2}\omega}\right)\right)-2T\left(\frac{-\frac{S}{2}-\xi}{\omega}, \alpha\right)\right).
\end{split}
\label{Formula:DualSkewSR1D}
\end{equation}
For rectangular targets, the overall 2D success rate is obtained by calculating $\sr_x$ (substituting width $W$ for $S$) and $\sr_y$ (substituting height $H$ for $S$) via \autoref{Formula:DualSkewSR1D} and multiplying them together (\autoref{formula:DualSkewRect}).
Since $\text{erf}(\cdot)$ and Owen's T-function are supported by most programming languages, the proposed model is easily implementable.
In this study, we utilized the \texttt{erf} and \texttt{owens\_t} functions from \texttt{scipy.special}, which employ high-precision approximation algorithms \citep{Johnson2012faddeeva,Patefield2000OwenT}.

\subsection{Parameters of the Skew-Normal Distribution}
\subsubsection{Parameter Conversion}
To calculate $\sr$, we need to determine the parameters of the skew-normal distribution for each axis: $\omega$ (scale), $\xi$ (location), and $\alpha$ (shape parameter representing the direction and magnitude of the skew).
While it is possible to estimate these parameters from measured distributions using maximum likelihood estimation or the method of moments, the estimates can become unstable when the underlying distribution is close to a normal distribution \citep{Pewsey2000ProblemOfAzzalini}.
In the proposed model, we assume the tap-coordinate distribution is normal when there is no influence from edges or corners.
Therefore, we did not employ these estimation methods.

Furthermore, if we were to formulate a model that directly predicts the unique skew-normal parameters ($\omega$, $\xi$, and $\alpha$), it would involve predicting parameters different from those in the baseline Dual Gaussian Distribution Model, which predicts $\sigma^2$ from the target size alone (\autoref{formula:Bi_sigma}).
This leads to a lack of consistency across the screen.
To maintain consistency with the existing model, we instead model the variance $\sigma^2$, mean $\mu$, and skewness $\gamma_1$—which can be calculated directly from the tap-coordinate distribution—and transform them using \autoref{Formula:Delta}--\autoref{Formula:Xi}:
\begin{gather}
\delta = \mathrm{sign}(\gamma_1) \times \min\left(0.999,\sqrt{\frac{\pi}{2}\frac{|\gamma_1|^{2/3}}{|\gamma_1|^{2/3}+\left(\frac{4-\pi}{2}\right)^{2/3}}}\right) \label{Formula:Delta}\\
\alpha = \frac{\delta}{\sqrt{1-\delta^2}} \label{Formula:Alpha}\\
\omega = \frac{\sigma}{\sqrt{1-\frac{2\delta^2}{\pi}}} \label{Formula:Omega}\\
\xi = \mu - \omega \times \delta \times \sqrt{\frac{2}{\pi}} \label{Formula:Xi}
\end{gather}
Here, we set the upper limit of $|\delta|$ to $0.999$, as $\alpha$ is undefined when $|\delta| \ge 1$.
While this limits the degree of skewness that can be accounted for, this threshold was exceeded in only one condition across our three experiments.
This occurred when the target was adjacent to the bezel.
In such cases, even if the skewness is underestimated, the distribution on the edge side remains within the target area, making the impact on $\sr$ prediction accuracy minimal.

When there is no skewness ($\gamma_1=0$), the parameters become $\delta=0, \alpha=0, \omega=\sigma,$ and $\xi=\mu$, causing the model to revert to the Dual Gaussian Distribution Model.
Through this transformation, we can estimate $\sr$ by predicting values that are directly calculable from the tap-coordinate distribution: $\sigma, \mu,$ and $\gamma_1$.

\subsubsection{Predictive Models for each Parameter}
\begin{figure}[ht]
\centering
\begin{minipage}[b]{0.24\textwidth}
    \centering
    \includegraphics[height = 5.5cm]{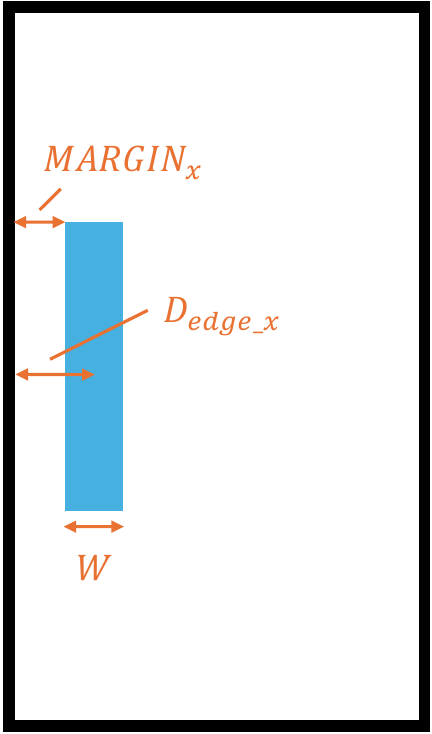}
    \subcaption{}
    \label{fig:ModelAndExperiment ParametersLeft}
\end{minipage}
\begin{minipage}[b]{0.24\textwidth}
    \centering
    \includegraphics[height = 5.5cm]{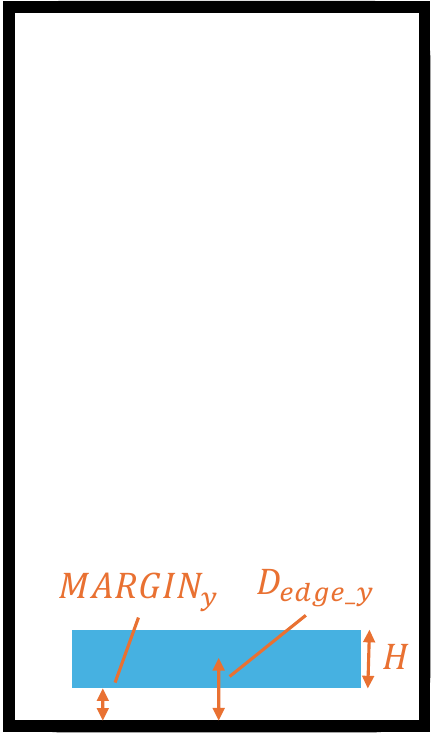}
    \subcaption{}
    \label{fig:ModelAndExperiment ParametersBottom}
\end{minipage}
\begin{minipage}[b]{0.24\textwidth}
    \centering
    \includegraphics[height = 5.5cm]{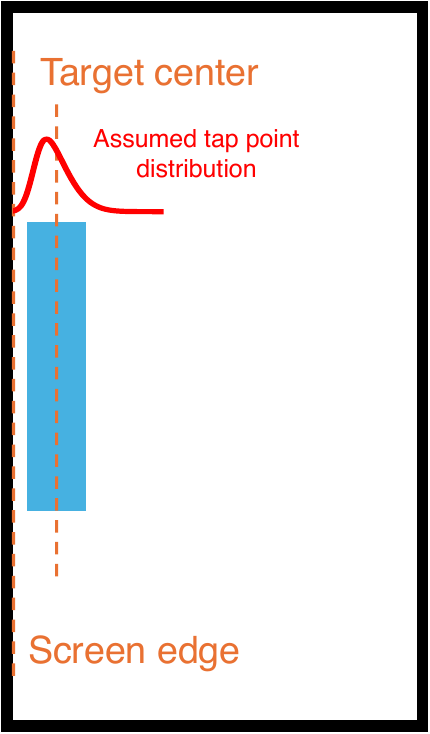}
    \subcaption{}
    \label{fig:EdgeAndTarget Left}
\end{minipage}
\begin{minipage}[b]{0.24\textwidth}
    \centering
    \includegraphics[height = 5.5cm]{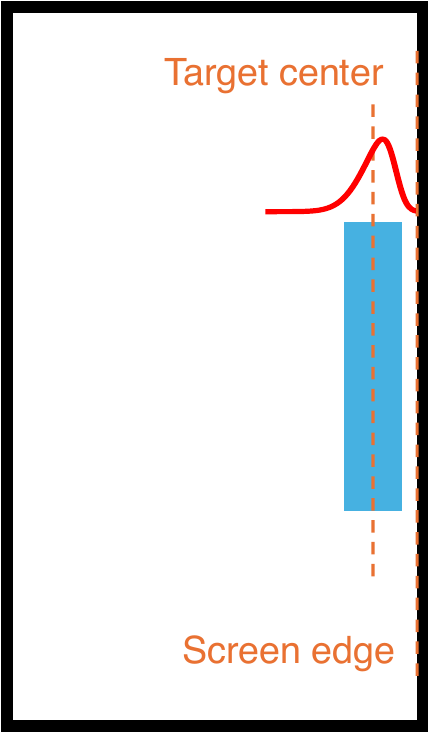}
    \subcaption{}
    \label{fig:EdgeAndTarget Right}
\end{minipage}
\caption{Parameters used in the proposed model and independent variables of the experiment: (a) For the $x$-axis, the distance from the screen edge to the target center is denoted as $\dedgex$, the distance from the edge to the target boundary as $\margin_x$, and the target width as $W$. (b) For the $y$-axis, $\dedgey$ and $\margin_y$ are defined analogously to $x$, and the target height as $H$. (c) When the screen edge is to the left of the target (negative direction), the distribution skews to the right (positive direction); the peak shifts left, and the tail extends to the right. (d) When the screen edge is to the right of the target (positive direction), the distribution skews to the left (negative direction); the peak shifts right, and the tail extends to the left. To encode the direction of skewness, the sign of $\gamma_1$ is calculated as $\text{sign}(\text{Target Center} - \text{Screen Edge})$.}
\label{fig:ModelAndExperimentParameters}
\end{figure}

As shown in \autoref{fig:ModelAndExperimentParameters}, the UI layout parameters are the target size ($W$ or $H$), the distance between the target and the screen edge ($\margin_x$, $\margin_y$, $\dedgex$, $\dedgey$), and the relative position between the target and the screen edge ($\text{Target Center} - \text{Screen Edge}$).
We model the $x$-axis parameters $\sigma^2_x$, $\mu_x$, and $\gamma_{1\_x}$, and the $y$-axis parameters $\sigma^2_y$, $\mu_y$, and $\gamma_{1\_y}$ as functions of these layout parameters.
In the following, for explanation, we describe the model derivations for the $x$-axis, and those for the $y$-axis can be developed in a similar manner.

\textbf{Skewness $\bm{\gamma_{1_x}}$}.
According to our model's assumption, the distribution skewness increases as the target approaches the screen edge and disappears when the target is sufficiently far from the edge (\autoref{fig:DistributionChange}).
Thus, as $\dedgex$ decreases, $|\gamma_{1\_x}|$ increases, and when $\dedgex$ exceeds a certain threshold, $\gamma_{1\_x}$ converges to 0.
Since the magnitude of the skewness should depend on the distance from the target center to the screen edge rather than the gap between the target boundary and the edge, we use $\dedgex$.
Additionally, the direction of the skewness changes based on the relative position of the adjacent screen edge.
Specifically, a positive $\gamma_{1}$ signifies positive skew where the peak shifts in the negative direction and the tail extends in the positive direction (\autoref{fig:EdgeAndTarget Left}), and a negative $\gamma_{1}$ signifies the opposite (\autoref{fig:EdgeAndTarget Right}).
Taking these factors into account, we modeled $\gamma_{1\_x}$ as
\begin{align}
\label{Formula:DualSkewNormalEstimateGamma1X}
\gamma_{1\_x} &= \mathrm{sign}(\text{Target Center} - \text{Screen Edge}) \times \max\left(0,\, c_x + d_x \times D_\mathit{edge\_x}\right),
\end{align}
where $c_x \ge 0$ and $d_x \le 0$ are regression constants.

\textbf{Variance $\bm{\sigma}^2_x$}.
Since tap events cannot occur off-screen, $\sigma^2_x$ should decrease as $\margin_x$ decreases when the target is near the screen edge.
Given that $\dedgex = \margin_x + W/2$, the influence of the distance to the screen edge is accounted for by $\margin_x$ to prevent the influence of $W$ from overlapping across multiple terms.
Furthermore, if there is no skewness in the distribution (i.e., the screen edge has no influence), the model should revert to the existing model (\autoref{formula:Bi_sigma}), which predicts $\sigma^2_x$ using only $W$.

Therefore, we set a threshold for $D_\mathit{edge\_x}$ at $-c_x/d_x$ (derived from \autoref{Formula:DualSkewNormalEstimateGamma1X} where $\gamma_{1\_x}=0$) and divided the model into two cases.
If $D_\mathit{edge\_x} < -\frac{c_x}{d_x}$, indicating that the distribution has skewness due to the screen edge, we use a model that accounts for the decrease in $\sigma^2_x$ caused by $\margin_x$.
If $D_\mathit{edge\_x} \ge -\frac{c_x}{d_x}$, indicating that the distribution is normal, we use the existing model as is.
We thus modeled $\sigma^2_x$ as follows.
\begin{equation}
\label{Formula:DualSkewNormalEstimateSigmaX}
\begin{split}
&D_\mathit{edge\_x} < -\frac{c_x}{d_x}:\ \ \sigma^2_x = e_x + f_x \times W^2 + g_x \times \margin_x\\
&D_\mathit{edge\_x} \ge -\frac{c_x}{d_x}:\ \ \sigma^2_x = h_x + i_x \times W^2
\end{split}
\end{equation}
Here, $e_x$, $f_x$, $g_x$, $h_x$, and $i_x$ are regression constants.

\textbf{Mean $\bm{\mu}_x$}.
In existing models, $\mu_x$ is approximated as $0$ because the average tap position is close to the target center \citep{Henze2011LargeExperiment}, although the mean tap-coordinate is influenced by various factors such as finger angle \citep{Holz2011UnderstandingTouch}, target position on the screen \citep{Azenkot2012TouchBehavior,Henze2011LargeExperiment}, and gripping posture \citep{Lehmann2018HowToHold}.
Following the existing model, we use $\mu_x=0$ as a simplification when $\dedgex \ge -\frac{c_x}{d_x}$, i.e., when the skewness can be discarded.

\begin{figure}[t]
\centering
    \includegraphics[width = 0.95\columnwidth]{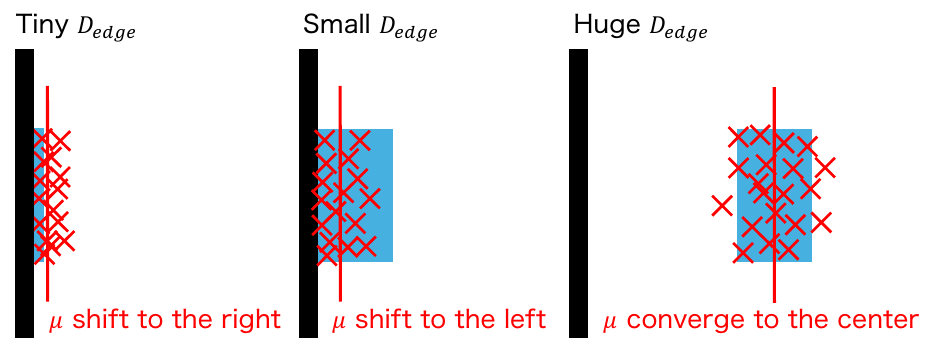}
\caption{Hypothesized effects of $\dedgex$ on $\mu_x$. (Left) When $\dedgex$ is minimal, $\mu_x$ shifts to the right (positive direction) relative to the target center. (Center) As $\dedgex$ increases with the expansion of $W$, $\mu_x$ shifts to the left (negative direction) relative to the target center, assuming users adopt a strategy of tapping the bezel and the target simultaneously to reduce errors on the opposite side. (Right) When $\dedgex$ increases sufficiently and the influence of the screen edge disappears, $\mu_x$ converges to the target center.}
\label{fig:ModelingMu}
\end{figure}

For cases where $\dedgex < -\frac{c_x}{d_x}$, we derived the $\mu_x$ model following the steps shown in \autoref{fig:ModelingMu}.
First, consider the case where $\dedgex$ is extremely small.
In this situation, tap events cannot occur off-screen; thus, $\mu_x$ inevitably shifts in the direction away from the screen edge (rightwards in \autoref{fig:ModelingMu} left).
Next, consider the case where $\dedgex$ increases as the target expands.
Assuming the scenario where users adopt a strategy of tapping the boundary between the screen edge and the target to reduce errors in the direction opposite to the edge, $\mu_x$ is thought to shift toward the screen edge relative to the target center (i.e., leftwards in \autoref{fig:ModelingMu} center).
Finally, as the target moves sufficiently far from the screen edge, $\mu_x$ should converge to the target center.

In summary, $\mu_x$ is expected to transition from a positive value through a negative value and finally converge to 0 as $\dedgex$ increases.
Based on this non-monotonic transition, we modeled $\mu_x$ using a quadratic function.
\begin{equation}
\label{Formula:DualSkewNormalEstimateMuX}
\begin{split}
&\dedgex < -\frac{c_x}{d_x}:\ \ \mu_x = j_x + k_x\left(\dedgex - l_x\right)^2 \\
&\dedgex \ge -\frac{c_x}{d_x}:\ \  \mu_x = 0
\end{split}
\end{equation}
Here, $j_x$, $k_x$, and $l_x$ are regression constants.

\subsection{Process to Use Our Model}
For a 1D target, the proposed model predicts $\sr_x$ through the following procedure.
First, estimate $\gamma_{1\_x}, \sigma_x^2,$ and $\mu_x$ using \autoref{Formula:DualSkewNormalEstimateGamma1X}--\autoref{Formula:DualSkewNormalEstimateMuX}.
These values are transformed into the parameters of the skew-normal distribution via \autoref{Formula:Delta}--\autoref{Formula:Xi} and substituted into \autoref{Formula:DualSkewSR1D} to calculate $\sr_x$.
For a 2D rectangular target, calculate $\sr_x$ and $\sr_y$ (analogously to the $x$-axis) and then determine $\sr$ using \autoref{formula:DualSkewRect}.

The apparent mathematical complexity arises from the parameter transformations and the calculation of the CDF, while the prediction models themselves are simple regression models.
Simple and easy-to-understand mathematical models are crucial for UI practitioners to build UIs theoretically.
For example, in the proposed model, by calculating $-c_x/d_x$ from the obtained regression constants, one can understand the distance $D_\mathit{edge\_x}$ at which the influence of the screen edge disappears ($\gamma_{1\_x}=0$), allowing practitioners to decide whether to avoid or utilize that area.
Furthermore, one can understand the influence of $\margin_x$ on the variance of tap-coordinates through $g_x$, or the distance at which the shift of the mean toward the screen edge is maximized through $l_x$.

\section{Experiment 1: 1D Pointing near Left Edge}
We conducted Experiment 1 to verify whether the proposed model could accurately predict $\sr_x$.
To isolate and examine the influence of a single screen edge, we employed a 1D pointing task with horizontal constraints targeting the left screen edge.

\subsection{Research Ethics}
This study involved only smartphone operation tasks and did not collect sensitive personal data.
In accordance with the ethics guidelines of Meiji University, we confirmed that such procedures do not require formal review by the institutional review board.
Participants provided verbal informed consent and were informed of their right to withdraw at any time without penalty.
Data were stored and analyzed in anonymized form.
The same ethical protocols and data handling procedures were followed in Experiments 2 and 3.

\subsection{Experimental Setup}
\subsubsection{Task}
\begin{figure}[ht]
    \centering
    \includegraphics[width = 0.6\columnwidth]{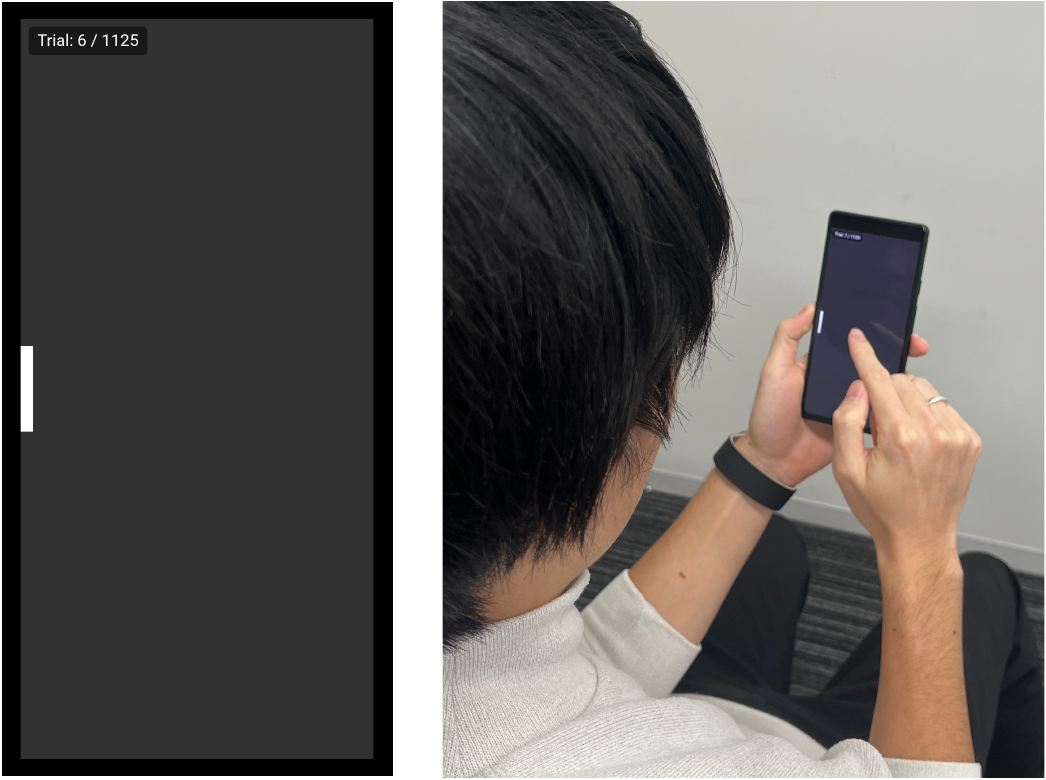}
    \caption{(Left) Screenshot of the experimental program used in Experiment 1. (Right) Participants performed the task while seated, holding the smartphone with their non-dominant hand and using the index finger of their dominant hand. Between each trial, they returned their dominant hand to their knee.}
    \label{fig:Ex1 Setup}
\end{figure}

The independent variables were the target width ($W$) and the distance from the left screen edge ($\margin_x$).
Participants performed an off-screen-start pointing task (a pointing task starting from outside the screen) to remain consistent with the initial validation of the Dual Gaussian Distribution Model \citep{Bi2016DualGaussian}.
They held the smartphone with their non-dominant hand and pointed with the index finger of their dominant hand (\autoref{fig:Ex1 Setup}).
We restricted the operation to the index finger of the dominant hand to avoid potential issues in one-handed thumb operation, e.g., reachability due to thumb length \citep{Perry2008Thumb,BergstromLehtovirta14}, which might prevent the isolation and verification of the influence exerted solely by the screen edges.

Following previous studies, for the first tap of each trial, pointing success or error was determined based on the coordinates at the moment the finger was lifted \citep{Bi2016DualGaussian,Yamanaka2020Rethinking}.
Participants were instructed to tap the target ``as quickly and accurately as possible.''
When an error occurred, the target flashed yellow, and the participant had to repeat the tap until successful.
Upon success, the screen went black for 0.5 seconds.
For each tap, audio feedback indicating success or error was played.

Participants were instructed to return their dominant hand to their knee during the blackout period and then tap the next target.
This instruction was given as a guideline to ensure that the hand was completely removed from the screen.
This might have increased three-dimensional hand movement compared to previous research, where participants were instructed to place their hand in a ``natural and comfortable position outside the screen'' \citep{Bi2016DualGaussian}.
However, in this study, we prioritized ensuring that the task was strictly an off-screen-start pointing task by controlling the starting position with an explicit guideline.

\subsubsection{Design}
The experiment followed a within-subjects design with $9\margin_x\times5W$.
The $\margin_x$ values were 0, 1.560, 3.119, 4.679, 7.798, 9.358, 12.477, 15.596, and 18.715 mm, and the $W$ values were 1.560, 2.339, 3.119, 4.679, and 7.798 mm\footnote{Integer pixel values were used for displaying targets, and millimeter values are reported here with necessary precision for reproducibility.\label{foot:ReasonMM}} (\autoref{fig:ModelAndExperimentParameters}).

The target height was fixed at 15.596 mm across all conditions.
Previous studies have shown that targets larger than 7.62 mm \citep{Bi2016DualGaussian} or 9 mm \citep{Yamanaka2020Rethinking} achieve an $\sr$ of over 99\%.
By setting the height larger than these thresholds, we simulated a pseudo-1D target with horizontal constraints.
While using a target of infinite height could have represented an ideal 1D target, the top and bottom screen edges might have introduced unintended influences.
For example, different participants might have tapped near the top or the center of the screen, potentially leading to inconsistent strategies.
Since it is known that sufficiently high targets can be treated as 1D targets on the $x$-axis \citep{Accot03BivariatePointing, fitts1954information}, we adopted a fixed height of 15.596 mm to validate the 1D model while excluding such unintended effects.
In fact, the number of errors above or below the target on the $y$-axis was only 29 out of 16,200 trials (0.179\%).

\subsubsection{Participants}
Fifteen computer science students (4 females and 11 males; mean age 21.3 years, SD 1.73 years; all right-handed) participated in Experiment 1.
The participation reward was 1,800 JPY (12.11 USD).
Participants received a briefing on the study and provided informed consent before beginning the experiment.

\subsubsection{Apparatus}
We used a Google Pixel 6a (body size: $152.2 \times 71.8 \times 8.9$ mm, display size: $142.5 \times 64.1$ mm, resolution: $2400 \times 1080$ px) without a case.
This smartphone features a thin bezel that curves toward the back, which we expected would not obstruct taps adjacent to the screen edges (\autoref{fig:FingerBezel}).
The experimental program was implemented using HTML/CSS/JavaScript and displayed in Google Chrome in full-screen mode.

\subsubsection{Procedure}
One \textit{set} consisted of 45 target conditions ($9 \margin_x \times 5 W$) presented in random order.
Participants completed 25 sets, with the first set being practice.
We recorded 16,200 data-collection trials ($9 \margin_x \times 5 W \times 24 \text{ sets} \times 15 \text{ participants}$).

Participants were instructed to sit and perform the task in a free posture, without resting their elbows on the desk or armrests.
Participants were allowed to take breaks at any time; additionally, a mandatory 30-second break was implemented every five sets.
After completing the task, we collected demographic data (gender and age) and open-ended feedback regarding their operational strategies.
All participants completed the experiment within 30 to 60 minutes.

\subsection{Results}
\subsubsection{Outliers}
Among the 16,200 trials, we excluded 29 trials (0.179\%) where errors occurred in the $y$-axis direction to focus on $\sr_x$.
Then, following previous research \citep{SOUKOREFF2004751}, we excluded 97 trials (0.600\%) with $x$ coordinates more than three standard deviations from the mean for each task condition and each participant.
The exclusion rate fell within the range reported in several previous studies (approximately 0.5\% to 7.0\%) \citep{Yamanaka2021Crowdsource, Yamanaka2024ISS, Komarov2013Crowdsourcing, Schwab2019PanZoom}.
The following analysis was conducted using the remaining 16,074 trials.

\subsubsection{Tests for the Normality of Tap-Coordinate Distributions.}
\begin{figure}[t]
    \centering
    \includegraphics[width = 0.55\columnwidth]{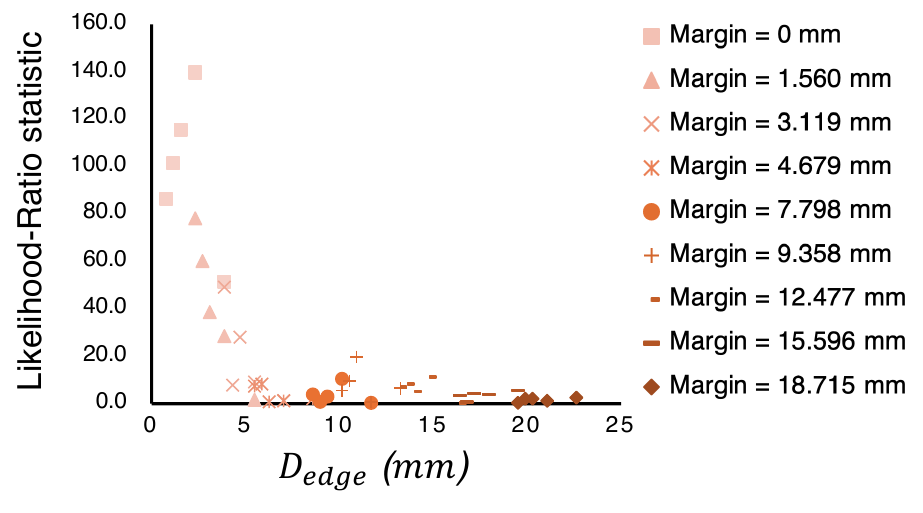}
    \caption{Likelihood ratio statistics in Experiment 1. As $D_\mathit{edge\_x}$ decreases (i.e., the target center approaches the screen edge), the tap-coordinate distribution follows a skew-normal distribution more closely than a normal distribution.}
    \label{fig:Ex1 LikelihoodRatio}
\end{figure}
While existing models assume a normal distribution for tap-coordinates, the Shapiro--Wilk tests at $\alpha=.05$ rejected normality in 41 out of 45 target conditions (91.1\%), suggesting limitations in the assumptions of existing models.
Furthermore, the likelihood ratio test comparing the fits of the normal and skew-normal distributions showed that the test statistics increased as $\dedgex$ decreased (\autoref{fig:Ex1 LikelihoodRatio}).
Specifically, the number of conditions showing a significant difference increased as $\margin_x$ decreased: one out of five width conditions at $\margin_x=7.798$ mm, two at $4.679$ mm, four at $3.119$ and $1.560$ mm, and all five at $\margin_x=0$ mm.
These results indicate that the fit for the skew-normal distribution improves as the target approaches the screen edge, supporting the assumption of the proposed model.

\subsubsection{Statistical Tests}
We conducted a two-way RM-ANOVA with independent variables of $\margin_x$ and $W$.
The dependent variables were success rate ($\sr_x$), standard deviation ($\sigma_x$, the square root of the variance $\sigma_x^2$), mean ($\mu_x$), and skewness ($\gamma_{1\_x}$).
When significant main effects or interactions were found with $\alpha=.05$, we performed pairwise comparisons using Bonferroni correction.
In the following figures, error bars represent 95\% confidence intervals.
Detailed ANOVA results are provided in the Supplementary Materials.
In the main text, we briefly summarize the effects of the independent variables on each dependent variable.

\begin{figure}[ht]
\centering
\begin{minipage}[b]{0.49\textwidth}
    \centering
    \includegraphics[width = 0.95\columnwidth]{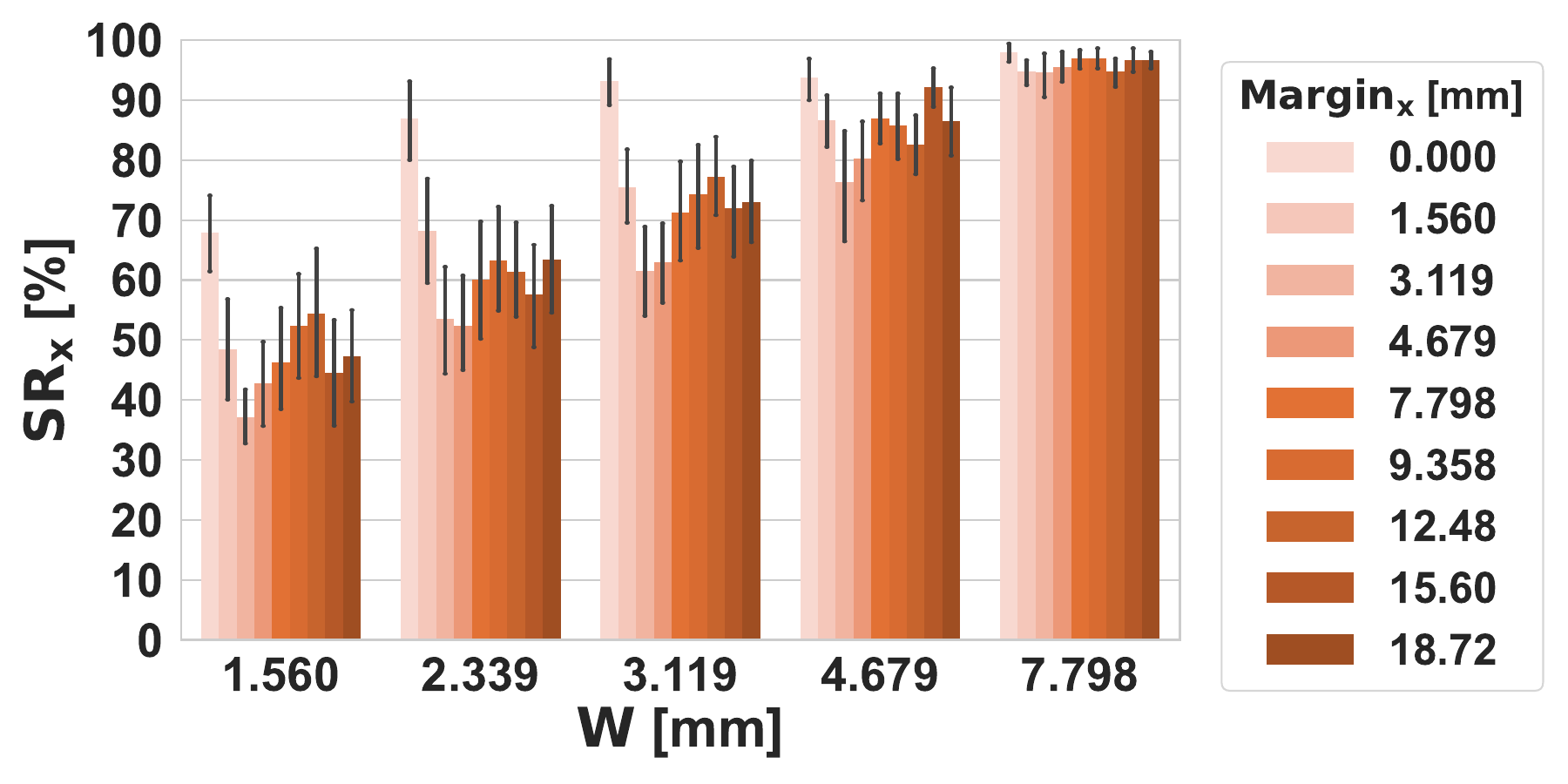}
    \subcaption{$\sr_x$}
    \label{fig:Ex1 SRx Interaction}
\end{minipage}
\begin{minipage}[b]{0.49\textwidth}
    \centering
    \includegraphics[width = 0.95\columnwidth]{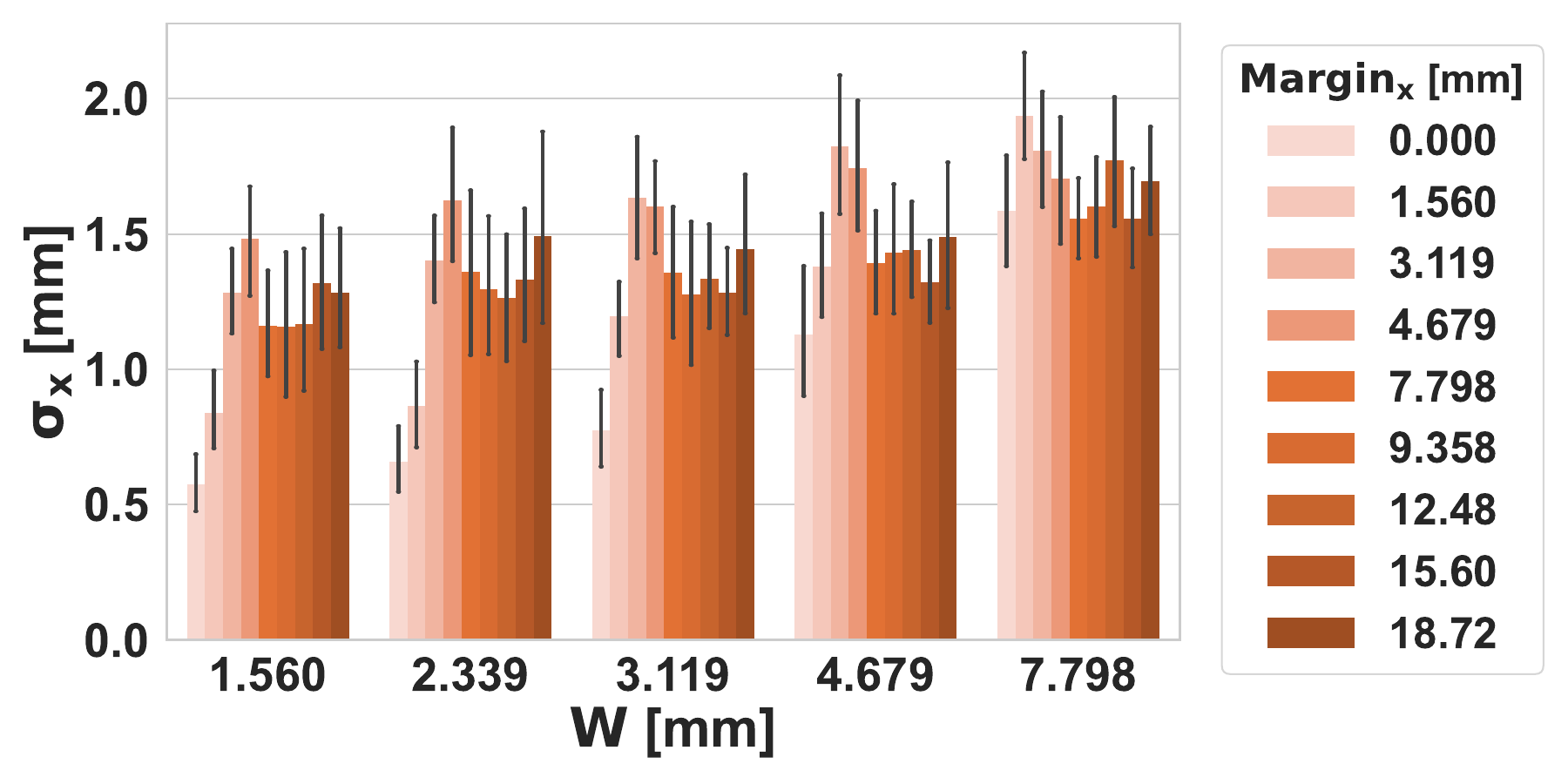}
    \subcaption{$\sigma_x$}
    \label{fig:Ex1 SigmaX Interaction}
\end{minipage}
\begin{minipage}[b]{0.49\textwidth}
    \centering
    \includegraphics[width = 0.95\columnwidth]{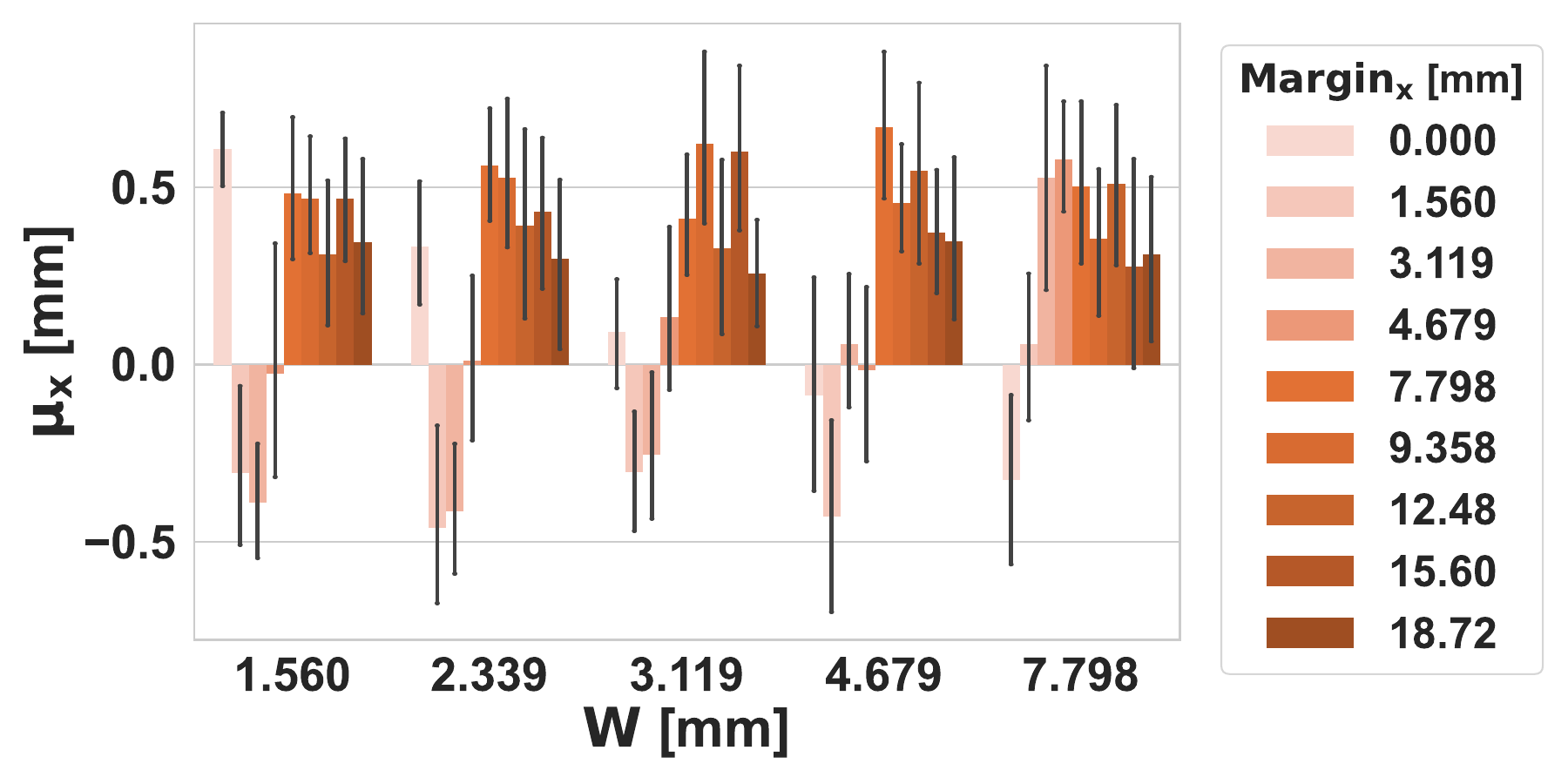}
    \subcaption{$\mu_x$}
    \label{fig:Ex1 MuX Interaction}
\end{minipage}
\begin{minipage}[b]{0.49\textwidth}
    \centering
    \includegraphics[width = 0.95\columnwidth]{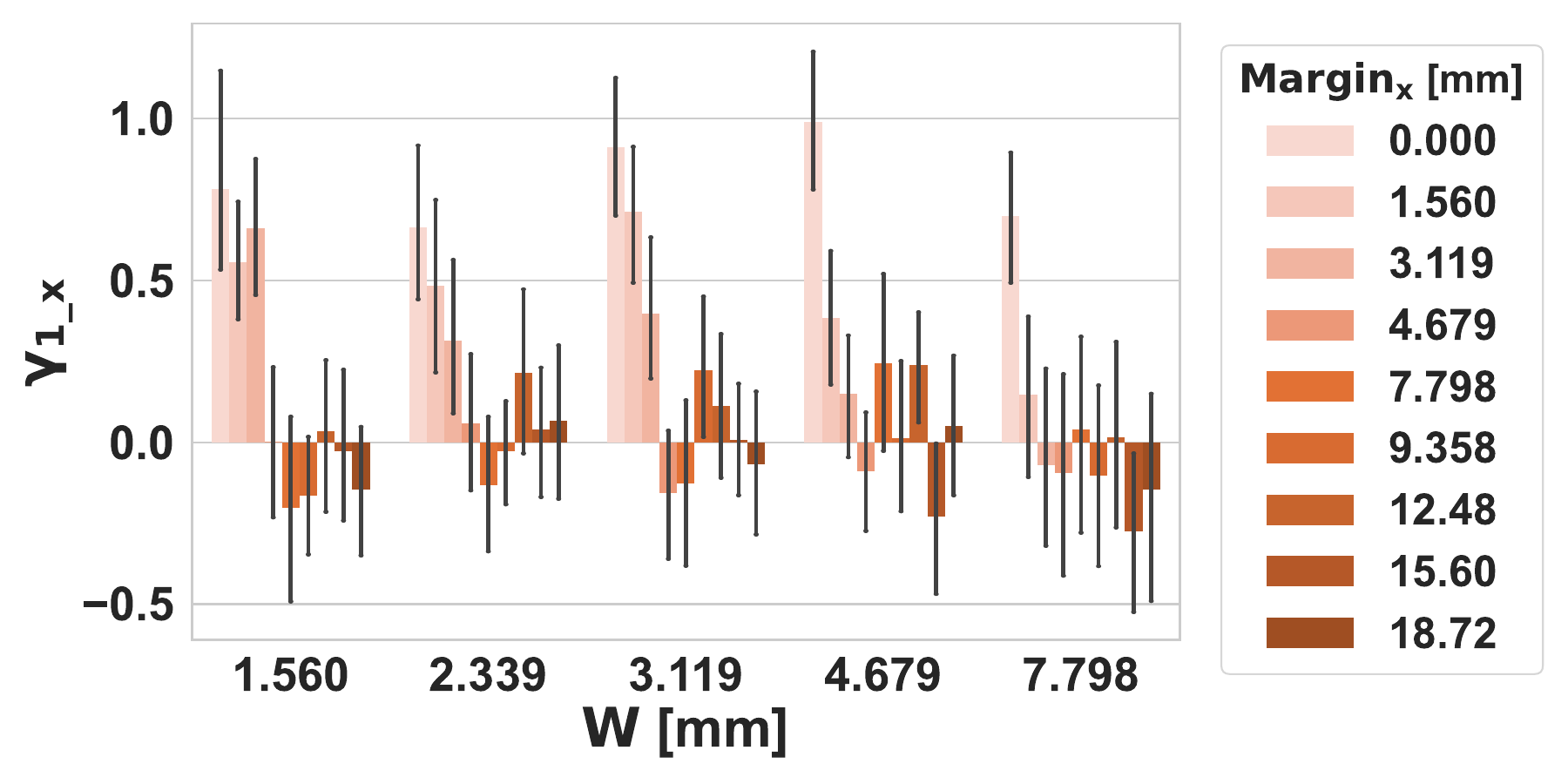}
    \subcaption{$\gammax$}
    \label{fig:Ex1 Gamma1X Interaction}
\end{minipage}
\caption{Effects of $\margin_x$ and $W$ on $\sr_x$, $\sigma_x$, $\mu_x$, and $\gammax$ in Experiment 1.}
\label{fig:Ex1_Interaction}
\end{figure}

All dependent variables were affected by $\margin_x$, suggesting limitations in existing models that estimate the $\sr_x$ based solely on $W$ (\autoref{fig:Ex1_Interaction}).
The high $\sr_x$ at $\margin_x = 0$ supports our hypothesis: for targets adjacent to the screen edge, users adopt a strategy of tapping both the bezel and the target simultaneously to minimize errors occurring in the direction opposite to the bezel (\autoref{fig:Ex1 SRx Interaction}, \autoref{fig:FingerBezelCircleRectangle}).
As $\margin_x$ decreased, $\sigma_x$ increased and then decreased, $\mu_x$ exhibited a quadratic trend, and $\gamma_{1\_x}$ increased.
All of these trends were consistent with the hypotheses of the proposed model.
However, contrary to the assumption that $\mu$ could be approximated to 0 in regions sufficiently far from the screen edge, positive values were observed.

\subsubsection{Prediction Accuracy of Tap Success Rate}
We compared our model's prediction accuracy for $\sr_x$ with that of the baseline Dual Gaussian Distribution Model.
We also examined  common machine learning (ML) models capable of learning complex non-linear relationships, including the influence of the screen edge.

\begin{table}[ht]
\caption{Regression constants and model accuracy metrics in Experiment 1.\protect\footnotemark\ The number of parameters in ML models refers to the quantity of parameters used to generate predictions.}
\centering
\resizebox{\textwidth}{!}{
\begin{tabular}{c|c|c|c|c|c|c|c|c}
\multicolumn{2}{c|}{} & \multicolumn{5}{c}{Regression Analysis} & \multicolumn{2}{|c}{LOOCV}\\
\hline
Model & Equation & Constants & $R^2$ & $\mae$ & $ \rmse$ & $\mape$ & $R^2$ & $\mae$\\
\hline
Dual Gauss. & $\sigma_x^2$ (\autoref{formula:Bi_sigma}) & $a_x = 1.50$, $b_x = 0.0236$ & $.437$ & $0.426$ & 0.574 & $37.6\%$ & $.390$ & $0.445$\\
\cline{2-9}
& $\sr_x$ (\autoref{formula:DualGaussian1D}) & - & $.816$ & $5.44$ & $7.75$ & $8.11\%$ & $.807$ & $5.56$\\
\hline
Skewed-Dual. & $\gamma_{1_x}$ (\autoref{Formula:DualSkewNormalEstimateGamma1X}) & $c_x = 1.09$, $d_x = -0.170$ & $.789$ & $0.123$ & $0.149$ & $192\%$ & $.761$ & $0.130$\\
\cline{2-9}
 & $\sigma_x$ (\autoref{Formula:DualSkewNormalEstimateSigmaX}) & \begin{tabular}{c}$e_x = 0.155$, $f_x = 0.0461$,\\$g_x = 0.466$, $h_x = 1.60$,\\ $i_x = 0.0205$\end{tabular} & $.882$ & $0.0753$ & $0.102$ & $5.67\%$ & $.851$ & $0.0855$\\
\cline{2-9}
& $\mu_x$ (\autoref{Formula:DualSkewNormalEstimateMuX}) & \begin{tabular}{c}$j_x = -0.393$, $k_x = 0.108$,\\ $l_x = 3.73$\end{tabular} & $.905$ & $0.0752$ & $0.0895$ & $104\%$ & $.847$ & $0.0938$\\
\cline{2-9}
& $\sr_x$ (\autoref{Formula:DualSkewSR1D}) & - & $.950$ & $3.23$ & $4.05$ & $4.85\%$ & $.944$ & $3.39$\\
\hline
\multicolumn{9}{c}{Machine Learning Models with Default Hyperparameters}\\
\hline
Lasso Regression & $\sr_x$ & number of parameters: 3 & .743 & 7.36 & 9.17 & 11.3\% & .706 & 7.88 \\
\hline
Random Forest & $\sr_x$ & number of parameters: 5,708 & .987 & 1.48 & 2.09 & 2.24\% & .903 & 3.94 \\
\hline
SVR & $\sr_x$ & number of parameters: 45 & .213 & 13.8 & 16.0 & 21.9\% & .137 & 14.6 \\
\hline
MLP Neural Net & $\sr_x$ & number of parameters: 401 & $-$2.65 & 28.2 & 34.5 & 38.2\% & $-$2.64 & 28.4 \\
\hline
\multicolumn{9}{c}{Machine Learning Models with Bayesian-Optimized Hyperparameters}\\
\hline
Lasso Regression & $\sr_x$ & number of parameters: 3 & .714 & 7.35 & 9.67 & 11.9\% & .693 & 7.61 \\
\hline
Random Forest & $\sr_x$ & number of parameters: 8,445 & .925 & 3.67 & 4.95 & 5.71\% & .857 & 4.74 \\
\hline
SVR & $\sr_x$ & number of parameters: 44 & .993 & 0.652 & 1.54 & 1.07\% & .968 & 2.42 \\
\hline
MLP Neural Net & $\sr_x$ & number of parameters: 7,999 & .999 & 0.230 & 0.325 & 0.349\% & .966 & 2.62 \\
\hline
\end{tabular}
}
\label{table:Ex1 Model Regression}
\end{table}

\vspace{3mm}
\noindent\textit{Existing Model: Dual Gaussian Distribution Model}
\begin{figure}[ht]
\centering
\begin{minipage}[b]{0.49\textwidth}
    \centering
    \includegraphics[width = 0.95\columnwidth]{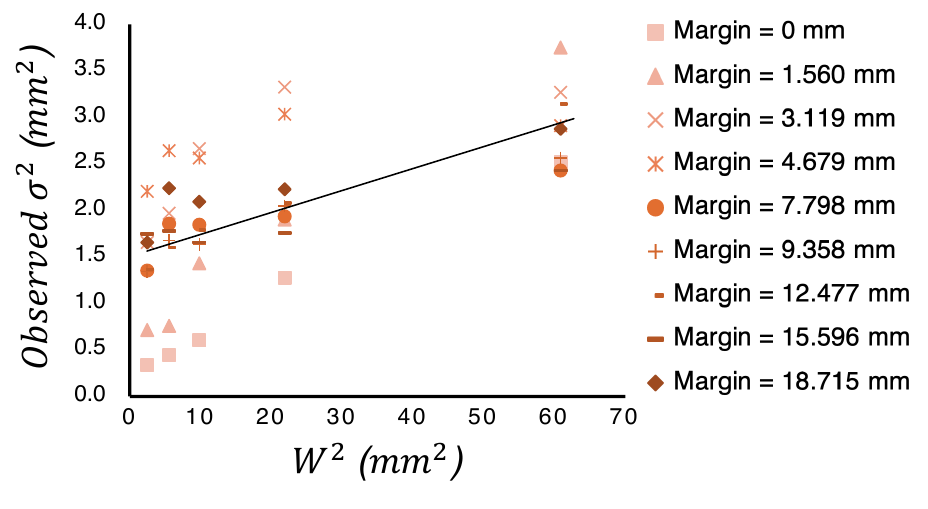}
    \subcaption{$W^2$ vs. $\sigma_x^2$}
    \label{fig:Ex1_DualGauss_W_vs_Sigma}
\end{minipage}
\begin{minipage}[b]{0.49\textwidth}
    \centering
    \includegraphics[width = 0.95\columnwidth]{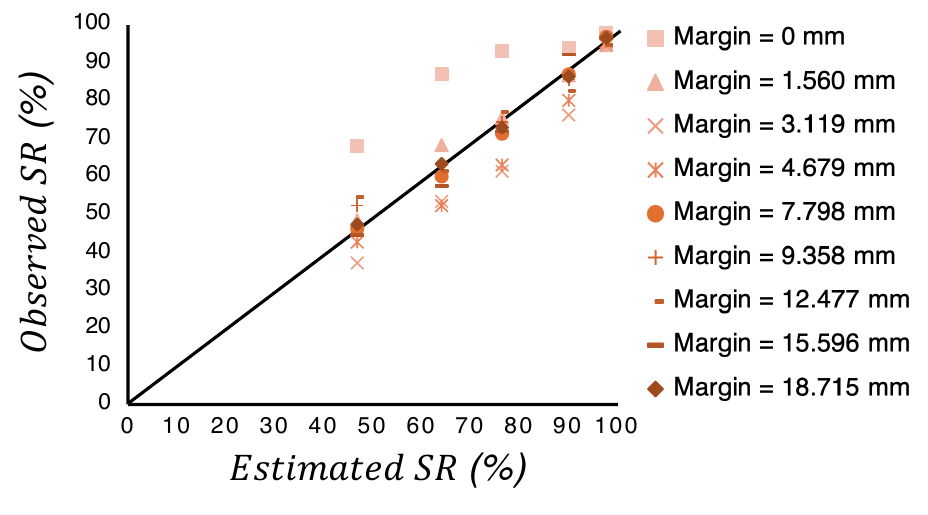}
    \subcaption{Predicted $\sr_x$ vs. Observed $\sr_x$}
    \label{fig:Ex1_DualGauss_SR_vs_SR}
\end{minipage}
\caption{Results regarding the $\sr_x$ prediction accuracy of the Dual Gaussian Distribution Model in Experiment 1. Discrepancies between observed values and model predictions were observed, particularly for small $\margin_x$, suggesting the need for a model that accounts for the distance from the screen edge. In (a), the line represents the regression line; in (b), the line represents the identity line where predicted and observed values match.}
\label{fig:Ex1_DualGauss}
\end{figure}
\\
\noindent As a baseline, we verified the $\sr_x$ prediction accuracy of the Dual Gaussian Distribution Model.
First, we performed a regression analysis on $\sigma_x^2$, averaged across participants for each condition, using \autoref{formula:Bi_sigma}.
The goodness of fit was moderate ($R^2=.437$, \autoref{table:Ex1 Model Regression}, \autoref{fig:Ex1_DualGauss_W_vs_Sigma}), with particularly large deviations observed in conditions where $\margin_x$ was small.
This suggests that a model considering only $W$ cannot fully capture the fluctuations in $\sigma_x^2$ caused by the screen edge.

\footnotetext[2]{While the $R^2$ of the MLP Neural Network is below $-1$, this is a possible outcome.\footnotemark MLP Neural Networks strongly require hyperparameter tuning and feature scaling; therefore, significantly low $R^2$ values may occur under default settings.\footnotemark}
\footnotetext[3]{\url{https://scikit-learn.org/stable/modules/generated/sklearn.metrics.r2_score.html}}
\footnotetext[4]{\url{https://scikit-learn.org/stable/modules/neural_networks_supervised.html}}

Next, we calculated $\sr_x$ by substituting the estimated $\sigma_x$ into \autoref{formula:DualGaussian1D} and conducted a regression analysis, yielding $R^2=.816$ (\autoref{table:Ex1 Model Regression}, \autoref{fig:Ex1_DualGauss_SR_vs_SR}).
Notably, at $\margin_x=0$, the observed $\sr_x$ significantly exceeded the predicted values.
These results indicate that the prediction accuracy of the existing model for $\sr_x$ decreases as the target approaches the screen edge, highlighting the necessity of a model adapted for predicting performance on edge-aligned targets.

To evaluate the generalization performance of the model, we conducted a Leave-One-Out Cross-Validation (LOOCV), where the model was trained on all but one task condition and then used to predict the excluded condition.
The $R^2$ and $\mae$ from the LOOCV were generally consistent with the results of the regression analysis using the full dataset, indicating that the model did not suffer from overfitting (\autoref{table:Ex1 Model Regression}).
This demonstrates that the model maintains a comparable level of predictive performance even for unverified conditions.

\vspace{3mm}
\noindent\textit{Proposed Model: Skewed Dual Normal Distribution Model}
\begin{figure}[ht]
\centering
\begin{minipage}[b]{0.49\textwidth}
    \centering
    \includegraphics[width = 0.95\columnwidth]{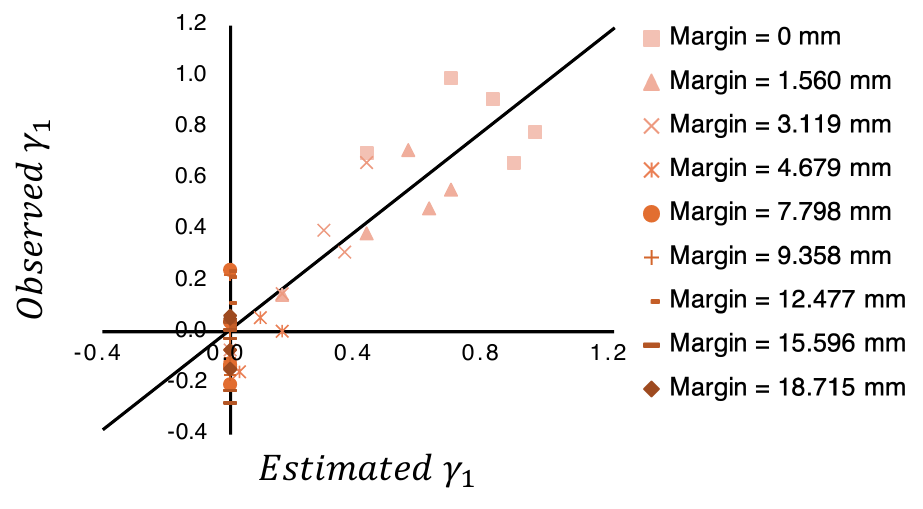}
    \subcaption{Predicted $\gamma_{1\_x}$ vs. Observed $\gamma_{1\_x}$}
    \label{fig:Ex1_DualSkew_Gamma}
\end{minipage}
\begin{minipage}[b]{0.49\textwidth}
    \centering
    \includegraphics[width = 0.95\columnwidth]{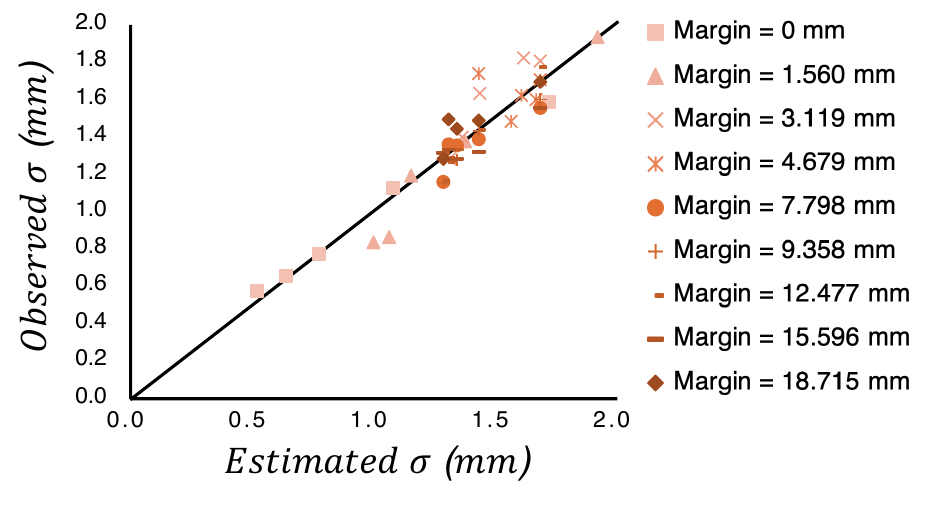}
    \subcaption{Predicted $\sigma_x$ vs. Observed $\sigma_x$}
    \label{fig:Ex1_DualSkew_Sigma}
\end{minipage}
\begin{minipage}[b]{0.49\textwidth}
    \centering
    \includegraphics[width = 0.95\columnwidth]{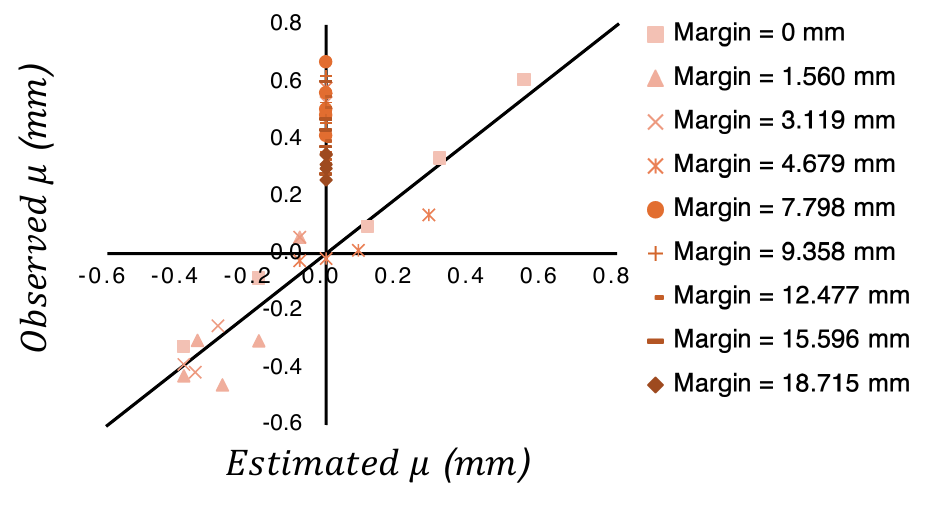}
    \subcaption{Predicted $\mu_x$ vs. Observed $\mu_x$}
    \label{fig:Ex1_DualSkew_Mu}
\end{minipage}
\begin{minipage}[b]{0.49\textwidth}
    \centering
    \includegraphics[width = 0.95\columnwidth]{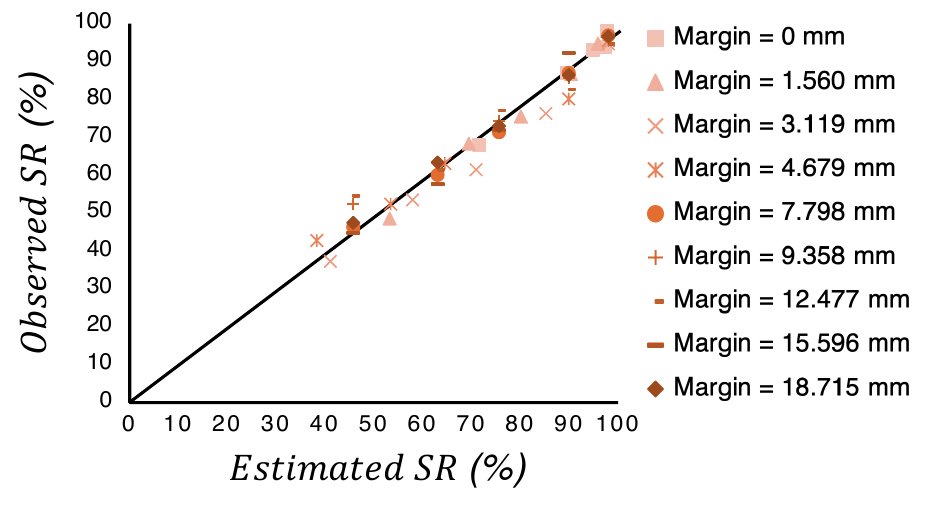}
    \subcaption{Predicted $SR_x$ vs. Observed $SR_x$}
    \label{fig:Ex1_DualSkew_SR}
\end{minipage}
\caption{Results of the Skewed Dual Normal Distribution Model's $\sr_x$ prediction accuracy in Experiment 1. The lines represent the ideal prediction ($X=Y$). (a) For small $\margin_x$, $\gamma_{1\_x}$ was well predicted. For large $\margin_x$, the results generally followed the assumption of $\gamma_{1\_x} \approx 0$, although slightly negative values were observed. (b) $\sigma_x$ was predicted accurately overall. (c) $\mu_x$ was accurately predicted when $\dedgex < -c_x/d_x$; however, contrary to the assumption of $\mu_x \approx 0$ elsewhere, positive $\mu_x$ values were observed. (d) $\sr_x$ was accurately predicted by the proposed model.}
\label{fig:Ex1_DualSkew}
\end{figure}

A regression analysis on $\gamma_{1\_x}$ using \autoref{Formula:DualSkewNormalEstimateGamma1X} yielded $R^2 = .789$ (\autoref{table:Ex1 Model Regression}, \autoref{fig:Ex1_DualSkew_Gamma}).
As hypothesized, $|\gamma_{1\_x}|$ increased as the target approached the screen edge (\autoref{fig:Ex1 Gamma1X Interaction}).
From the regression constants $c_x$ and $d_x$, we calculated the X-intercept of the regression line ($-c_x/d_x$) to be $6.40$.
Thus, our model predicts a skewed tap-coordinate distribution due to screen edge effects when $\dedgex$ is less than $6.40$ mm, and a normal distribution when $\dedgex$ exceeds this value.
This result is consistent with the likelihood ratio test comparing the fit of normal and skew-normal distributions (\autoref{fig:Ex1 LikelihoodRatio}).

Next, using \autoref{Formula:DualSkewNormalEstimateSigmaX} with $-c_x/d_x = 6.40$, we performed a regression analysis on $\sigma_x$, yielding $R^2 = .882$ (\autoref{table:Ex1 Model Regression}, \autoref{fig:Ex1_DualSkew_Sigma}).
In the vicinity of the screen edge, $\sigma_x$ tended to decrease as the target moved closer to the edge (\autoref{fig:Ex1 SigmaX Interaction}); the proposed model proved capable of modeling $\sigma_x$ by accounting for this variation.
Furthermore, a regression analysis restricted to the non-skewed region ($\dedgex \ge 6.40$), which utilizes the same formula as the existing model, resulted in $R^2 = .705$.
This performance is superior to the fit of the existing model across the entire dataset ($R^2 = .437$), suggesting that the structural design of the proposed model—switching between an edge-aware model and the existing model—functions effectively.

For $\mu_x$, regression analysis using \autoref{Formula:DualSkewNormalEstimateMuX} on data where $\dedgex < -c_x/d_x$ yielded $R^2 = .905$ (\autoref{table:Ex1 Model Regression}, \autoref{fig:Ex1_DualSkew_Mu})\footnote{Regression analysis was not performed for data where $\dedgex \ge -c_x/d_x$, which was approximated as $\mu_x = 0$.}.
This indicates that the proposed model effectively captures the quadratic variation in $\mu_x$.
However, in the region where $\dedgex \ge -c_x/d_x$ (where both proposed and existing models assume $\mu_x \approx 0$), observed positive values of $\mu_x$ (approximately $0.2$ to $0.6$) deviated from the model predictions (\autoref{fig:Ex1 MuX Interaction}).

Finally, the $\sr_x$ prediction resulted in $R^2 = .950$ (\autoref{table:Ex1 Model Regression}, \autoref{fig:Ex1_DualSkew_SR}).
This demonstrates that the proposed model maintains high prediction accuracy for $\sr_x$ even for targets near the screen edge, suggesting greater generalizability than the existing model.
LOOCV results for $R^2$ and $\mae$ were comparable to the full-dataset regression, confirming that the model did not suffer from overfitting and possesses high predictive power for unverified conditions (\autoref{table:Ex1 Model Regression}).

\vspace{3mm}
\noindent\textit{Machine Learning Models}
\\
\noindent We compared our model with several established ML models.
We used $W$ and $\margin_x$ as inputs to directly predict $\sr_x$.
The comparisons included Lasso regression, Random Forest, Support Vector Regression (SVR), and Multi-Layer Perceptron (MLP) neural networks.
Two sets of hyperparameters were evaluated: the default values from the scikit-learn library and values obtained through Bayesian optimization using the Optuna library.
In the Bayesian optimization, we searched for hyperparameters that minimized the $\mathit{RMSE}$ using 5-fold cross-validation.

With default settings, only Random Forest showed a higher $R^2$ ($0.987$) and lower $\mae$ ($1.48$) than the proposed model. With Bayesian-optimized settings, SVR and MLP neural networks outperformed the proposed model with higher $R^2$ ($0.993$ and $0.999$, respectively) and lower $\mae$ ($0.652$ and $0.230$, respectively) (\autoref{table:Ex1 Model Regression}).
In the LOOCV analysis, Random Forest with default settings showed a lower $R^2$ than the proposed model, while optimized SVR and MLP neural networks maintained higher $R^2$ values ($0.968$ and $0.966$).
However, the differences from our model were marginal ($\le 0.024$).
These results indicate that while the proposed model is slightly inferior to the ML models in simple $\sr_x$ prediction for targets within the training dataset, it possesses a comparable level of predictive accuracy for held-out conditions within the tested design space.

\subsubsection{Analyzing Tap-Coordinate Distributions}
\begin{figure}[ht]
\centering
\begin{minipage}[b]{0.49\columnwidth}
    \centering
    \includegraphics[width=0.95\columnwidth]{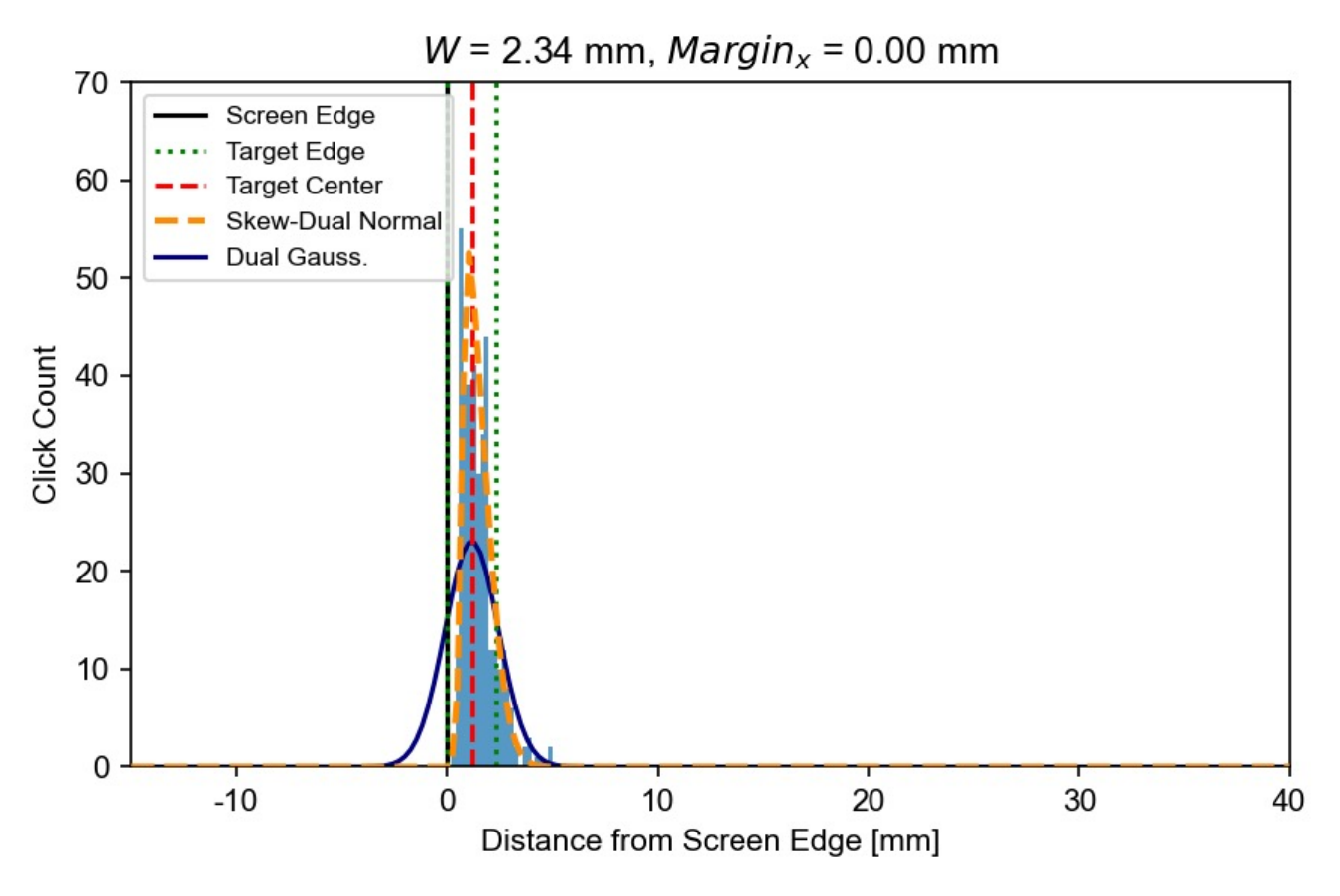}
    \subcaption{\(\margin_x = 0\) condition}
    \label{fig:Ex1 Distribution No Margin}
\end{minipage}
\begin{minipage}[b]{0.49\columnwidth}
    \centering
    \includegraphics[width=0.95\columnwidth]{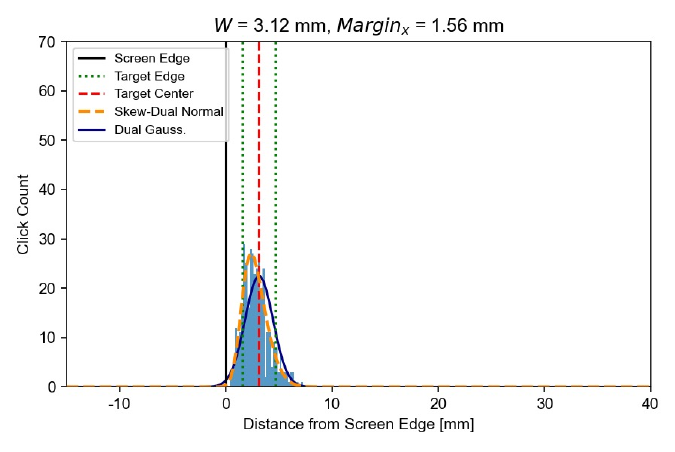}
    \subcaption{Small \(\margin_x\) condition}
    \label{fig:Ex1 Distribution Small Margin}
\end{minipage}
\begin{minipage}[b]{0.49\columnwidth}
    \centering
    \includegraphics[width=0.95\columnwidth]{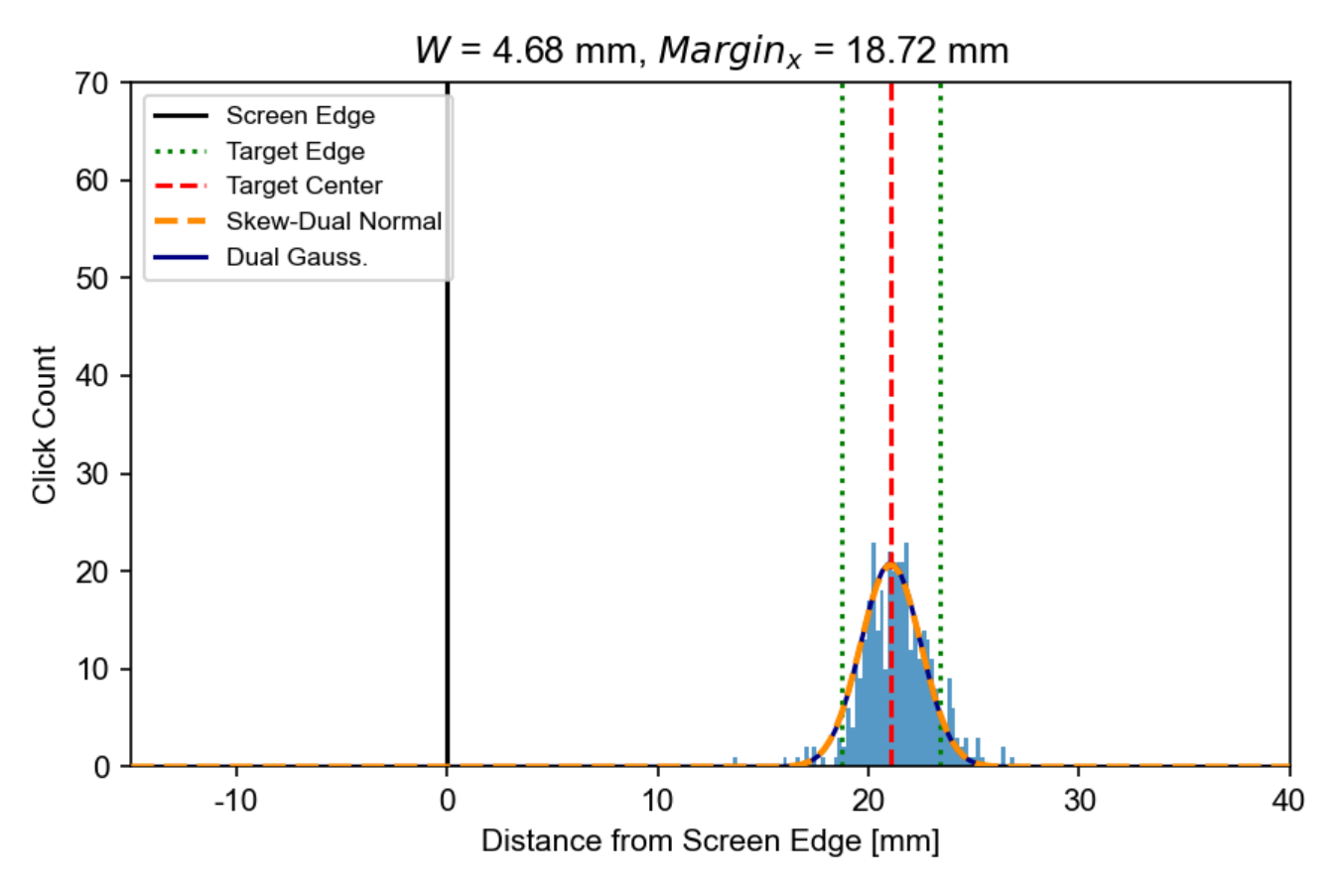}
    \subcaption{Condition with sufficient \(\margin_x\)}
    \label{fig:Ex1 Distribution Enough Margin}
\end{minipage}
\begin{minipage}[b]{0.49\columnwidth}
    \centering
    \includegraphics[width=0.95\columnwidth]{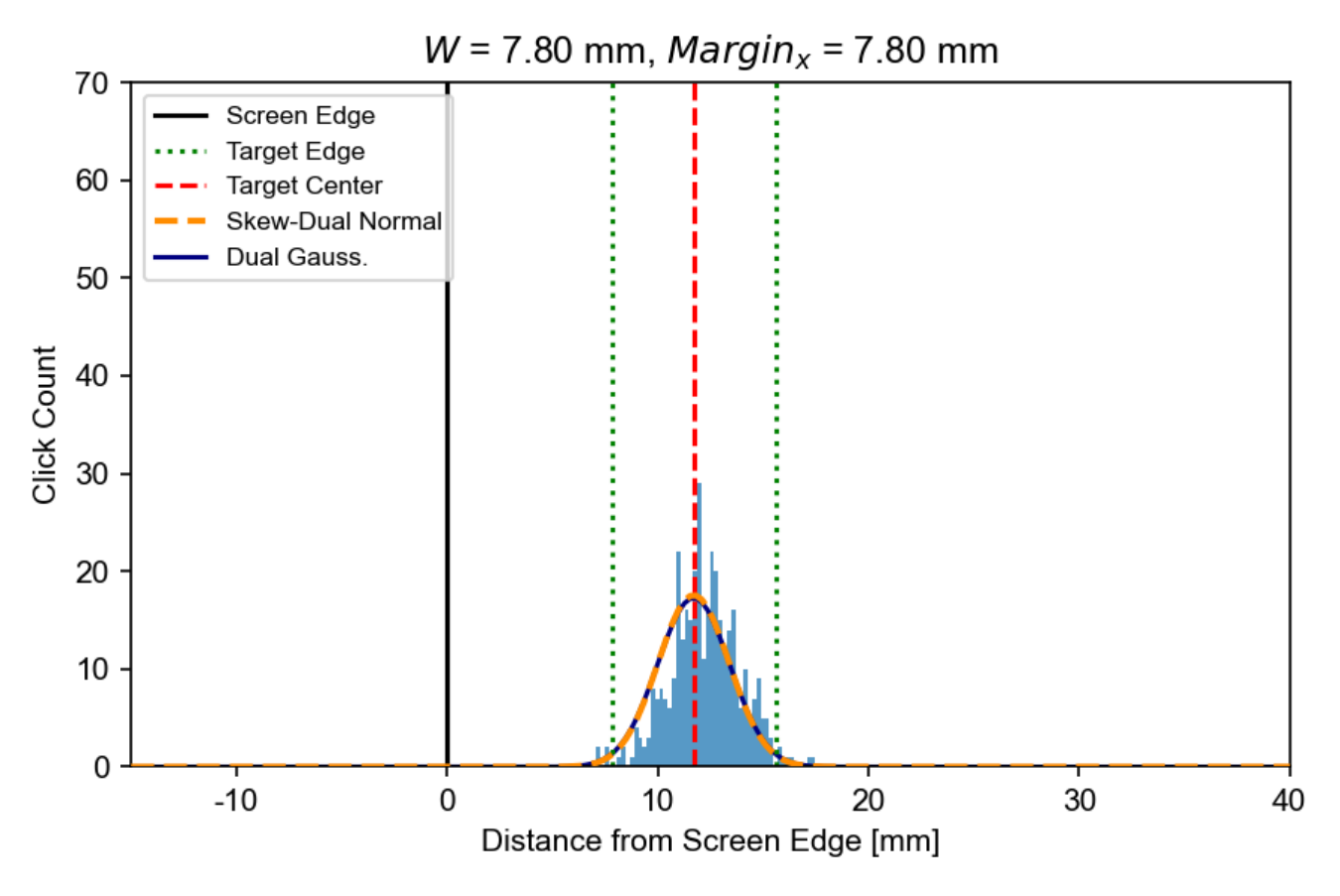}
    \subcaption{Exceptional case shifted to the right}
    \label{fig:Ex1 Distribution Exception}
\end{minipage}
\caption{Observed and predicted tap-coordinate distributions in Experiment 1. The proposed model successfully captures (a) the extreme skewness when the target is adjacent to the screen edge and (b) the gradual skewness when near the edge. (c) When the target is sufficiently far from the screen edge, the model regresses to the Dual Gaussian Distribution Model, enabling comprehensive prediction of tap-coordinate distributions for targets ranging from the edge to the center of the screen. (d) In exceptional cases, the mean of the distribution shifted to the right. However, since almost the entire shifted distribution remained within the target area, the impact on $\sr_x$ prediction accuracy was limited.}
\label{Fig:Ex1 TapDistribution}
\end{figure}

We analyzed the distributions of observed tap-coordinates and compared them with the distributions predicted by the existing and proposed models.
ML models were excluded from this analysis as they directly predict $\sr_x$.
The proposed model successfully captures the skewness that arises as the target approaches the screen edge (\autoref{Fig:Ex1 TapDistribution}).
The skewness predicted by the proposed model effectively modeled both the extreme skewness observed when the target is adjacent to the screen edge (\autoref{fig:Ex1 Distribution No Margin}) and the gradual skewness when the target is near the edge (\autoref{fig:Ex1 Distribution Small Margin}).
In addition, when there is a sufficient distance between the screen edge and the target, the tap-coordinate distribution can be explained by the Dual Gaussian Distribution Model (\autoref{fig:Ex1 Distribution Enough Margin}).
However, exceptional cases were observed where the mean of the distribution shifted to the right side of the target (\autoref{fig:Ex1 Distribution Exception}).

\subsubsection{Participant Questionnaire}
In the free-response comments, three participants reported intentionally tapping both the target and the bezel simultaneously when the target was adjacent to the screen edge.
This supports the user strategy assumed in our proposed model: tapping the bezel to avoid errors in the direction opposite to the edge.
The concentration of tap-coordinates near the screen edge (\autoref{fig:Ex1 Distribution No Margin}) suggests that other participants may have implicitly adopted this strategy.
Additionally, diverse individual strategies were reported, such as tilting the smartphone, tapping with the side of the finger, or applying firm pressure.

\subsection{Discussion}
\subsubsection{Distribution Skewness and Operation Strategies near the Screen Edge}
We showed that as the target approaches the screen edge, the tap-coordinate distribution deviates from a normal distribution and fits a skew-normal distribution (\autoref{fig:Ex1 LikelihoodRatio}).
This supports the hypothesis that the proposed model can estimate tap success rates by predicting the distribution via a skew-normal model (\autoref{fig:Ex1_DualSkew_SR}).

Notably, as $\margin_x$ approaches zero under a certain $W$ condition, $\sr_x$ tended to increase rather than decrease (\autoref{fig:Ex1 SRx Interaction}).
While this contradicts some prior studies \citep{Avrahami2015EdgeTouch,Usuba2023EdgeTarget,Henze2011LargeExperiment}, it may be explained by changes in operation strategies due to the use of flat bezels and rectangular targets, as we discussed (\autoref{fig:FingerBezelCircleRectangle}).
This is further evidenced by participants explicitly reporting this intentional strategy in the free-response comments.
These findings indicate that the proposed model effectively captures changes in both tap distributions and user strategies near the screen edge.

\subsubsection{Deviations from the Proposed Model}
Observed deviations included positive $\mu_x$ values (approx. $0.2$--$0.6$) in the region where the model assumes $\mu_x \approx 0$ ($\dedgex \ge -c_x/d_x$) (\autoref{fig:Ex1 MuX Interaction}, \autoref{fig:Ex1 Distribution Exception}).
These shifts are likely attributable to the starting position \citep{Henze2011LargeExperiment}, finger angle \citep{Azenkot2012TouchBehavior,Holz2011UnderstandingTouch}, and grip \citep{Lehmann2018HowToHold}.
Specifically, since all participants were right-handed and the starting position was fixed to their right knee, taps may have skewed rightward.

While this suggests that the $\mu_x \approx 0$ approximation is not perfectly accurate, the impact on $\sr_x$ was minimal because these deviations occurred primarily with large targets, where the distribution remains mostly within the target boundaries.
Since $\mu_x$ tends to approach zero in large-scale experiments as individual differences are neutralized \citep{Henze2011LargeExperiment}, we maintain the $\mu_x=0$ approximation.

\subsubsection{Comparison with Machine Learning Models}
In terms of pure $\sr_x$ prediction, ML models (particularly Bayesian-optimized SVR and MLP) outperformed the proposed model.
However, the performance gap was marginal ($R^2$ difference of $0.024$), and the proposed analytical model offers superior advantages for HCI modeling and UI design support in two key areas:

First is interpretability. 
ML models are black boxes where the mechanism linking $W$ or $\margin_x$ to success rates is unclear. 
In contrast, our model describes how variance changes ($g_x$) and the degree of skewness ($c_x, d_x$) via regression constants. 
For instance, the derived threshold ($-c_x/d_x \approx 6.40$~mm) provides a concrete guideline: designers must account for distribution skewing within 6.4~mm of the screen edge. 
Furthermore, while ML models only output $\sr_x$, the proposed model predicts the distribution shape, allowing designers to visualize user behavior behind the success rates.

Second is scalability. 
By geometrically modeling the distribution shape, our model can be extended to various contexts. 
Similar to the Dual Gaussian model, it could predict 2D success rates from 1D data \citep{Usuba2022ER1Dto2D} or adapt to moving targets \citep{Huang2018Moving1D,Huang2019MovingTarget2D} and arbitrary shapes \citep{Zhang2020ArbitraryShapedMovingTarget,Zhang2023ArbitraryShape}.
Because the proposed model targets the geometric influence of the edge on the distribution rather than just the correlation between $(W, \margin_x)$ and $\sr_x$, it remains applicable across different target dimensions and shapes as long as the edge effect on skewness remains constant.

\section{Experiment 2: 1D Pointing near Bottom Edge}
\subsection{Overview}
We conducted Experiment 2 to verify whether the proposed model generalizes across different axes.
Almost all conditions were identical to those in Experiment 1, such as the apparatus, instructions, compensation, and data analysis.
Only the points of difference are described hereafter.

We changed the axis to a vertically constrained 1D pointing task targeting the bottom screen edge.
The target height $H$ and distance from the bottom screen edge $\margin_y$ were used as independent variables.
The specific values of $H$ and $\margin_y$ were identical to $W$ and $\margin_x$ in Experiment 1, respectively.
Fifteen computer science students (5 females and 10 males; mean age 20.9 years, SD 1.94 years; all right-handed) participated independently of Experiment 1. 

\subsection{Results}
\subsubsection{Outliers}
We excluded 58 trials ($0.358\%$) where errors occurred in the $x$-axis direction to focus on $\sr _y$.
For each participant and condition, 70 outlier trials ($0.433\%$) were removed based on the 3SD rule.
The following analyses were conducted using the remaining 16,072 trials.

\subsubsection{Normality Test of Tap-Coordinate Distributions and Statistical Test}
\begin{figure}[ht]
    \centering
    \includegraphics[width=0.55\columnwidth]{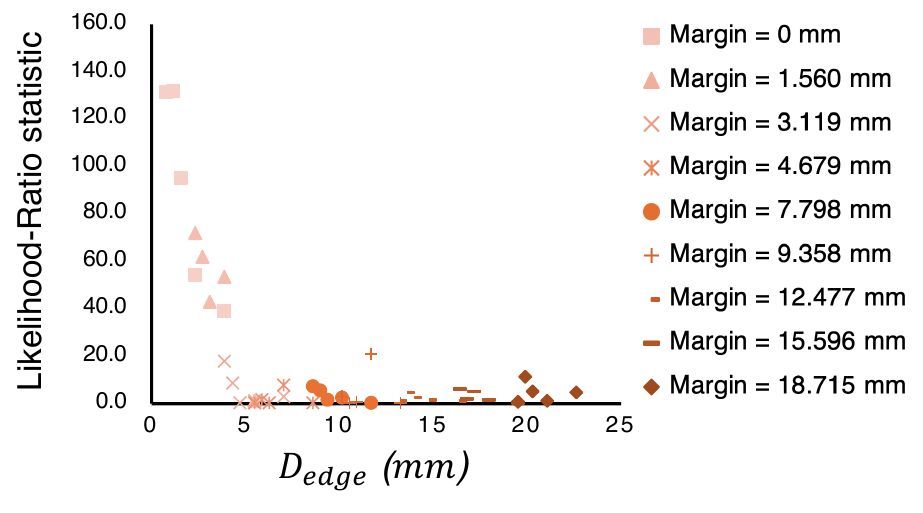}
    \caption{Likelihood ratio statistics in Experiment 2. As $D_\mathit{edge\_y}$ decreases, the tap-coordinate distribution increasingly follows a skew-normal distribution rather than a normal distribution.}
    \label{fig:Ex2 LikelihoodRatio}
\end{figure}
\noindent The Shapiro–Wilk test results showed that normality was rejected in 23 out of 45 target conditions ($51.1\%$). 
In the likelihood ratio test comparing the fit of the normal and skew-normal distributions, the test statistic increased as $\dedgey$ decreased (\autoref{fig:Ex2 LikelihoodRatio}). 
As in Experiment 1, these results demonstrate that the fit to the skew-normal distribution significantly improves as the target approaches the screen edge, thereby supporting the assumptions of the proposed model.

\begin{figure}[ht]
\centering
\begin{minipage}[b]{0.49\textwidth}
    \centering
    \includegraphics[width = 0.95\columnwidth]{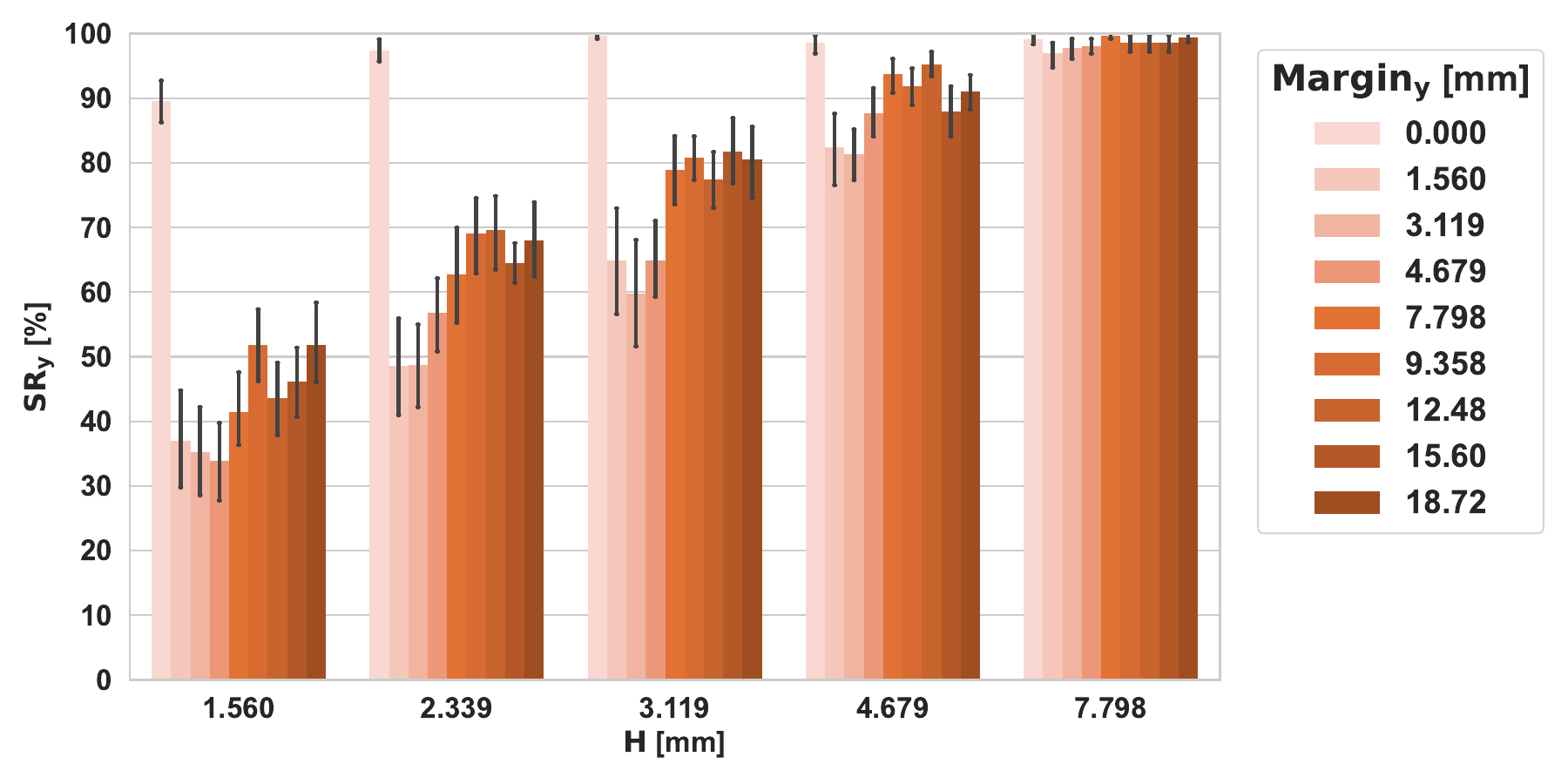}
    \subcaption{$\sr_y$}
    \label{fig:Ex2 SRy Interaction}
\end{minipage}
\begin{minipage}[b]{0.49\textwidth}
    \centering
    \includegraphics[width = 0.95\columnwidth]{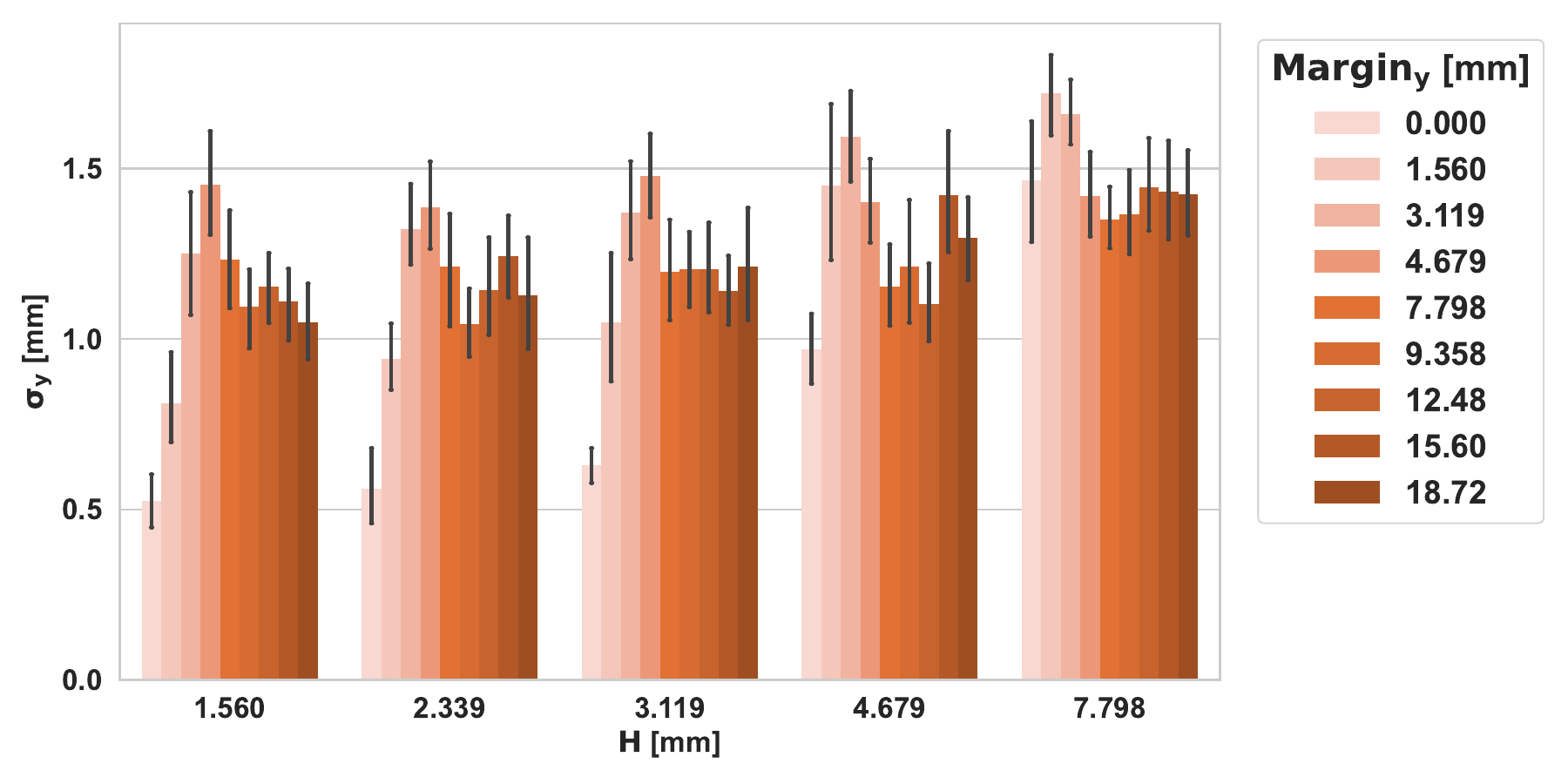}
    \subcaption{$\sigma_y$}
    \label{fig:Ex2 SigmaY Interaction}
\end{minipage}
\begin{minipage}[b]{0.49\textwidth}
    \centering
    \includegraphics[width = 0.95\columnwidth]{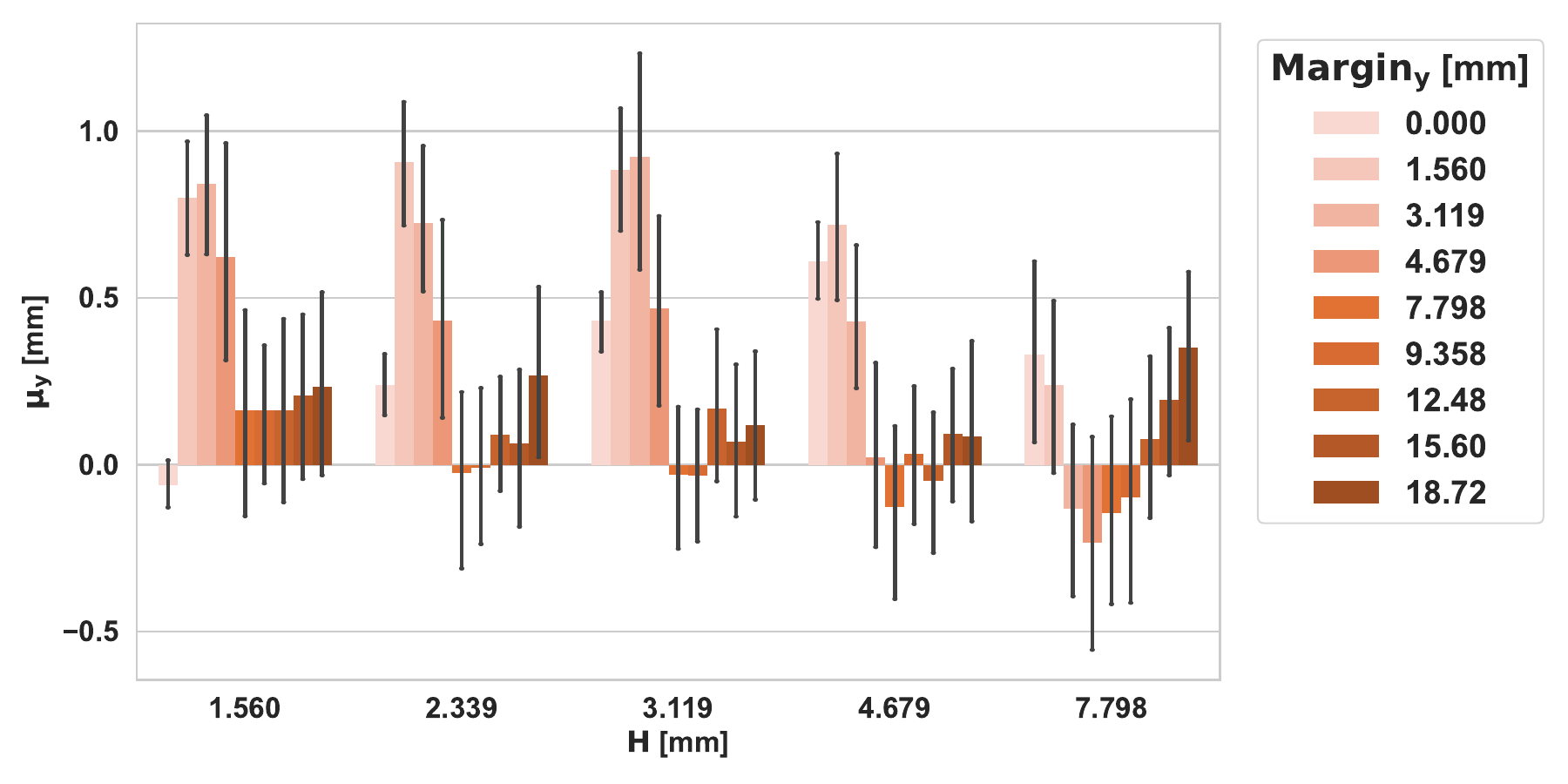}
    \subcaption{$\mu_y$}
    \label{fig:Ex2 MuY Interaction}
\end{minipage}
\begin{minipage}[b]{0.49\textwidth}
    \centering
    \includegraphics[width = 0.95\columnwidth]{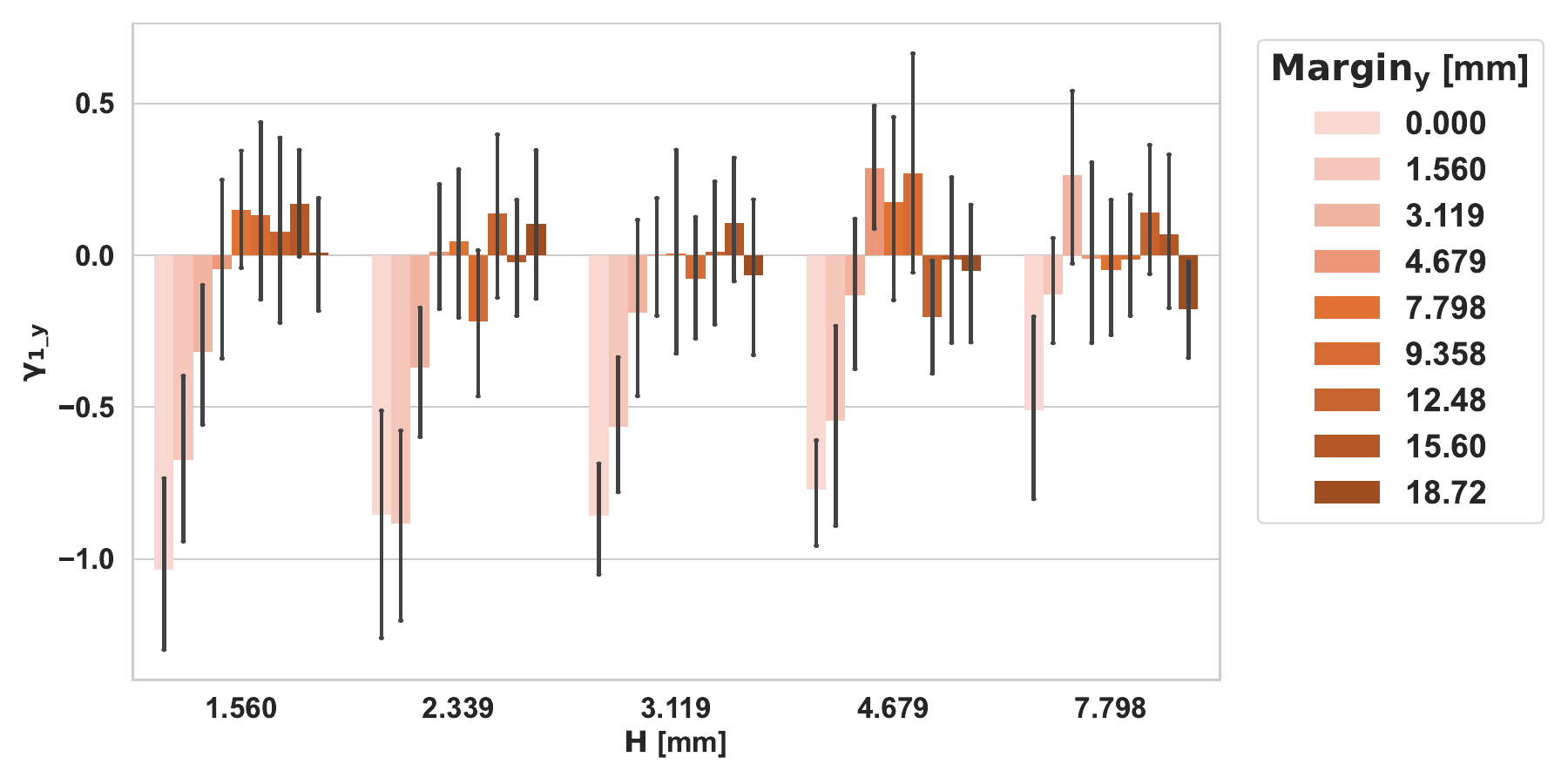}
    \subcaption{$\gammay$}
    \label{fig:Ex2 Gamma1Y Interaction}
\end{minipage}
\caption{Effects of $\margin_y$ and $H$ on $\sr_y$, $\sigma_y$, $\mu_y$, and $\gammay$ in Experiment 2.}
\label{fig:Ex2_Interaction}
\end{figure}
The results of two-way RM-ANOVAs showed that, as in Experiment 1, $\margin_y$ affected each dependent variable, demonstrating the limitations of existing models while simultaneously supporting the assumptions of the proposed model regarding the changes in these variables (\autoref{fig:Ex2_Interaction}).

\subsubsection{Prediction Accuracy of Tap Success Rate}
\begin{table}[ht]
\caption{Regression constants and model accuracy metrics in Experiment 2.}
\centering
\resizebox{\textwidth}{!}{
\begin{tabular}{c|c|c|c|c|c|c|c|c}
\multicolumn{2}{c|}{} & \multicolumn{5}{c|}{Regression Analysis} & \multicolumn{2}{|c}{LOOCV}\\
\hline
Model & Formula & Regression Constants & $R^2$ & $\mae$ & $ \rmse$ & $\mape$ & $R^2$ & $\mae$\\
\hline
Dual Gauss. & $\sigma_y^2$ (\autoref{formula:Bi_sigma}) & $a_y = 1.23$, $b_y = 0.0164$ & $.370$ & $0.355$ & 0.459 & $39.1\%$ & $.315$ & $0.371$\\
\cline{2-9}
& $\sr_y$ (\autoref{formula:DualGaussian1D}) & - & $.699$ & $7.51$ & $11.6$ & $12.5\%$ & $.691$ & $7.61$\\
\hline
Skewed-Dual. & \begin{tabular}{c}$\gamma_{1_y}$ (\autoref{Formula:DualSkewNormalEstimateGamma1X}\\ using $\dedgey$)\end{tabular} & $c_y = 1.20$, $d_y = -0.199$ & $.873$ & $0.0915$ & $0.120$ & $81.6\%$ & $.867$ & $0.0944$\\
\cline{2-9}
 & \begin{tabular}{c}$\sigma_y$ (\autoref{Formula:DualSkewNormalEstimateSigmaX}\\ using $H$ \& $\margin_y$)\end{tabular} & \begin{tabular}{c}$e_y = 0.123$, $f_y = 0.0371$,\\$g_y = 0.415$, $h_y = 1.31$,\\ $i_y = 0.0130$\end{tabular} & $.871$ & $0.0675$ & $0.0915$ & $5.61\%$ & $.841$ & $0.0761$\\
\cline{2-9}
& \begin{tabular}{c}$\mu_y$ (\autoref{Formula:DualSkewNormalEstimateMuX} \\using $\dedgey$)\end{tabular} & \begin{tabular}{c}$j_y = 0.804$, $k_y = -0.0961$,\\ $l_y = 3.60$\end{tabular} & $.641$ & $0.124$ & $0.167$ & $36.0\%$ & $.484$ & $0.153$\\
\cline{2-9}
& $\sr_y$ (\autoref{Formula:DualSkewSR1D}) & - & $.953$ & $3.13$ & $4.56$ & $5.24\%$ & $.947$ & $3.35$\\
\hline
\multicolumn{9}{c}{Machine learning models using default hyperparameters}\\
\hline
Lasso Regression & $\sr_y$ & number of parameters: 3 & .608 & 10.5 & 13.2 & 16.7\% & .558 & 11.1 \\
\hline
Random Forest & $\sr_y$ & number of parameters: 5,528 & .967 & 2.24 & 3.81 & 3.27\% & .799 & 5.82 \\
\hline
SVR & $\sr_y$ & number of parameters: 46 & .106 & 16.2 & 19.9 & 28.9\% & .065 & 16.7 \\
\hline
MLP Neural Net & $\sr_y$ & number of parameters: 401 & -2.49 & 32.1 & 39.4 & 41.4\% & -2.08 & 29.9 \\
\hline
\multicolumn{9}{c}{Machine learning models using Bayesian-optimized hyperparameters}\\
\hline
Lasso Regression & $\sr_y$ & number of parameters: 3 & .597 & 10.3 & 13.4 & 17.1\% & .565 & 10.7 \\
\hline
Random Forest & $\sr_y$ & number of parameters: 8,171 & .967 & 2.54 & 3.80 & 3.81\% & .806 & 5.67 \\
\hline
SVR & $\sr_y$ & number of parameters: 44 & .835 & 3.70 & 8.56 & 4.72\% & .753 & 5.59 \\
\hline
MLP Neural Net & $\sr_y$ & number of parameters: 21,058 & .999 & 0.180 & 0.307 & 0.295\% & .943 & 3.75 \\
\hline
\end{tabular}
}
\label{table:Ex2 Model Regression}
\end{table}

\vspace{3mm}
\noindent\textit{Existing Model: Dual Gaussian Distribution Model}
\begin{figure}[ht]
\centering
\begin{minipage}[b]{0.49\textwidth}
    \centering
    \includegraphics[width = 0.95\columnwidth]{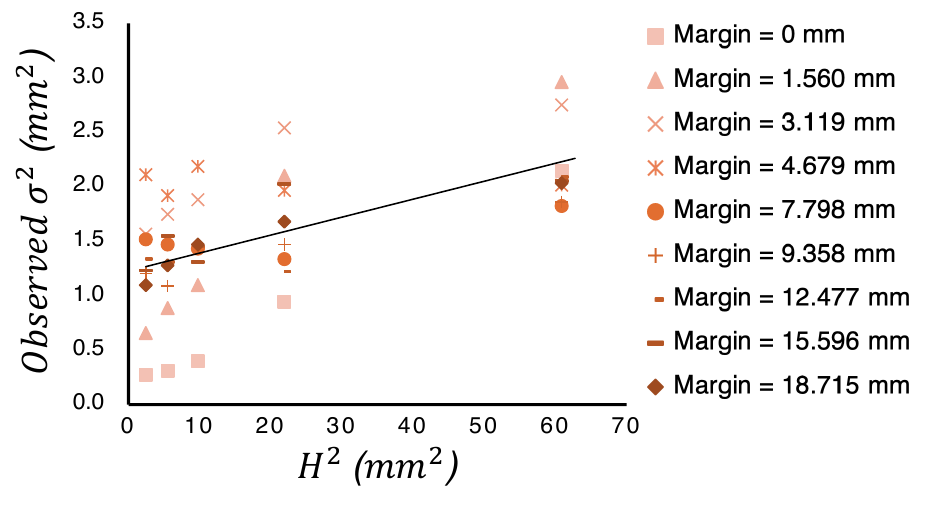}
    \subcaption{$H^2$ vs. $\sigma_y^2$}
    \label{fig:Ex2_DualGauss_H_vs_Sigma}
\end{minipage}
\begin{minipage}[b]{0.49\textwidth}
    \centering
    \includegraphics[width = 0.95\columnwidth]{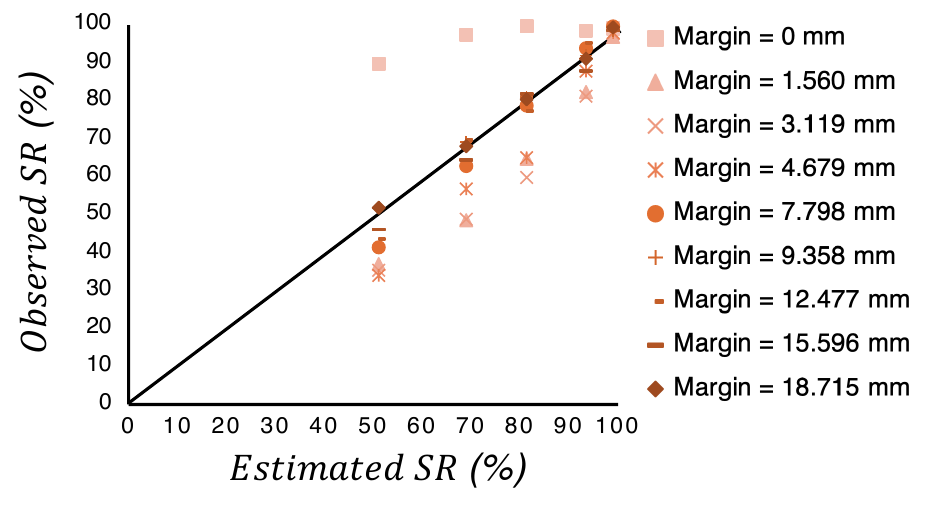}
    \subcaption{Estimated $\sr_y$ vs. Observed $\sr_y$}
    \label{fig:Ex2_DualGauss_SR_vs_SR}
\end{minipage}
\caption{Results regarding the $\sr_y$ prediction accuracy of the Dual Gaussian Distribution Model in Experiment 2. As in Experiment 1, discrepancies between observed values and model predictions were found when $\margin_y$ was small. In (a), the line represents the regression line; in (b), the line represents the identity line where predicted and observed values match.}
\label{fig:Ex2_DualGauss}
\end{figure}
\\
\noindent The $\sigma_y^2$ prediction model (\autoref{formula:Bi_sigma}) showed moderate fit ($R^2 = .370$; \autoref{table:Ex2 Model Regression}, \autoref{fig:Ex2_DualGauss_H_vs_Sigma}), with particularly large deviations observed in conditions where $\margin_y$ was small. 
This suggests that a model considering only $H$ cannot fully capture the fluctuations in $\sigma_y^2$ caused by the screen edge.

The prediction accuracy of $\sr_y$ resulted in $R^2 = .699$ (\autoref{table:Ex2 Model Regression}, \autoref{fig:Ex2_DualGauss_SR_vs_SR}). 
As in Experiment 1, at $\margin_y = 0$, the observed $\sr_y$ significantly exceeded the predicted values. 
The $R^2$ and $\mae$ values from LOOCV were generally consistent with the results of the regression analysis using the full dataset.

\vspace{3mm}
\noindent\textit{Proposed Model: Skewed Dual Normal Distribution Model}
\begin{figure}[ht]
\centering
\begin{minipage}[b]{0.49\textwidth}
    \centering
    \includegraphics[width = 0.95\columnwidth]{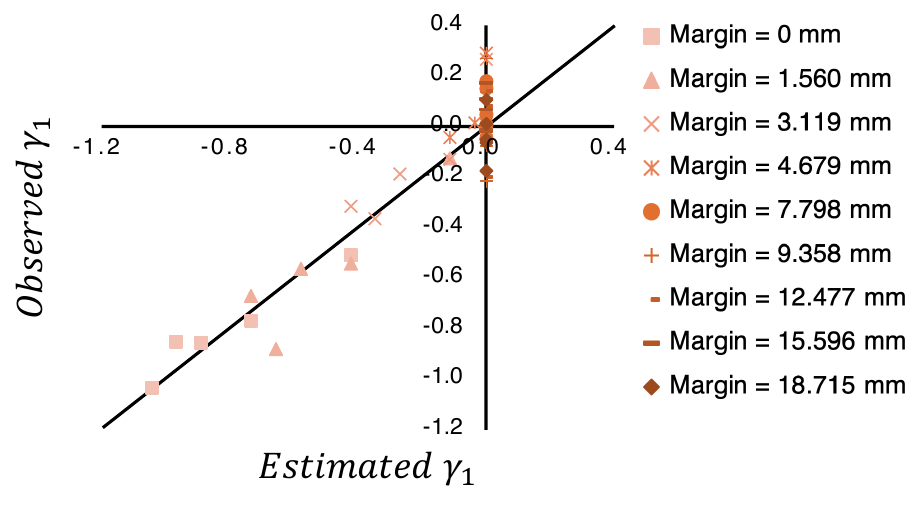}
    \subcaption{Predicted $\gamma_{1\_y}$ vs. Observed $\gamma_{1\_y}$}
    \label{fig:Ex2_DualSkew_Gamma}
\end{minipage}
\begin{minipage}[b]{0.49\textwidth}
    \centering
    \includegraphics[width = 0.95\columnwidth]{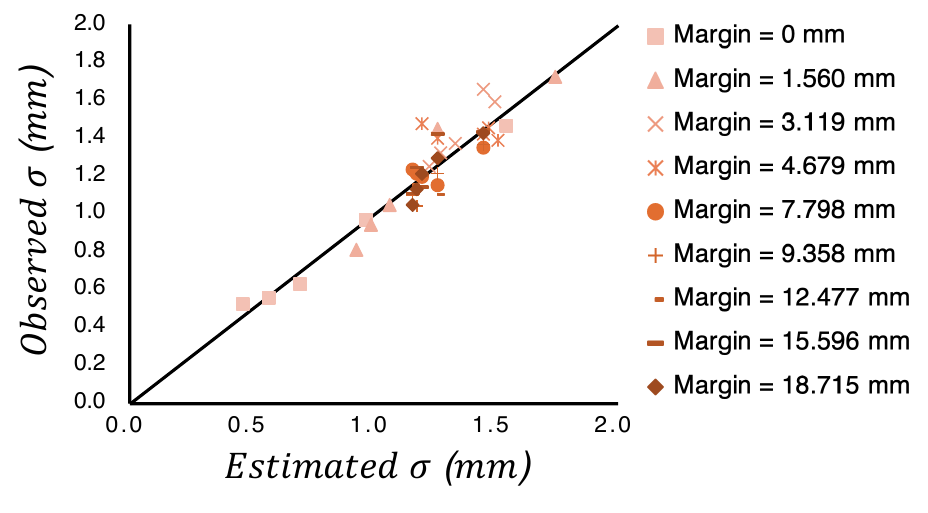}
    \subcaption{Predicted $\sigma_y$ vs. Observed $\sigma_y$}
    \label{fig:Ex2_DualSkew_Sigma}
\end{minipage}
\begin{minipage}[b]{0.49\textwidth}
    \centering
    \includegraphics[width = 0.95\columnwidth]{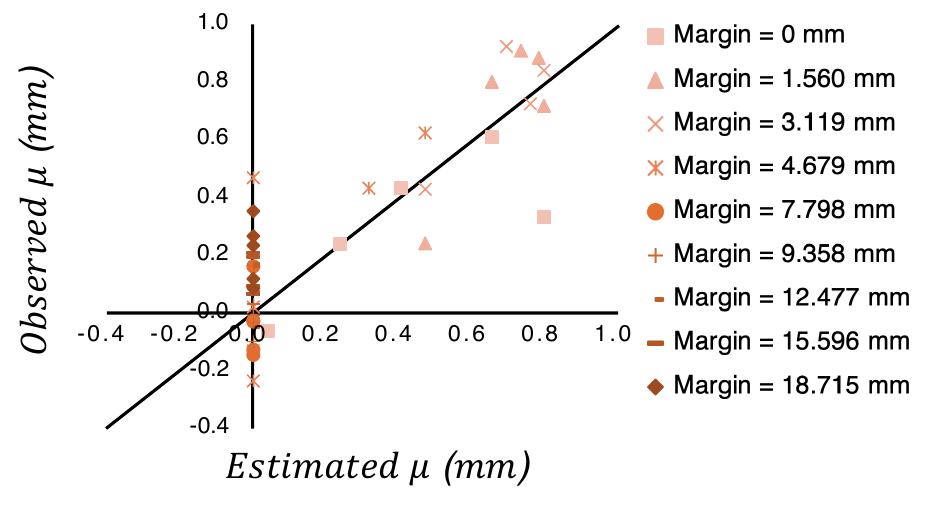}
    \subcaption{Predicted $\mu_y$ vs. Observed $\mu_y$}
    \label{fig:Ex2_DualSkew_Mu}
\end{minipage}
\begin{minipage}[b]{0.49\textwidth}
    \centering
    \includegraphics[width = 0.95\columnwidth]{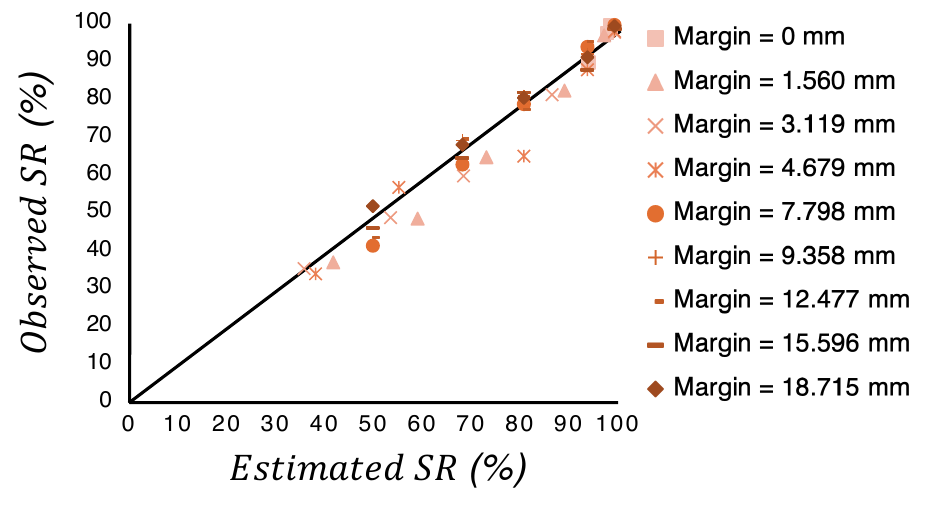}
    \subcaption{Predicted $\sr_y$ vs. Observed $\sr_y$}
    \label{fig:Ex2_DualSkew_SR}
\end{minipage}
\caption{Results regarding the $\sr_y$ prediction accuracy of the Skewed Dual Normal Distribution Model in Experiment 2. The lines represent the identity lines. (a) $\gamma_{1\_y}$ was well-predicted when $\margin_y$ was small. For large $\margin_y$, the results generally followed the assumption that $\gamma_{1\_y} \approx 0$. (b) $\sigma_y$ was accurately predicted by the proposed model. (c) $\mu_y$ was accurately predicted when $\dedgey < -c_y/d_y$. Otherwise, while generally following the $\mu_y \approx 0$ assumption, observed $\mu_y$ values ranged from $-$0.2 to 0.4 depending on the condition. (d) $\sr_y$ was accurately predicted by the proposed model.}
\label{fig:Ex2_DualSkew}
\end{figure}
\\
\noindent A regression analysis on $\gammay$ (\autoref{Formula:DualSkewNormalEstimateGamma1X} using $\dedgey$ instead of $\dedgex$) yielded $R^2 = .873$ (\autoref{table:Ex2 Model Regression}, \autoref{fig:Ex2_DualSkew_Gamma}). 
As in Experiment 1, $|\gamma_{1\_y}|$ increased as the target approached the screen edge. 
In Experiment 2, $\gammay$ was negative because the edge was in a positive position (bottom) relative to the target (\autoref{fig:Ex2 Gamma1Y Interaction}, \autoref{fig:Ex2_DualSkew_Gamma}, \autoref{fig:EdgeAndTarget Left}, \autoref{fig:EdgeAndTarget Right}).

The X-intercept of the regression line ($-c_y/d_y$) was $6.02$. 
Thus, our model predicts distribution skewness when $\dedgey < 6.02$~mm, and assumes a normal distribution otherwise.
This is consistent with Experiment 1 ($-c_x/d_x = 6.40$) and the likelihood ratio test results (\autoref{fig:Ex2 LikelihoodRatio}).

The fit for $\sigma_y$ resulted in $R^2 = .871$ (\autoref{table:Ex2 Model Regression}, \autoref{fig:Ex2_DualSkew_Sigma}). 
A regression analysis restricted to the non-skewed region ($\dedgey \ge 6.02$), where the existing model's formula is applied, yielded $R^2 = .564$—outperforming the existing model's overall fit ($R^2 = .370$). 

A regression analysis for $\mu_y$ on data where $\dedgey < -c_y/d_y$ yielded $R^2 = .641$ (\autoref{table:Ex2 Model Regression}, \autoref{fig:Ex2_DualSkew_Mu})\footnote{Regression analysis was not performed for $\dedgey \ge -c_y/d_y$ where $\mu_y \approx 0$ is assumed.}. 
While this accuracy was lower than in Experiment 1 ($R^2 = .905$), deviations from the model in the region where $\mu_y \approx 0$ ($\dedgey \ge -c_y/d_y$) were reduced compared to Experiment 1 (\autoref{fig:Ex2 MuY Interaction}).

Finally, the prediction accuracy for $\sr_y$ was $R^2 = .953$ (\autoref{table:Ex2 Model Regression}, \autoref{fig:Ex2_DualSkew_SR}). 
The LOOCV analysis yielded $R^2$ and $\mae$ values comparable to those from the full-dataset regression (\autoref{table:Ex2 Model Regression}). 
These results mirror those of Experiment 1, demonstrating that the proposed model accurately predicts $\sr_y$ even for the bottom edge.

\vspace{3mm}
\noindent\textit{Machine Learning Models}
\\
\noindent With default hyperparameters, only the Random Forest outperformed the proposed model, yielding $R^2=0.967$ and $\mae = 2.24$. 
Under Bayesian-optimized hyperparameters, both the Random Forest and the MLP neural network achieved higher $R^2$ ($0.967$ and $0.999$, respectively) and lower $\mae$ ($2.54$ and $0.180$) than the proposed model (\autoref{table:Ex2 Model Regression}). 
However, unlike Experiment 1, the LOOCV analysis revealed that all ML models exhibited lower $R^2$ values than the proposed model.
Thus, our model possesses superior predictive performance for new target conditions.

\subsubsection{Analyzing Tap-Coordinate Distributions}
\begin{figure}[ht]
\centering
\begin{minipage}[b]{0.49\columnwidth}
    \centering
    \includegraphics[width=0.95\columnwidth]{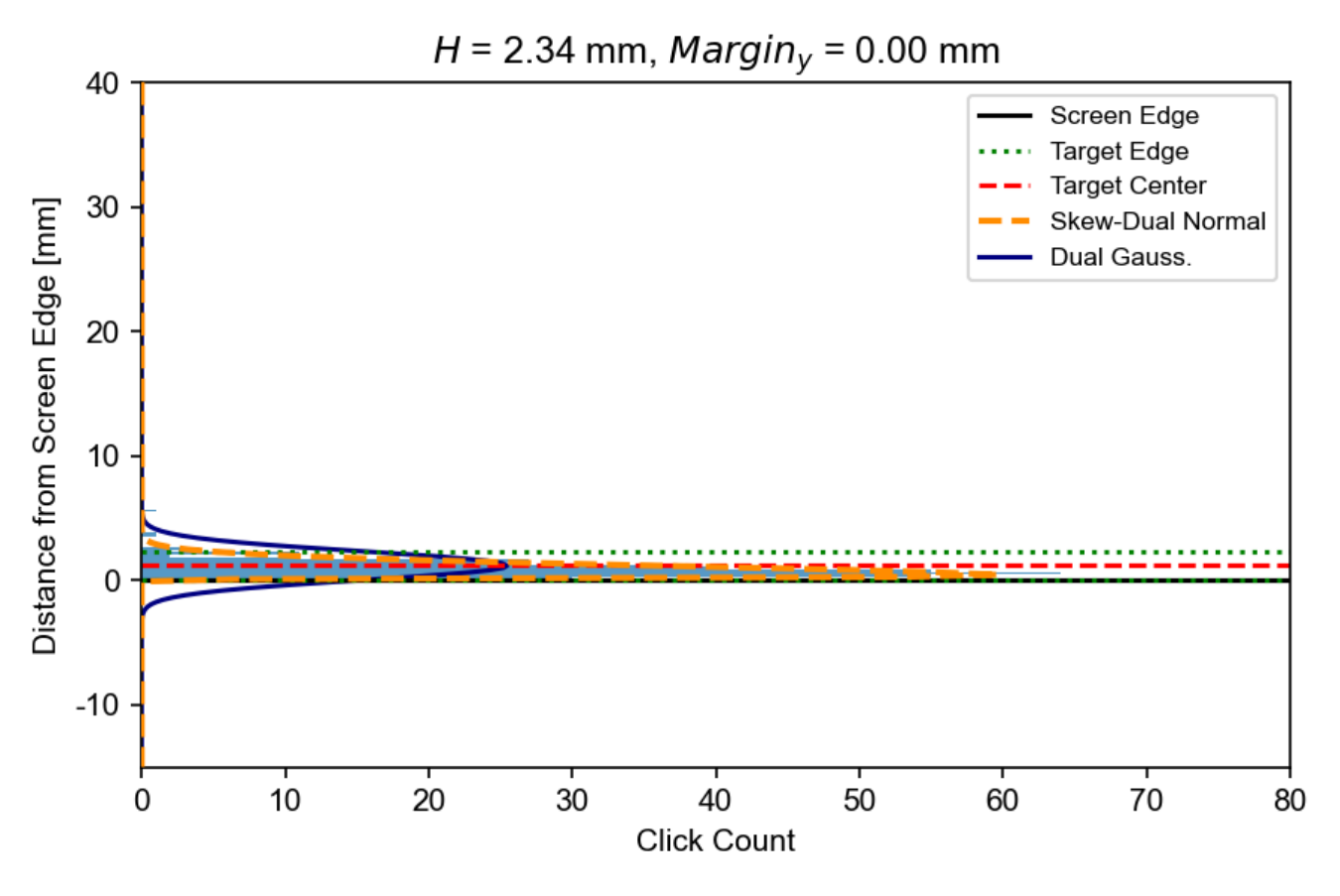}
    \subcaption{\(\margin_y = 0\) condition}
    \label{fig:Ex2 Distribution No Margin}
\end{minipage}
\begin{minipage}[b]{0.49\columnwidth}
    \centering
    \includegraphics[width=0.95\columnwidth]{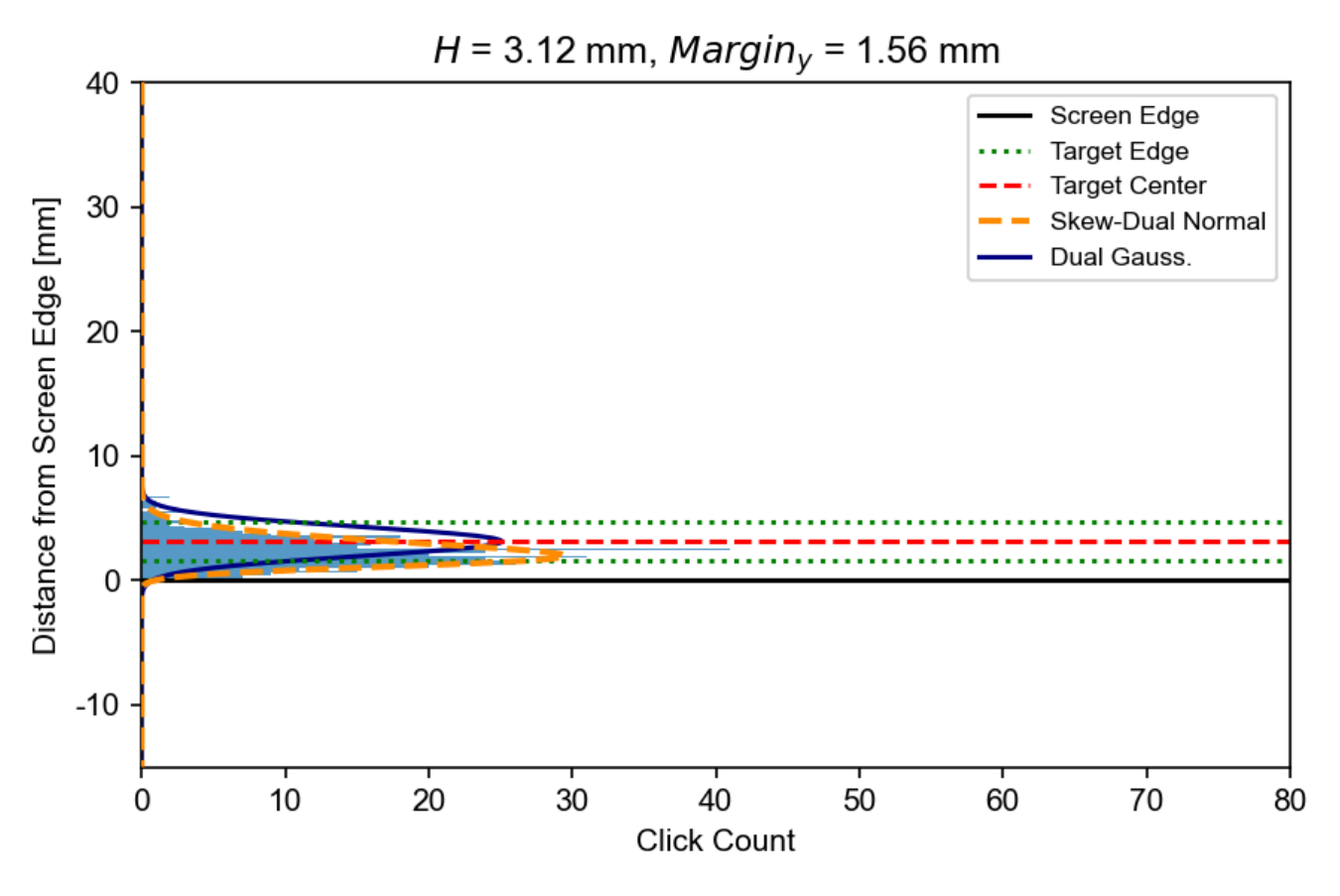}
    \subcaption{Small \(\margin_y\) condition}
    \label{fig:Ex2 Distribution Small Margin}
\end{minipage}
\begin{minipage}[b]{0.49\columnwidth}
    \centering
    \includegraphics[width=0.95\columnwidth]{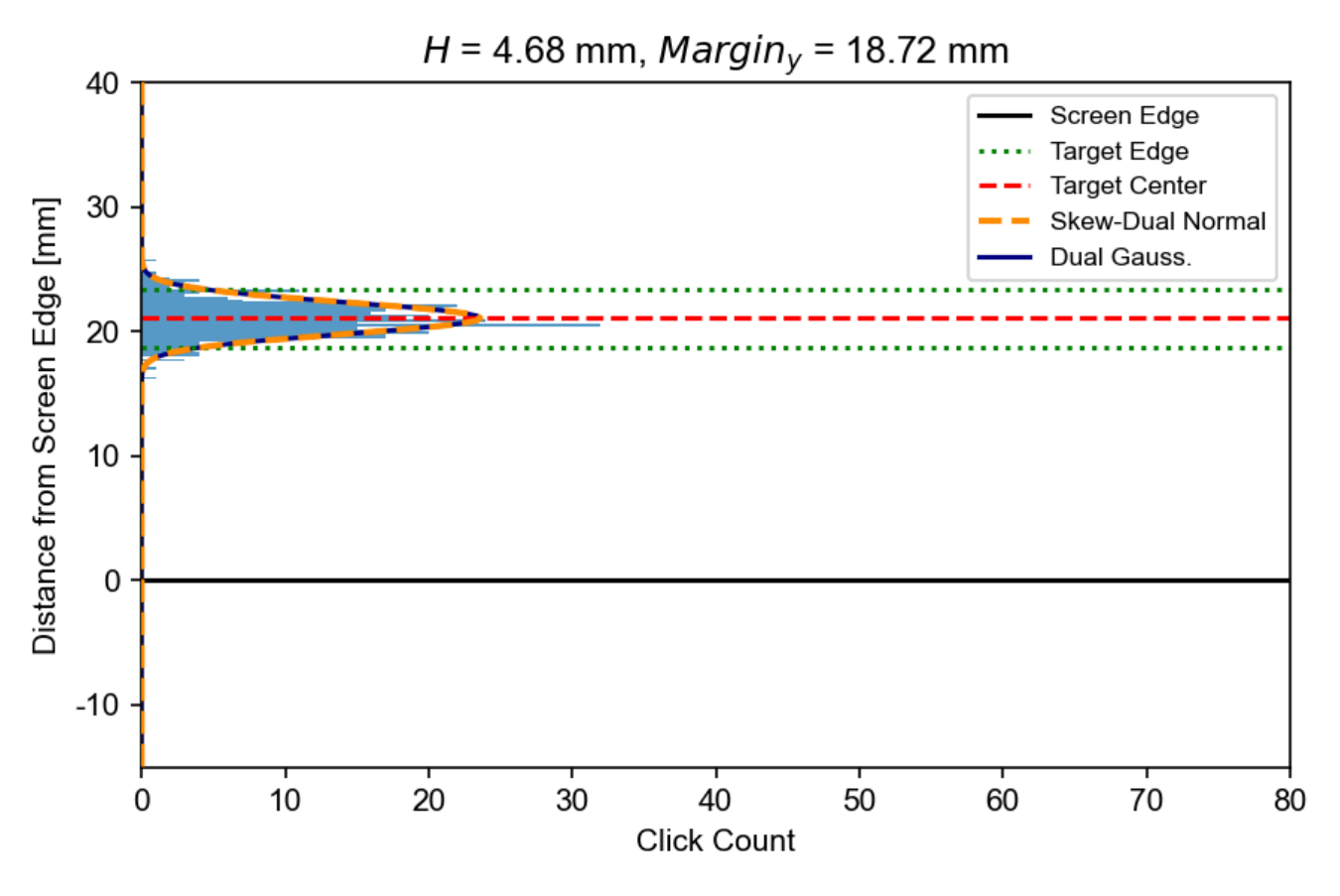}
    \subcaption{Condition with sufficient \(\margin_y\)}
    \label{fig:Ex2 Distribution Enough Margin}
\end{minipage}
\begin{minipage}[b]{0.49\columnwidth}
    \centering
    \includegraphics[width=0.95\columnwidth]{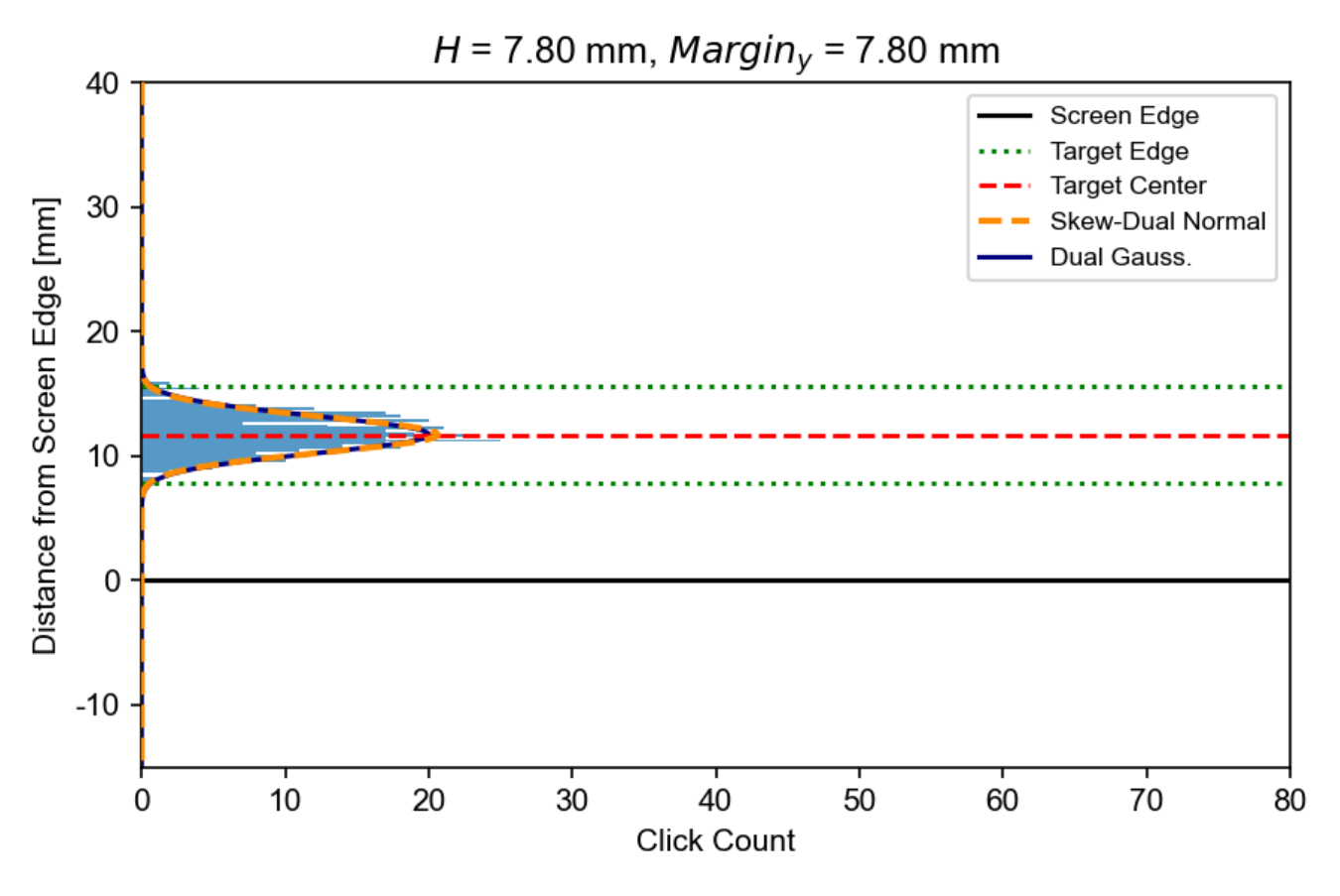}
    \subcaption{No exceptional distribution shifts observed}
    \label{fig:Ex2 Distribution Exception}
\end{minipage}
\caption{Observed and predicted tap-coordinate distributions in Experiment 2. As in Experiment 1, the proposed model appropriately captures (a) extreme skewness when the target is in contact with the screen edge and (b) gradual skewness when near the edge. By regressing to the Dual Gaussian Distribution Model when the target is sufficiently far from the edge (c), the proposed model comprehensively predicts tap distributions from the screen edge to the center. (d) Unlike Experiment 1, no exceptional distribution shifts were observed in Experiment 2.}
\label{Fig:Ex2 TapDistribution}
\end{figure}

\noindent Even along the $y$-axis, the proposed model can appropriately capture the skewness that arises as a target approaches the screen edge (\autoref{Fig:Ex2 TapDistribution}). 
The skewness predicted by the proposed model effectively accounted for both the extreme skewness observed when the target is in contact with the edge (\autoref{fig:Ex2 Distribution No Margin}) and the gradual skewness seen when the target is near the edge (\autoref{fig:Ex2 Distribution Small Margin}). 
Furthermore, the results suggest that when there is a sufficient distance between the screen edge and the target, the tap-coordinate distribution can be explained by the Dual Gaussian Distribution Model (\autoref{fig:Ex2 Distribution Enough Margin}). 
Notably, the exceptional cases where the distribution mean shifted away from the target center—as seen in Experiment 1 (\autoref{fig:Ex1 Distribution Exception})—were not observed in Experiment 2 (\autoref{fig:Ex2 Distribution Exception}).

\subsubsection{Participant Questionnaire}
As in Experiment 1, three participants reported that they intentionally tapped both the target and the bezel simultaneously when the target was in contact with the screen edge. 
However, six participants reported that they aimed for the upper part of the target because the contact area of their finger was lower than the actual fingertip. 
Other diverse strategies were also reported, such as tilting the smartphone, pressing the finger firmly against the screen, or performing the task without any specific conscious strategy. 
Despite these variations in subjective operational strategies, the observed tap-coordinates showed trends consistent with Experiment 1, with a bias toward the screen edge for targets located near the boundary (\autoref{fig:Ex2 Distribution No Margin}, \autoref{fig:Ex2 Distribution Small Margin}).

\subsection{Discussion}
\subsubsection{Consistency with Experiment 1}
Experiment 2 suggested that the proposed model is also applicable to $\sr_y$ and tap-coordinate distribution predictions along the $y$-axis. 
As in Experiment 1, the tap-coordinate distribution showed a higher fit to the skew-normal distribution as the target approached the screen edge (\autoref{fig:Ex2 LikelihoodRatio}), and the proposed model successfully captured this change (\autoref{fig:Ex2_DualSkew_SR}). 
An increase in $\sr_y$ as $\margin_y$ approaches zero was also observed (\autoref{fig:Ex2 SRy Interaction}). 
This suggests that even when the screen edge is positioned along the $y$-axis, participants may employ a strategy of simultaneously tapping the bezel and the target to avoid errors in the direction opposite to the screen edge.

\subsubsection{Differences from Experiment 1}
In Experiment 1, positive $\mu_x$ values were observed in the region $\dedgex \ge -c_x/d_x$ where the model assumed $\mu_x \approx 0$ (\autoref{fig:Ex1 MuX Interaction}, \autoref{fig:Ex1 Distribution Exception}); however, no such deviations were observed in Experiment 2 (\autoref{fig:Ex2 MuY Interaction}, \autoref{fig:Ex2 Distribution Exception}). 
This difference can likely be attributed to the influence of the starting position and the participants' operational strategies. 
In Experiment 1, the fixed starting position on the right knee combined with targets being displayed relative to the left edge of the screen likely caused a rightward shift in the tap distribution as participants sought to shorten the movement distance. 
In Experiment 2, although the starting position remained, the targets were displayed at the bottom edge.
Hence, the distance from the right knee was shorter, potentially reducing the need for participants to shorten their movement distance.

Furthermore, according to the open-ended comments, six participants reported that they ``intentionally tapped the upper part of the target because the contact area of the finger was lower than the fingertip.'' 
This suggests that participants prioritized an error-avoidance strategy (aiming higher) over a movement-minimization strategy (aiming lower), which may have resulted in $\mu_y$ values closer to zero. 
As a result, the assumption that ``the mean of the tap-coordinate distribution regresses to the target center when the target is sufficiently far from the screen edge'' was supported in Experiment 2, contrary to Experiment 1. 
This is consistent with prior research indicating that in large-scale experiments, the effects of starting position variability and individual differences are absorbed, leading the mean tap-coordinates to approach the target center \citep{Henze2011LargeExperiment}.

Moreover, while ML models outperformed the proposed model in LOOCV for $\sr_x$ in Experiment 1, the proposed model achieved higher predictive accuracy for $\sr_y$ in Experiment 2. 
This indicates that the proposed model is more accurate in predicting performance for unknown conditions, suggesting that its utility is even more robust than shown in Experiment 1. 
Specifically, while Experiment 1 concluded that the proposed model was superior in terms of interpretability and extensibility, Experiment 2 suggested that it also excels in predictive performance for unknown conditions.

\section{Experiment 3: 2D Pointing near the Top-Right Corner}
\label{sec:exp3}
Experiment 3 was conducted to verify whether the proposed model is applicable for predicting the tap success rate for 2D rectangular targets near a screen corner, where they are influenced by two screen edges simultaneously (\autoref{fig:Ex3 Setup}).

\subsection{Experimental Setup}
Except for the target shape and its location, almost all conditions were identical to those used in Experiments 1 and 2.
Aspects of the task and experimental design were modified to accommodate the transition to 2D rectangular pointing.

\subsubsection{Task}
\begin{figure}[ht]
    \centering
    \includegraphics[width=\textwidth]{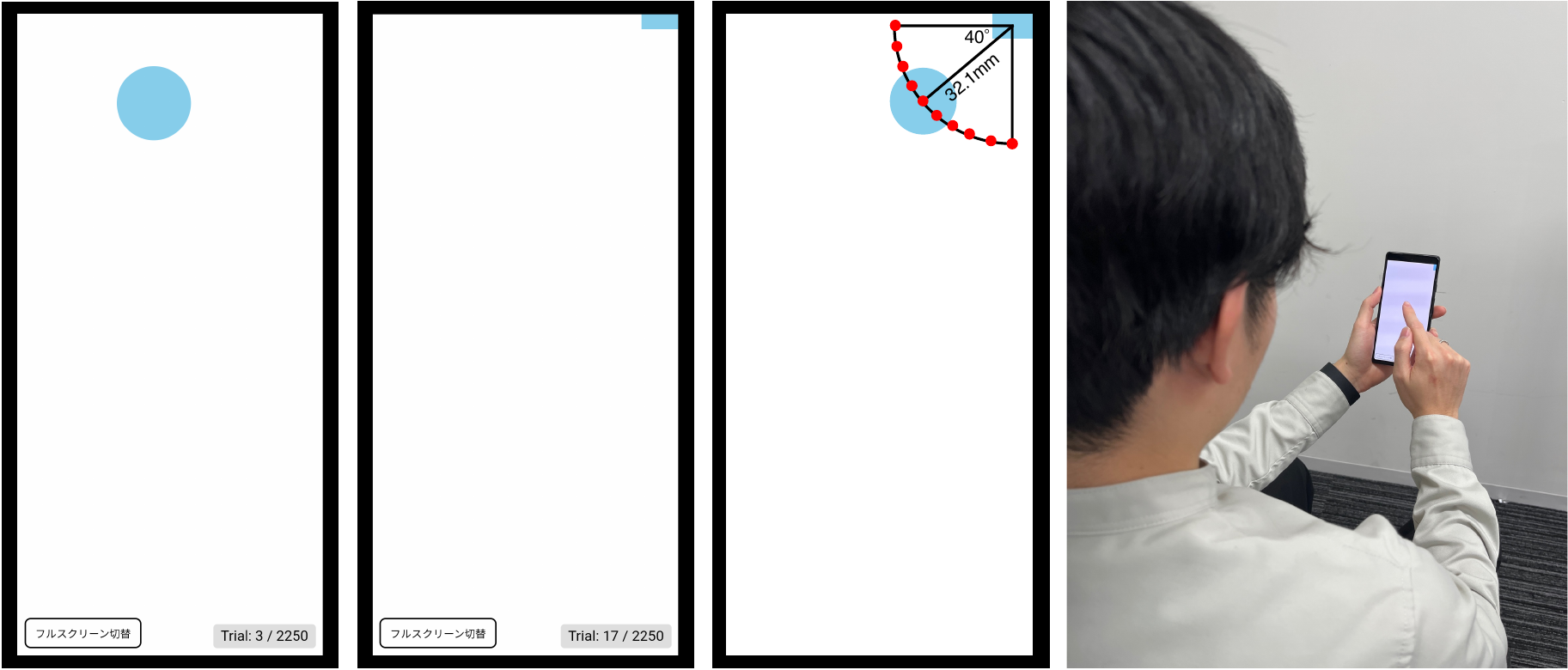}
    \caption{(Left) Experiment 3 employed an on-screen-start task. After tapping a large circular starting target, participants tapped (Middle Left) a rectangular target in the top-right corner. (Middle Right) The starting target was positioned along a circular arc with a radius of 32.1 mm, related to the rectangular target. (Right) Participants did not return their hand to their knee but started each trial from the starting target.}
    \label{fig:Ex3 Setup}
\end{figure}

The independent variables were $W$, $H$, $\margin_x$, and $\margin_y$.
In Experiments 1 and 2, the initial hand position affected the tap-coordinate distributions.
Thus, in Experiment 3, we balanced the angle at which the finger approached the target by modifying the task to an on-screen-start pointing (where a starting target within the screen serves as the starting point), requiring participants to return their hand to the starting target after each trial.
Since it is known that the Dual Gaussian Distribution Model can estimate tap success rates with high accuracy even for on-screen-start tasks \citep{Yamanaka2020Rethinking}, we verified whether the same holds true for our model.

The starting targets were positioned at 10-degree intervals (0°, 10°, 20°, 30°, 40°, 50°, 60°, 70°, 80°, and 90°) along a circular arc with a radius of 32.1 mm, located to the lower-left of the next target to be displayed (\autoref{fig:Ex3 Setup} (Middle Right)).
The ten angle conditions were presented within each block, and the experiment consisted of ten such repeating \textit{sets} (all angles were presented over the course of 10 repeating sets).
The size of the starting target was set to 15.60 mm in diameter.

\subsubsection{Design}
The experiment followed a $5 \margin_x \times 5 \margin_y \times 3 W \times 3 H$ within-subjects design. 
Their specific values were as follows: $\margin_x$ was 0, 1.560, 3.900, 7.018, and 11.70 mm, $\margin_y$ was 0, 1.560, 3.900, 7.018, and 11.70 mm, $W$ was 3.119, 5.459, and 7.798 mm, and $H$ was 3.119, 5.459, and 7.798 mm (\autoref{fig:ModelAndExperimentParameters}). 
One set consisted of 225 trials presented in a randomized order ($5 \margin_x \times 5 \margin_y \times 3 W \times 3 H$), and this was repeated across 10 sets, totaling 2,250 trials ($5 \margin_x \times 5 \margin_y \times 3 W \times 3 H \times 10$ sets).

\subsubsection{Participants}
Thirty-six computer science students (11 females and 25 males; mean age 20.5 years, SD 1.89 years; 32 right-handed and 4 left-handed) independently participated in the experiment. 
They received 2,000 JPY (13.51 USD) as compensation.

\subsection{Results}
\subsubsection{Outliers}
In Experiment 3, we applied a data removal criterion different from those used in Experiments 1 and 2. 
If the same criteria had been applied, we would have excluded the first set of 225 trials, which would be excessive compared to Experiments 1 and 2 (45 trials). 
Furthermore, the 3SD criterion using only the 10 repetitions available for each condition presents a mathematical issue: because of the small sample size, no trials were identified as outliers.

To address these issues, we excluded the first 45 trials for each participant as practice and identified trial-level outliers using the Interquartile Range (IQR) method.
Although the target conditions for practice trials varied across participants, the ordering effect was reduced by randomizing them.
Furthermore, since the data were averaged within each condition and subsequently aggregated across participants, these variations are unlikely to change our conclusions.
Specifically, we computed the distribution of tap-coordinates for each participant and each condition, and then excluded any trials located beyond 3 IQR from the quartiles.
As a result, 1,620 practice trials ($45 \text{ trials} \times 36 \text{ participants}$) and 1,637 outlier trials (2.06\%) were removed. 
The subsequent analyses were conducted using the remaining 77,743 trials.

\subsubsection{Tests for the Normality of Tap-Coordinate Distributions}
\begin{figure}[ht]
\centering
\begin{minipage}[b]{0.49\textwidth}
    \centering
    \includegraphics[width = 0.7\columnwidth]{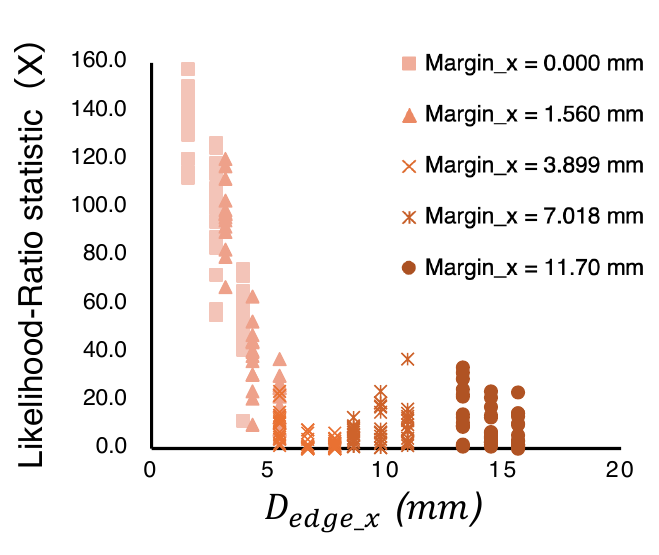}
    \subcaption{$x$-axis}
    \label{fig:Ex3 LikelihoodRatio X}
\end{minipage}
\begin{minipage}[b]{0.49\textwidth}
    \centering
    \includegraphics[width = 0.7\columnwidth]{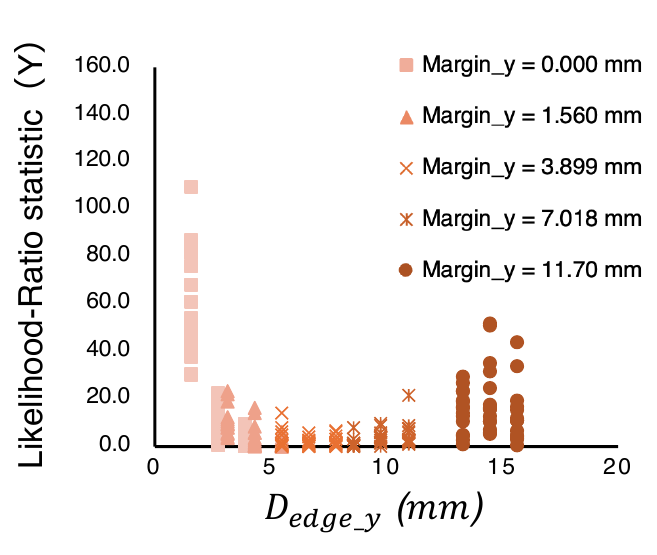}
    \subcaption{$y$-axis}
    \label{fig:Ex3 LikelihoodRatio Y}
\end{minipage}
\caption{Likelihood ratio statistics in Experiment 3.}
\label{fig:Ex3 LikelihoodRatio}
\end{figure}

\noindent The Shapiro--Wilk test revealed that normality was rejected in 179 out of 225 target conditions ($79.6\%$) for the $x$-axis and 133 conditions ($59.1\%$) for the $y$-axis.
Consistent with Experiments 1 and 2, these results show the limitations of the existing model's assumption of normality.

In the likelihood ratio test comparing the fit between normal and skew-normal distributions, the statistics increased as $\dedgex$ and $\dedgey$ approached zero (\autoref{fig:Ex3 LikelihoodRatio}). 
Similar to Experiments 1 and 2, the fit to the skew-normal distribution became significantly higher as the target approached the screen edge, supporting the assumptions of the proposed model. 
However, unlike Experiment 2, for the $y$-axis, the likelihood ratio statistics did not show a clear increase except for the $\margin_y = 0$ condition, and a tendency to follow the skew-normal distribution was observed even in conditions where $\dedgey$ was large.

\subsubsection{Statistical Tests}
\begin{figure}[p]
\centering
\begin{minipage}[b]{0.32\textwidth}
    \centering
    \includegraphics[width = 0.9\columnwidth]{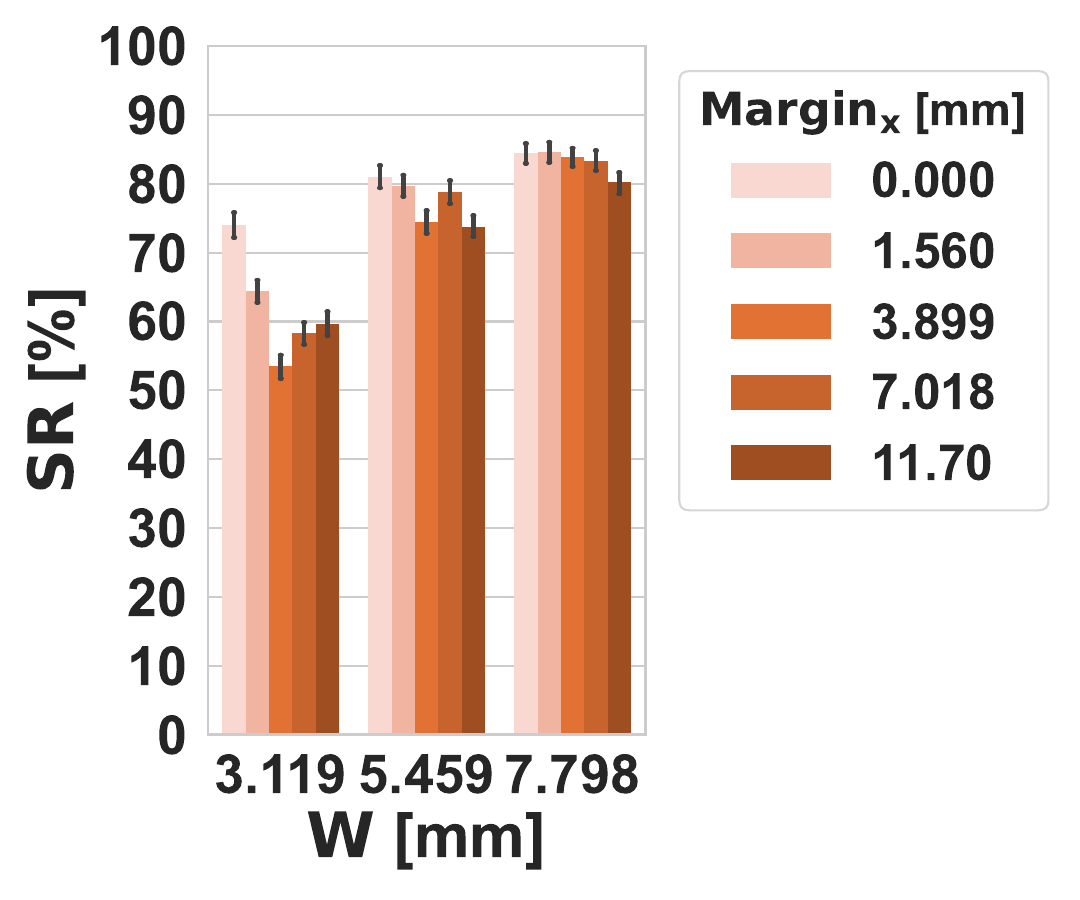}
    \subcaption{$\sr$: $x$-axis factors}
    \label{fig:Ex3 SR Interaction X}
\end{minipage}
\begin{minipage}[b]{0.32\textwidth}
    \centering
    \includegraphics[width = 0.9\columnwidth]{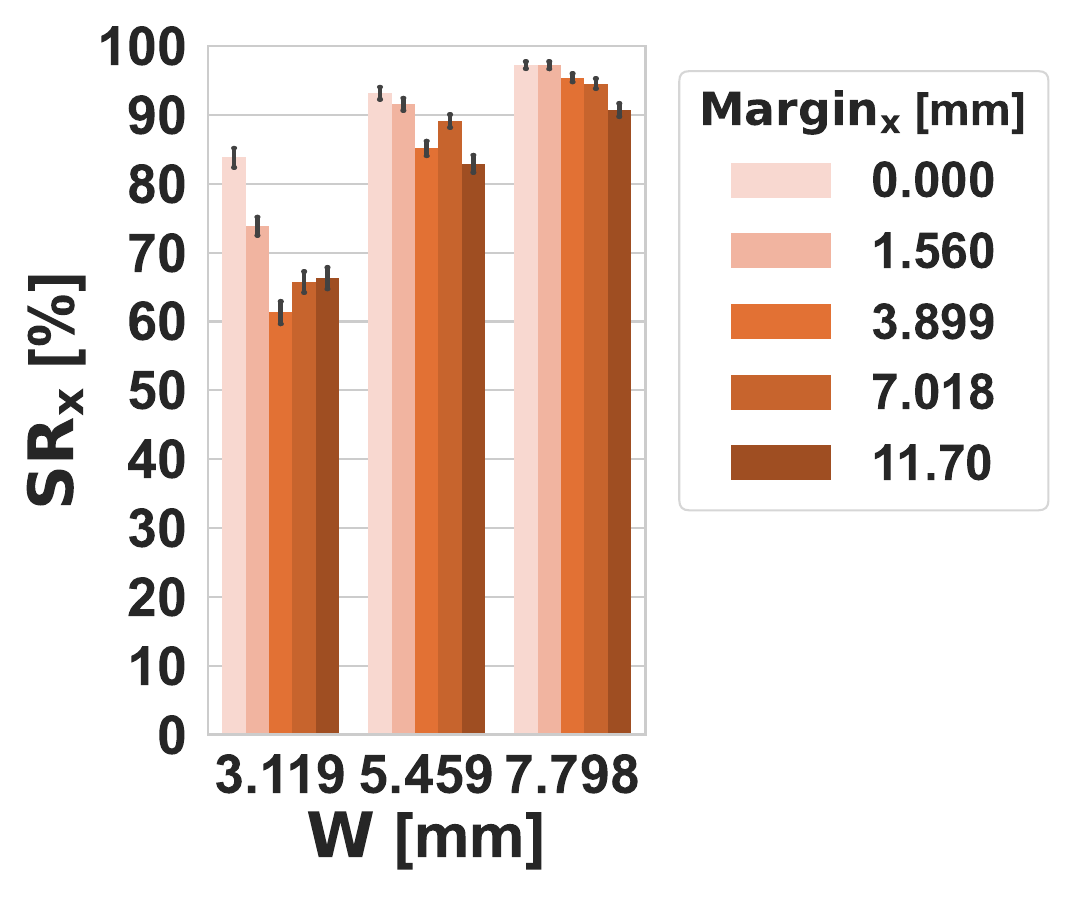}
    \subcaption{$\sr_x$: $x$-axis factors}
    \label{fig:Ex3 SRx Interaction X}
\end{minipage}
\begin{minipage}[b]{0.32\textwidth}
    \centering
    \includegraphics[width = 0.9\columnwidth]{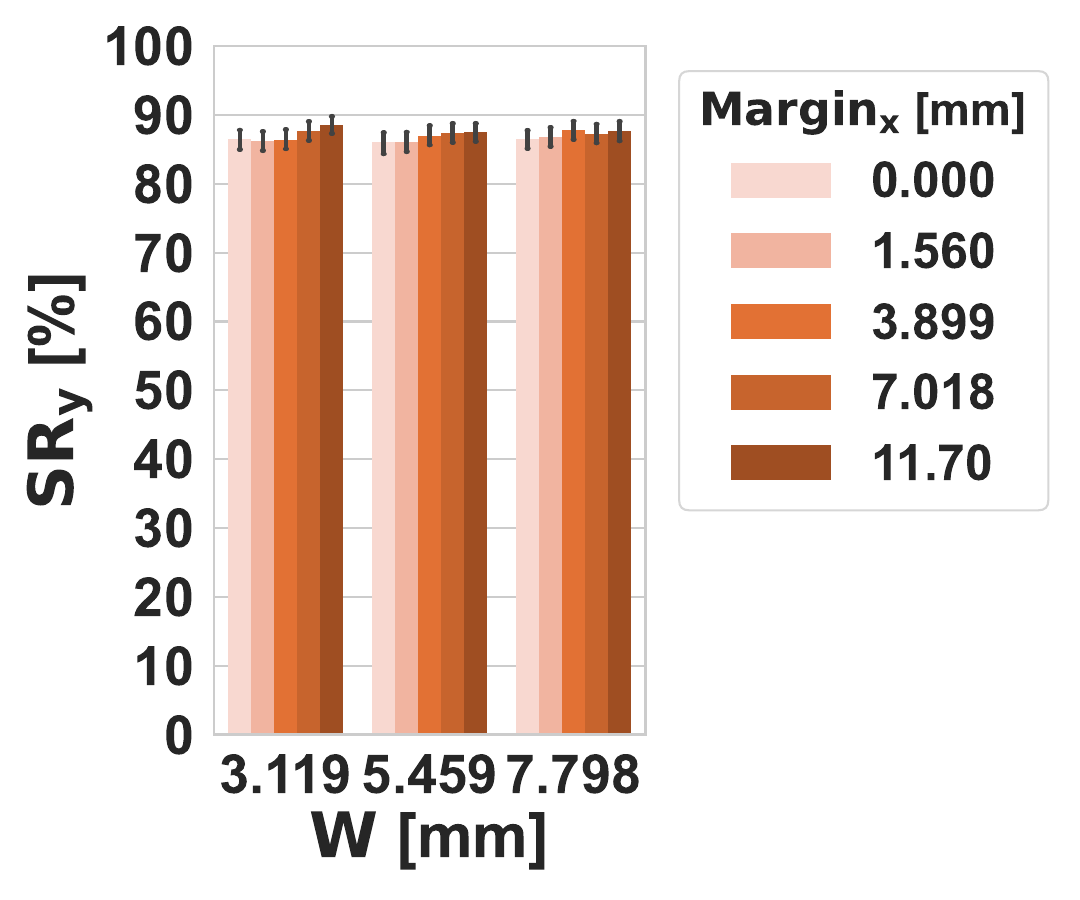}
    \subcaption{$\sr_y$: $x$-axis factors}
    \label{fig:Ex3 SRy Interaction X}
\end{minipage}
\begin{minipage}[b]{0.32\textwidth}
    \centering
    \includegraphics[width = 0.9\columnwidth]{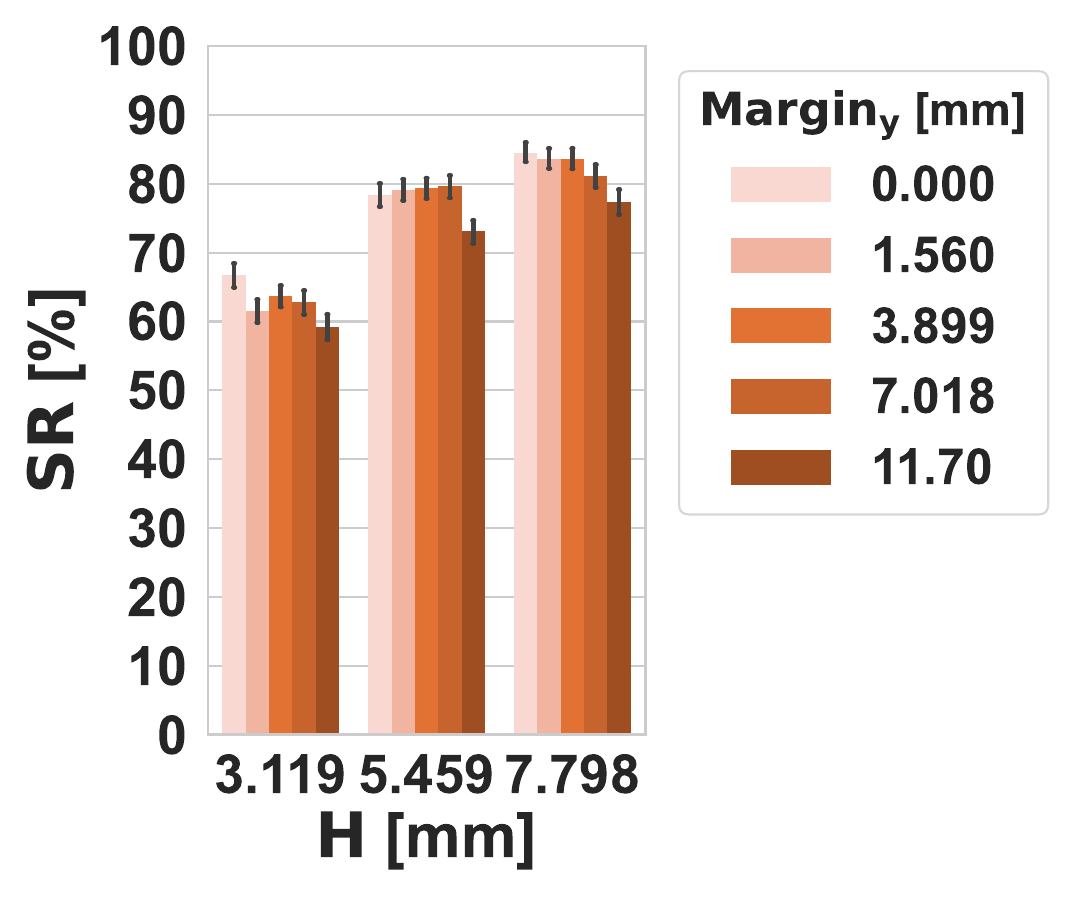}
    \subcaption{$\sr$: $y$-axis factors}
    \label{fig:Ex3 SR Interaction Y}
\end{minipage}
\begin{minipage}[b]{0.32\textwidth}
    \centering
    \includegraphics[width = 0.9\columnwidth]{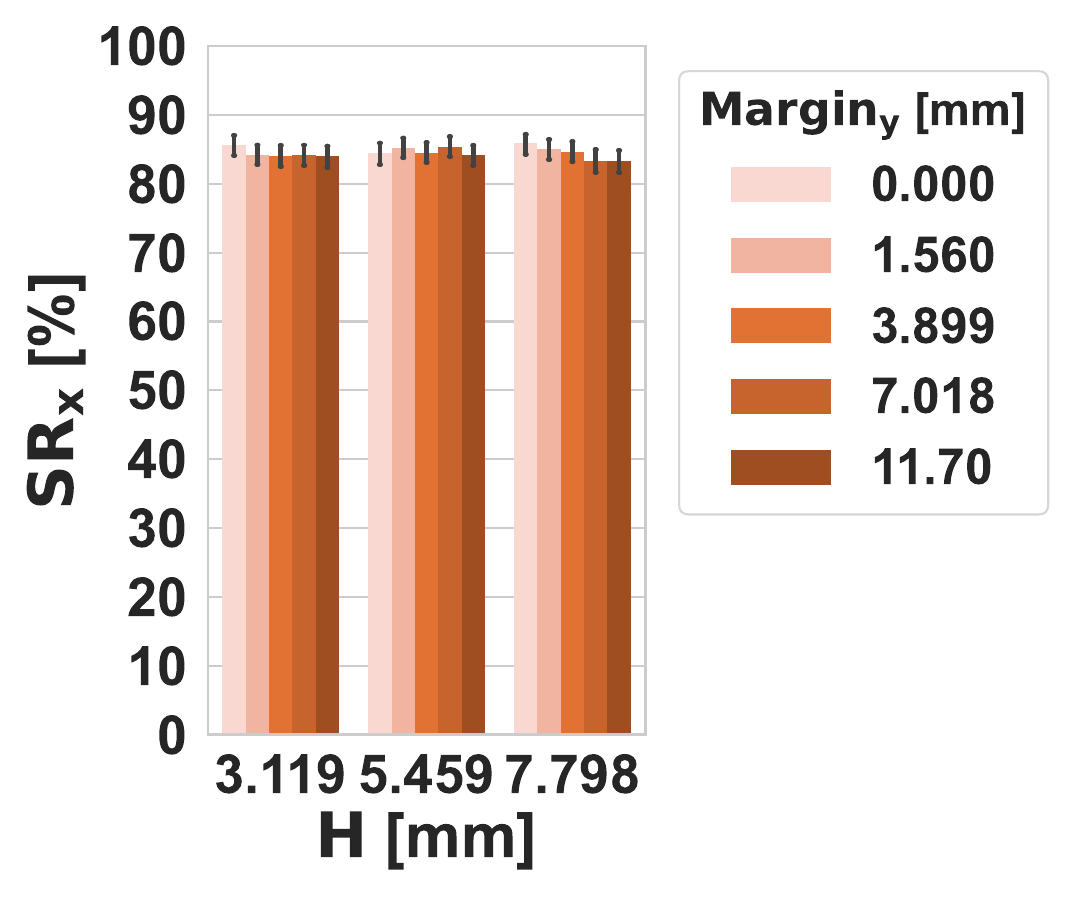}
    \subcaption{$\sr_x$: $y$-axis factors}
    \label{fig:Ex3 SRx Interaction Y}
\end{minipage}
\begin{minipage}[b]{0.32\textwidth}
    \centering
    \includegraphics[width = 0.9\columnwidth]{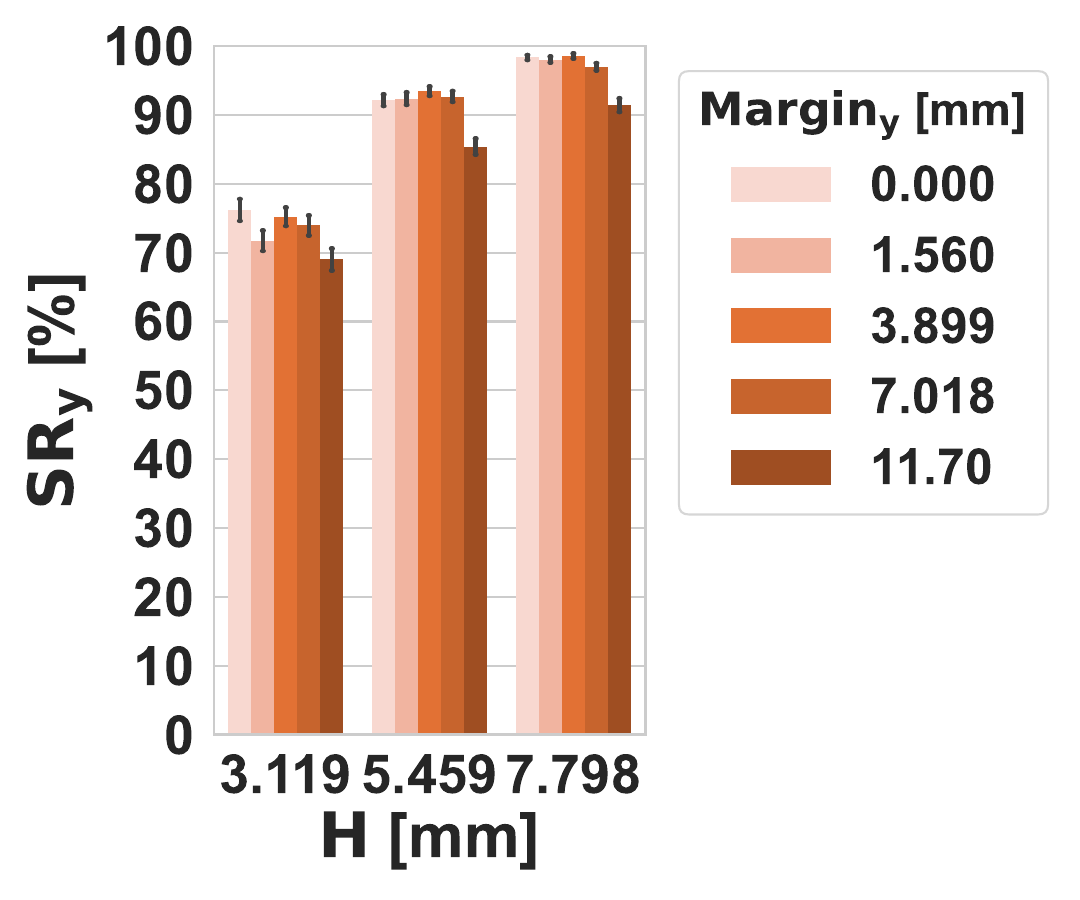}
    \subcaption{$\sr_y$: $y$-axis factors}
    \label{fig:Ex3 SRy Interaction Y}
\end{minipage}
\begin{minipage}[b]{0.24\textwidth}
    \centering
    \includegraphics[width = 0.9\columnwidth]{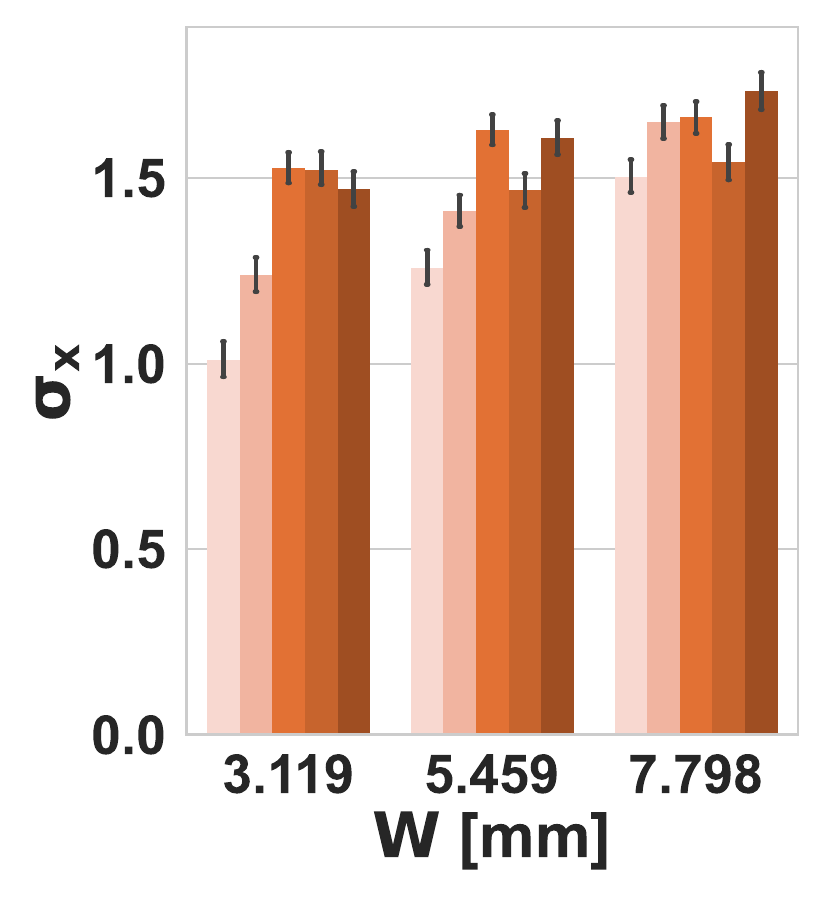}
    \subcaption{$\sigma_x$: $x$-axis factors}
    \label{fig:Ex3 SigmaX Interaction X}
\end{minipage}
\begin{minipage}[b]{0.24\textwidth}
    \centering
    \includegraphics[width = 0.9\columnwidth]{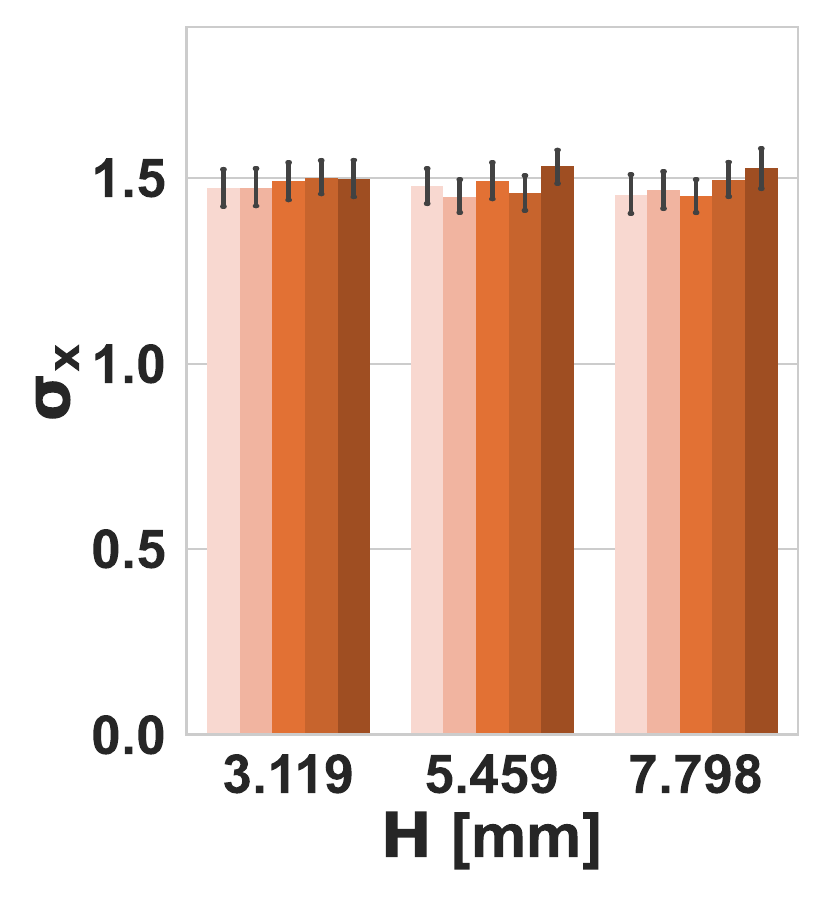}
    \subcaption{$\sigma_x$: $y$-axis factors}
    \label{fig:Ex3 SigmaX Interaction Y}
\end{minipage}
\begin{minipage}[b]{0.24\textwidth}
    \centering
    \includegraphics[width = 0.9\columnwidth]{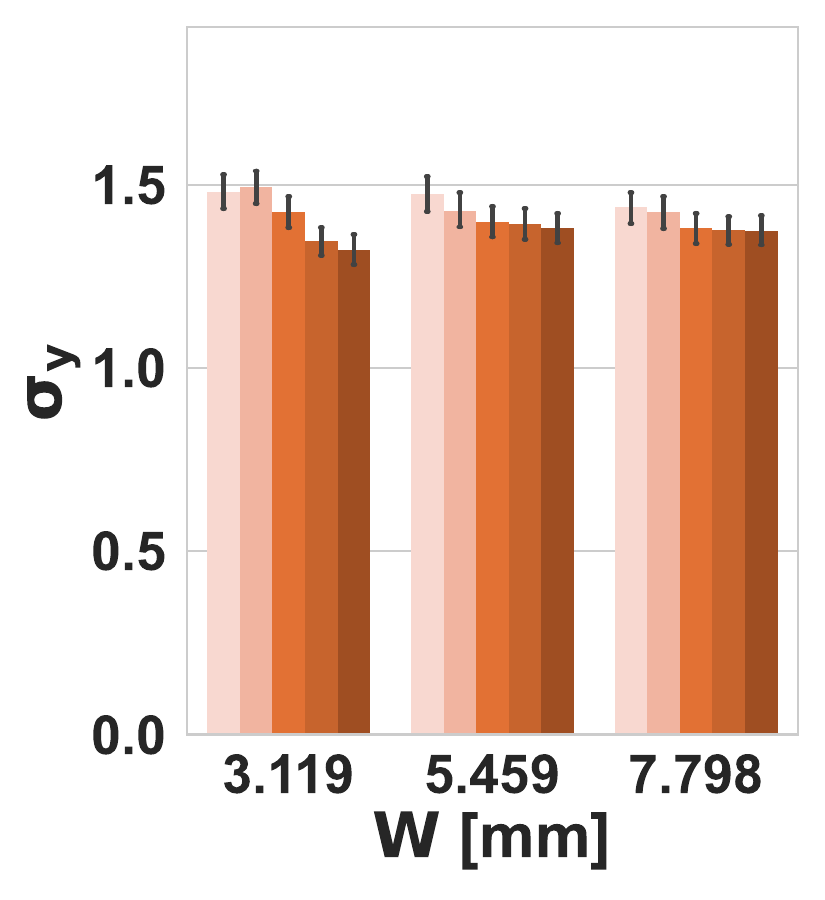}
    \subcaption{$\sigma_y$: $x$-axis factors}
    \label{fig:Ex3 SigmaY Interaction X}
\end{minipage}
\begin{minipage}[b]{0.24\textwidth}
    \centering
    \includegraphics[width = 0.9\columnwidth]{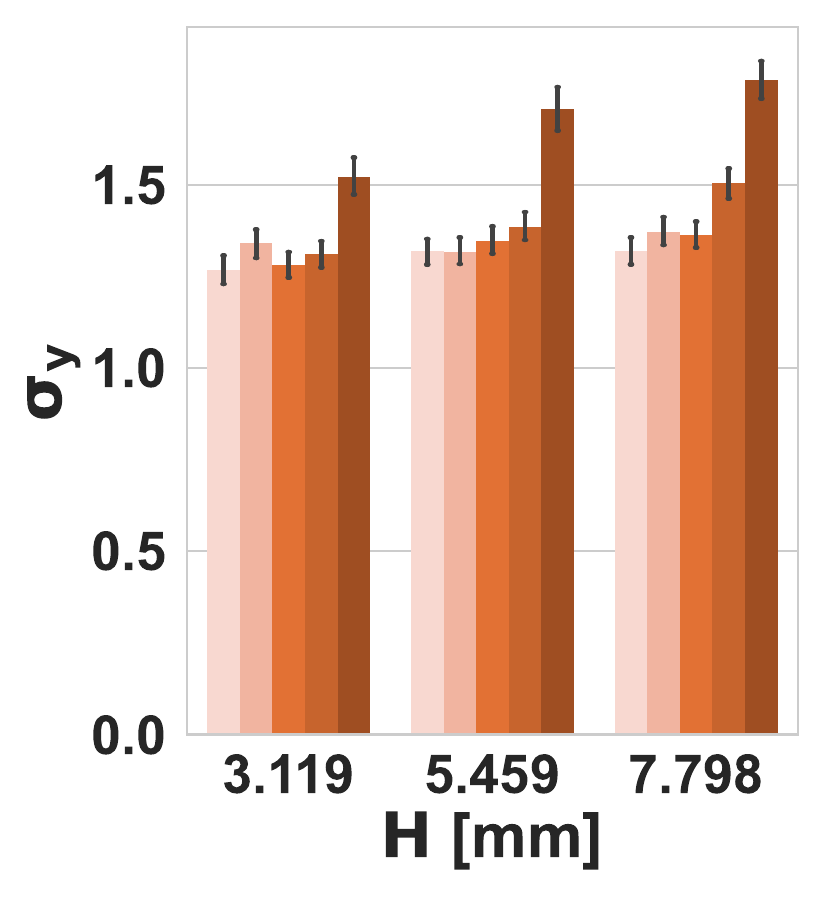}
    \subcaption{$\sigma_y$: $y$-axis factors}
    \label{fig:Ex3 SigmaY Interaction Y}
\end{minipage}
\begin{minipage}[b]{0.24\textwidth}
    \centering
    \includegraphics[width = 0.9\columnwidth]{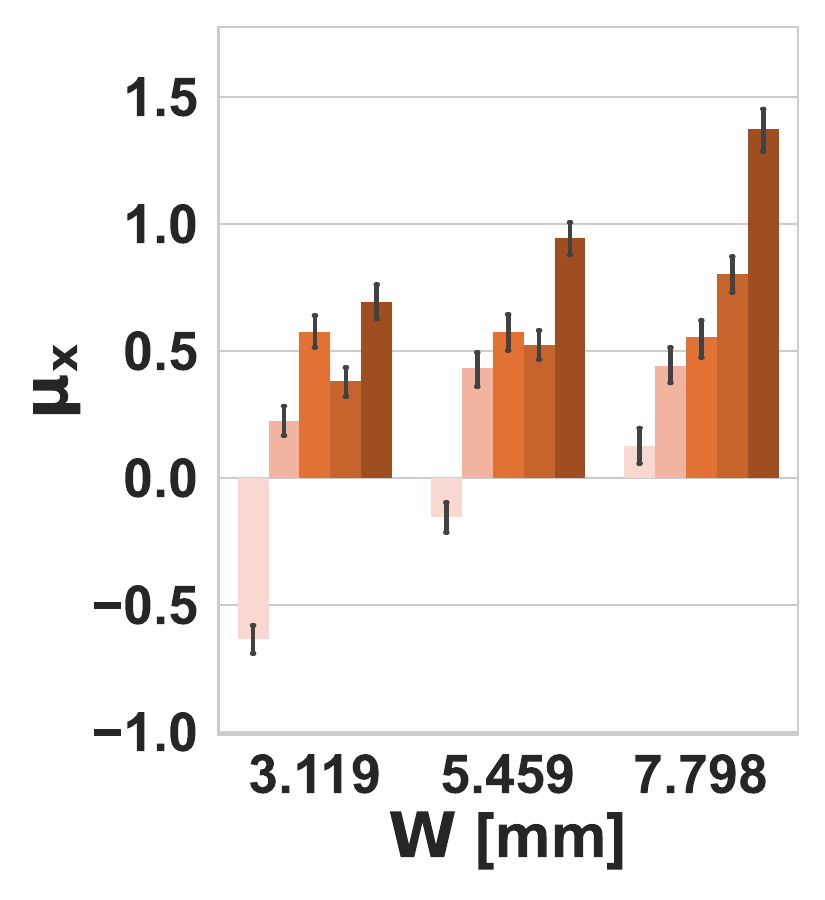}
    \subcaption{$\mu_x$: $x$-axis factors}
    \label{fig:Ex3 MuX Interaction X}
\end{minipage}
\begin{minipage}[b]{0.24\textwidth}
    \centering
    \includegraphics[width = 0.9\columnwidth]{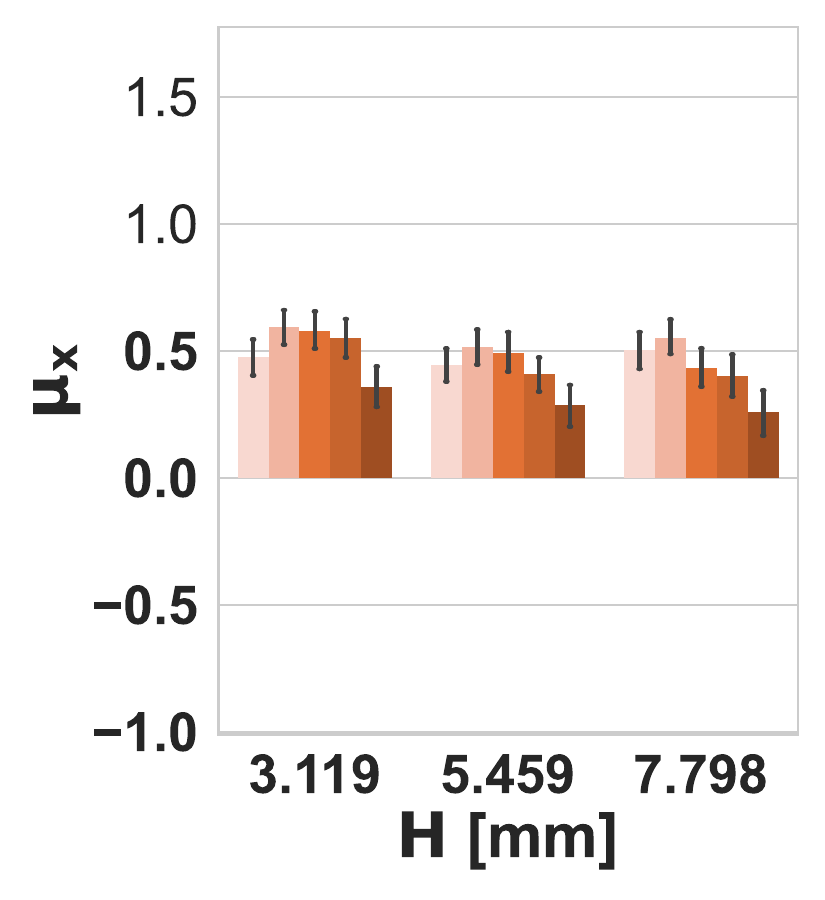}
    \subcaption{$\mu_x$: $y$-axis factors}
    \label{fig:Ex3 MuX Interaction Y}
\end{minipage}
\begin{minipage}[b]{0.24\textwidth}
    \centering
    \includegraphics[width = 0.9\columnwidth]{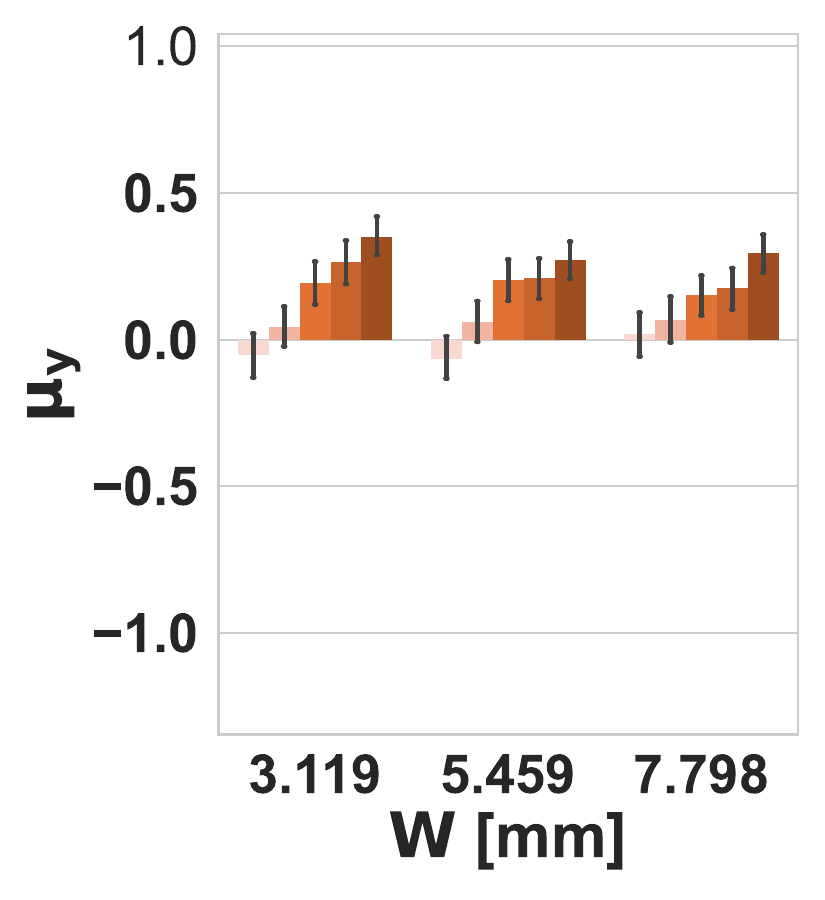}
    \subcaption{$\mu_y$: $x$-axis factors}
    \label{fig:Ex3 MuY Interaction X}
\end{minipage}
\begin{minipage}[b]{0.24\textwidth}
    \centering
    \includegraphics[width = 0.9\columnwidth]{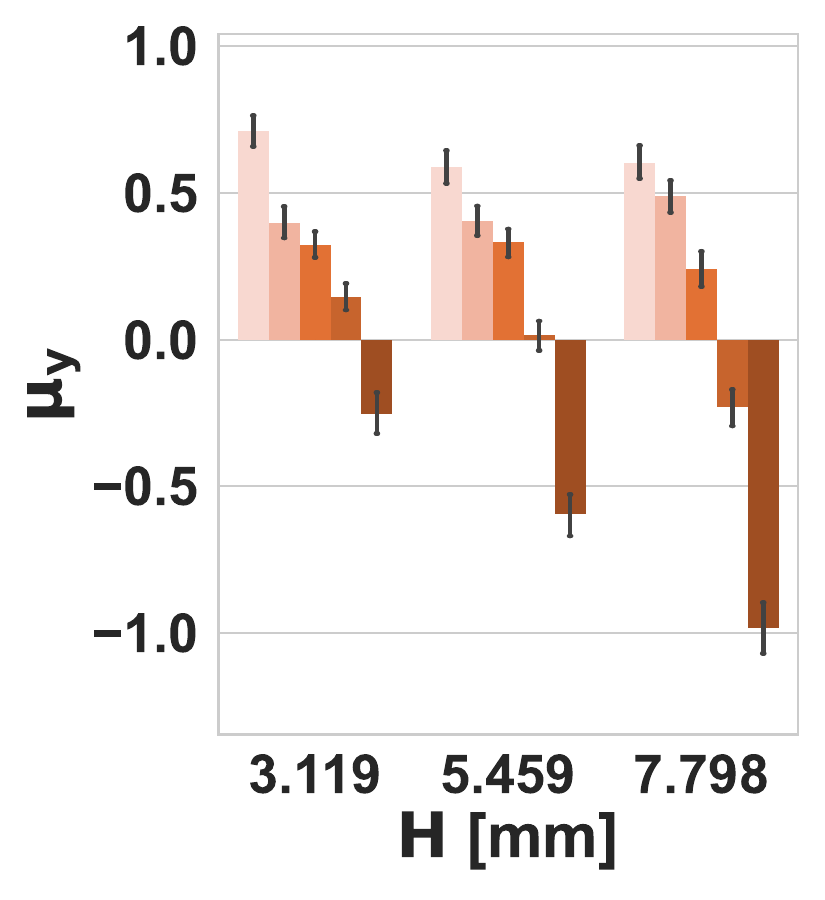}
    \subcaption{$\mu_y$: $y$-axis factors}
    \label{fig:Ex3 MuY Interaction Y}
\end{minipage}
\begin{minipage}[b]{0.24\textwidth}
    \centering
    \includegraphics[width = 0.9\columnwidth]{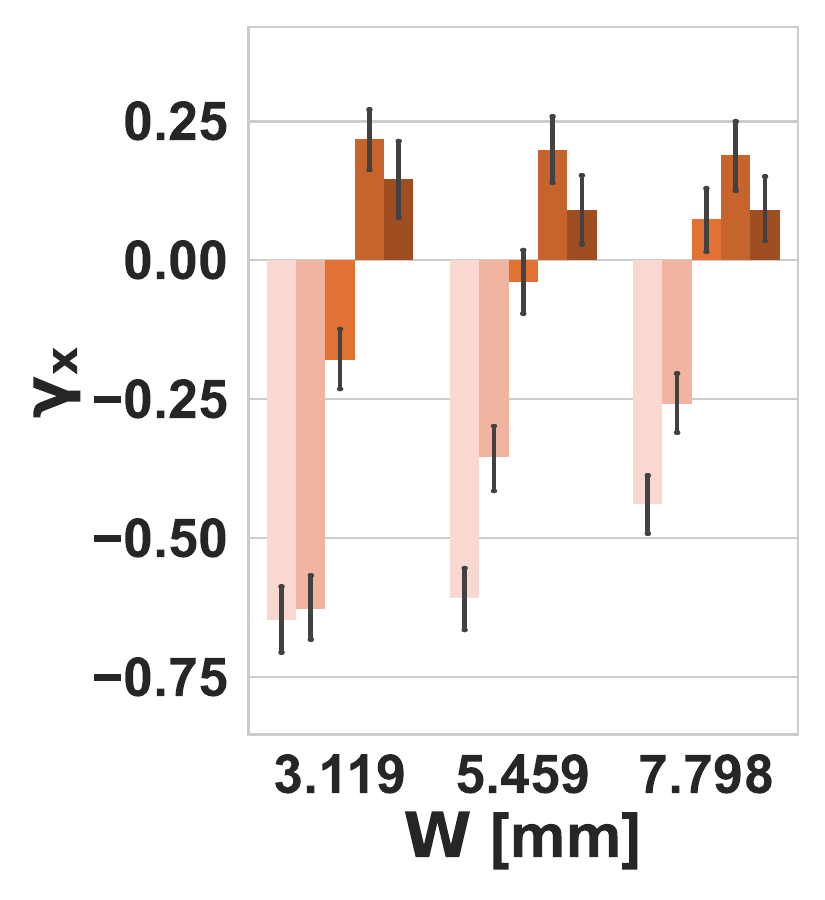}
    \subcaption{$\gammax$: $x$-axis factors}
    \label{fig:Ex3 GammaX Interaction X}
\end{minipage}
\begin{minipage}[b]{0.24\textwidth}
    \centering
    \includegraphics[width = 0.9\columnwidth]{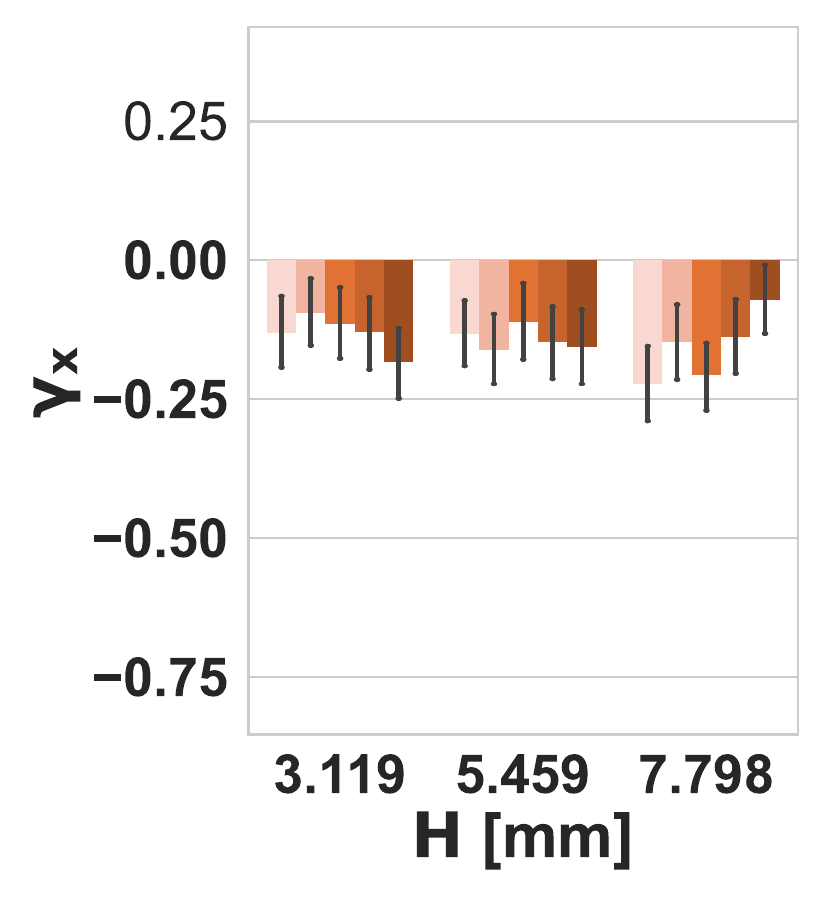}
    \subcaption{$\gammax$: $y$-axis factors}
    \label{fig:Ex3 GammaX Interaction Y}
\end{minipage}
\begin{minipage}[b]{0.24\textwidth}
    \centering
    \includegraphics[width = 0.9\columnwidth]{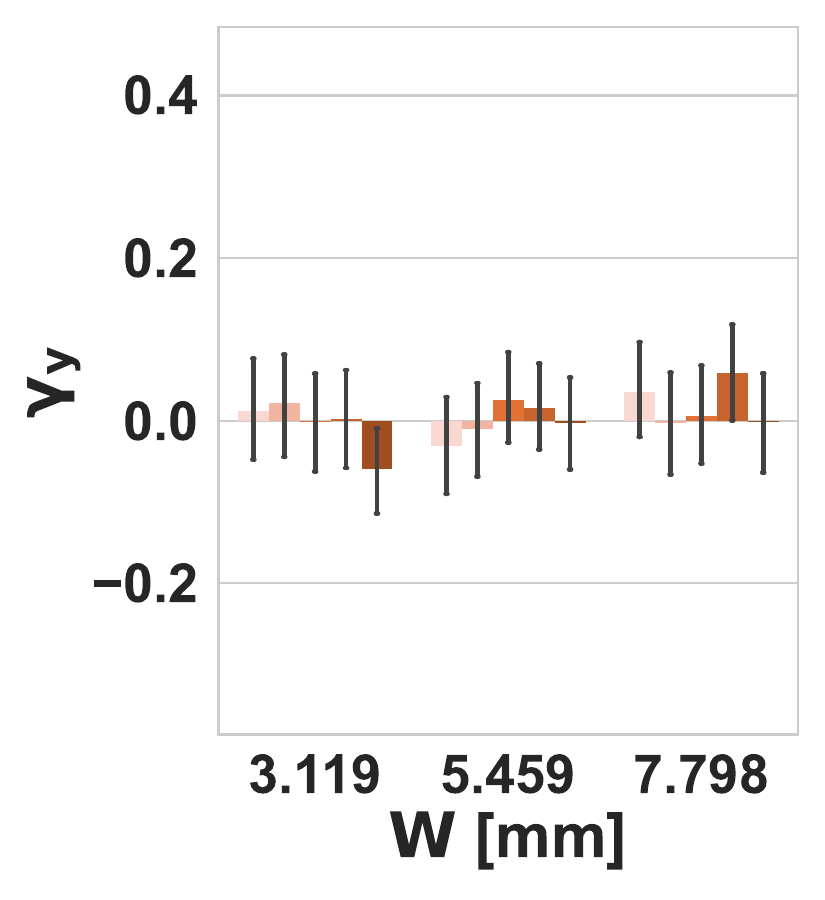}
    \subcaption{$\gammay$: $x$-axis factors}
    \label{fig:Ex3 GammaY Interaction X}
\end{minipage}
\begin{minipage}[b]{0.24\textwidth}
    \centering
    \includegraphics[width = 0.9\columnwidth]{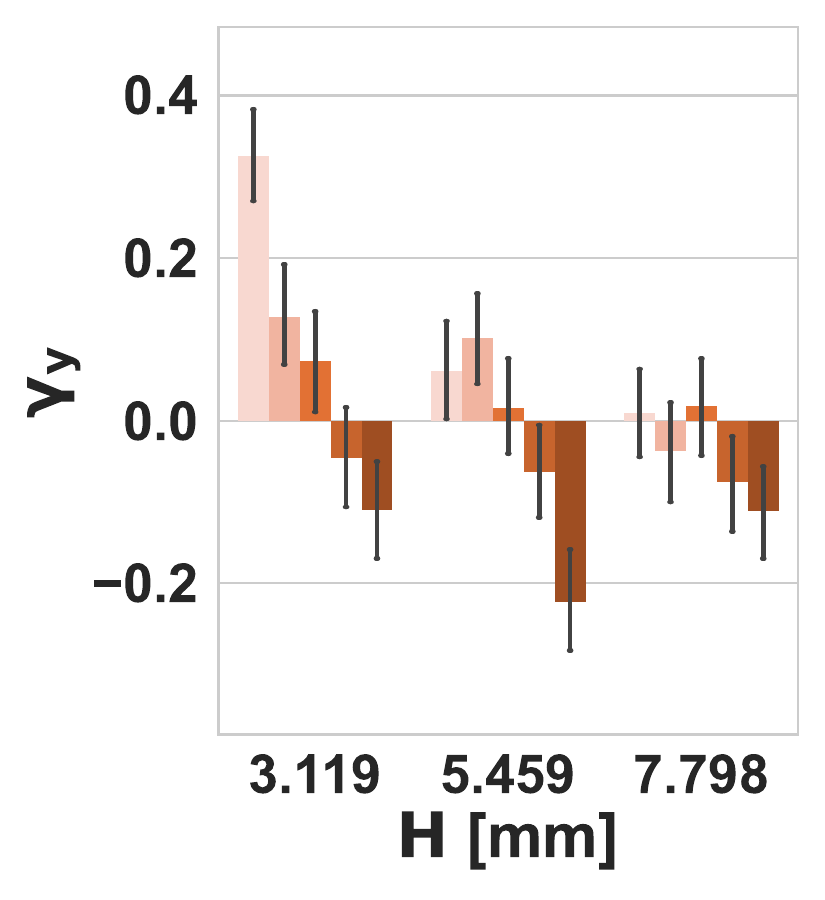}
    \subcaption{$\gammay$: $y$-axis factors}
    \label{fig:Ex3 GammaY Interaction Y}
\end{minipage}
\caption{Influence of factors ($W$, $H$, $\margin_x$, $\margin_y$) on the $\sr$, $\sr_x$, $\sr_y$, $\sigma_x$, $\sigma_y$, $\mu_x$, $\mu_y$, $\gammax$, and $\gammay$ in Experiment 3.}
\label{fig:Ex3 Sigma Interaction}
\end{figure}
\noindent We performed a four-way RM-ANOVA using $\margin_x$, $\margin_y$, $W$, and $H$.
Similar to Experiment 1, $\margin_x$ affected the tap success rate ($\sr$); however, $\margin_y$ had less impact compared to $\margin_x$ and Experiment 2.
The influence of factors from one axis on the success rate of the other was minimal, suggesting the validity of a 2D model that independently calculates and then multiplies $\sr_x$ and $\sr_y$.
Regarding $\sigma$, it first increased and then decreased as the target approached the screen edge along the $x$-axis (as in Experiment 1), but mostly decreased along the $y$-axis.
For $\mu$, a quadratic trend was not clearly observed in either axis.
For $\gamma_1$, both axes exhibited a trend to support our model: its absolute value increased as it approached the edge.

\subsubsection{Prediction Accuracy of Tap Success Rate}
\begin{table}[ht]
\caption{Regression constants and model accuracy metrics in Experiment 3}
\centering
\resizebox{\textwidth}{!}{
\begin{tabular}{c|c|c|c|c|c|c|c|c}
\multicolumn{2}{c|}{} & \multicolumn{5}{c|}{Regression Analysis} & \multicolumn{2}{c}{LOOCV}\\
\hline
Model & Formula & Regression Constants & $R^2$ & $\mae$ & $ \rmse$ & $\mape$ & $R^2$ & $\mae$\\
\hline
Dual Gauss. & $\sigma_x^2$ (\autoref{formula:Bi_sigma}) & $a_x=0.0175$, $b_x=2.30$ & $.314$ & $0.441$ & $0.543$ & $17.7\%$ & $.301$ & $0.445$\\
\cline{2-9}
& $\sr_x$ (\autoref{formula:DualGaussian1D}) & - & $.750$ & $4.36$ & $6.01$ & $5.48\%$ & $.747$ & $4.38$\\
\cline{2-9}
& $\sigma_y^2$ (\autoref{formula:Bi_sigma}) & $a_y=0.0107$, $b_y=2.15$ & $.089$ & $0.534$ & $0.723$ & $20.2\%$ & $.072$ & $0.539$\\
\cline{2-9}
& $\sr_y$ (\autoref{formula:DualGaussian1D}) & - & $.865$ & $2.80$ & $3.90$ & $3.42\%$ & $.863$ & $2.82$\\
\cline{2-9}
& $\sr$ (\autoref{formula:DualGaussianRectangle}) & - & $.784$ & $4.71$ & $6.35$ & $6.64\%$ & $.782$ & $4.73$\\
\hline
Skewed-Dual. & $\gamma_{1_x}$ (\autoref{Formula:DualSkewNormalEstimateGamma1X}) & $c_x=2.48$, $d_x=-0.370$ & $.806$ & $0.272$ & $0.363$ & $63.5\%$ & $.805$ & $0.274$\\
\cline{2-9}
& $\sigma_x$ (\autoref{Formula:DualSkewNormalEstimateSigmaX}) & \begin{tabular}{c}$e_x=1.42$, $f_x=0.0249$\\$g_x=0.295$, $h_x=2.69$\\$i_x=0.0128$\end{tabular} & $.679$ & $0.093$ & $0.114$ & $5.64\%$ & $.665$ & $0.095$\\
\cline{2-9}
& $\mu_x$ (\autoref{Formula:DualSkewNormalEstimateMuX}) & \begin{tabular}{c}$j_x=0.560$, $k_x=-0.0548$\\$l_x=6.20$\end{tabular} & $.730$ & $0.179$ & $0.232$ & $133\%$ & $.718$ & $0.183$\\
\cline{2-9}
& $\sr_x$ (\autoref{Formula:DualSkewSR1D}) & - & $.917$ & $2.60$ & $3.46$ & $3.18\%$ & $.915$ & $2.62$\\
\cline{2-9}
& \begin{tabular}{c}$\gamma_{1_y}$ (\autoref{Formula:DualSkewNormalEstimateGamma1X}\\ using $\dedgey$)\end{tabular} & $c_y=1.16$, $d_y=-0.210$ & $.417$ & $0.278$ & $0.363$ & $124\%$ & $.413$ & $0.280$\\
\cline{2-9}
& \begin{tabular}{c}$\sigma_y$ (\autoref{Formula:DualSkewNormalEstimateSigmaX}\\using $H$ \& $\margin_y$)\end{tabular} & \begin{tabular}{c}$e_y=2.07$, $f_y=0.00304$\\$g_y=-0.0299$, $h_y=2.38$\\$i_y=0.0127$\end{tabular} & $.305$ & $0.140$ & $0.183$ & $8.70\%$ & $.269$ & $0.145$\\
\cline{2-9}
& \begin{tabular}{c}$\mu_y$ (\autoref{Formula:DualSkewNormalEstimateMuX}\\ using $\dedgey$)\end{tabular} & \begin{tabular}{c}$j_y=0.408$, $k_y=0.0158$\\$l_y=5.86$\end{tabular} & $.226$ & $0.144$ & $0.177$ & $42.4\%$ & $.186$ & $0.149$\\
\cline{2-9}
& $\sr_y$ (\autoref{Formula:DualSkewSR1D}) & - & $.881$ & $2.76$ & $3.65$ & $3.34\%$ & $.874$ & $2.85$\\
\cline{2-9}
& $\sr$ (\autoref{formula:DualSkewRect}) & - & $.888$ & $3.52$ & $4.57$ & $4.92\%$ & $.883$ & $3.59$\\
\hline
\multicolumn{9}{c}{Machine learning models using default hyperparameters}\\
\hline
Lasso Regression & $\sr$ & number of parameters: 5 & .795 & 4.93 & 6.19 & 7.15\% & .786 & 5.04 \\
\hline
Random Forest & $\sr$ & number of parameters: 28,312 & .991 & 0.979 & 1.28 & 1.43\% & .938 & 2.64 \\
\hline
SVR & $\sr$ & number of parameters: 225 & .663 & 6.56 & 7.93 & 10.0\% & .620 & 7.12 \\
\hline
MLP Neural Net & $\sr$ & number of parameters: 601 & -.015 & 11.0 & 13.8 & 15.1\% & .076 & 10.5 \\
\hline
\multicolumn{9}{c}{Machine learning models using Bayesian-optimized hyperparameters}\\
\hline
Lasso Regression & $\sr$ & number of parameters: 5 & .798 & 4.90 & 6.13 & 7.07\% & .790 & 5.01 \\
\hline
Random Forest & $\sr$ & number of parameters: 49,217 & .982 & 1.39 & 1.82 & 2.07\% & .938 & 2.60 \\
\hline
SVR & $\sr$ & number of parameters: 181 & .973 & 1.57 & 2.26 & 2.31\% & .948 & 2.43 \\
\hline
MLP Neural Net & $\sr$ & number of parameters: 41,401 & .960 & 2.19 & 2.73 & 3.11\% & .937 & 2.73 \\
\hline
\end{tabular}
}
\label{table:Ex3 Model Regression}
\end{table}

For the proposed model and the Dual Gaussian Distribution Model, $\sr_x$ and $\sr_y$ were predicted first and then multiplied to compute $\sr$.
In contrast, the ML models directly predicted $\sr$ using the four factors ($\margin_x, \margin_y, W, H$) as input. 
In Experiments 1 and 2, regression analysis was performed on the mean parameter values calculated after determining parameters for each participant and condition.
However, in Experiment 3, since the number of repetitions per participant was smaller, regression analysis was conducted on parameters calculated from the aggregate tap-coordinate distributions of all participants.

\vspace{3mm}
\noindent\textit{Existing Model: Dual Gaussian Distribution Model}
\begin{figure}[ht]
\centering
\begin{minipage}[b]{0.49\textwidth}
    \centering
    \includegraphics[width = 0.95\columnwidth]{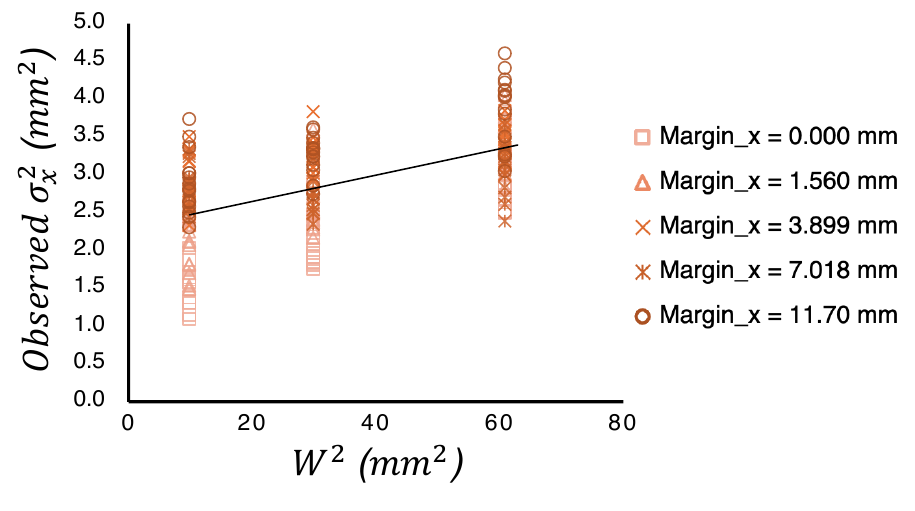}
    \subcaption{$W^2$ vs $\sigma_x^2$}
    \label{fig:Ex3_DualGauss_W_vs_Sigma}
\end{minipage}
\begin{minipage}[b]{0.49\textwidth}
    \centering
    \includegraphics[width = 0.95\columnwidth]{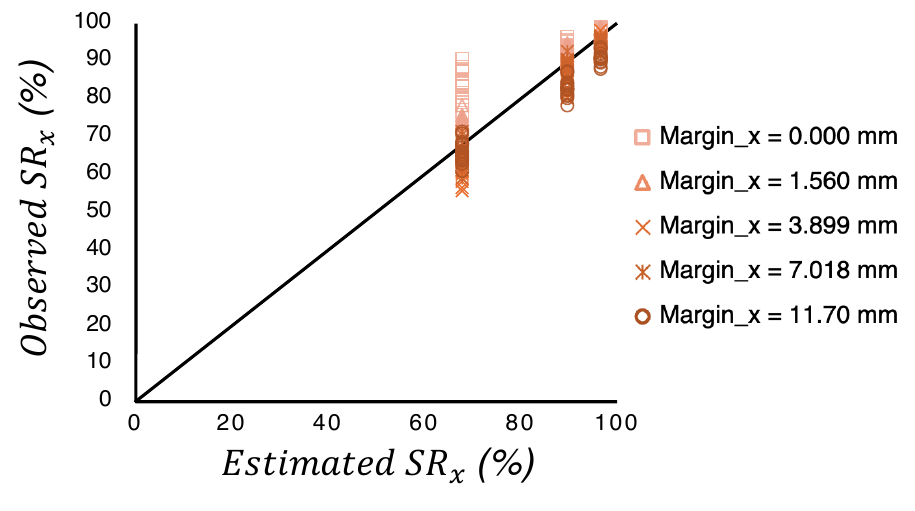}
    \subcaption{Predicted $SR_x$ vs. Observed $SR_x$}
    \label{fig:Ex3_DualGauss_SRx_vs_SRx}
\end{minipage}
\caption{Results regarding the prediction accuracy of the Dual Gaussian Distribution Model for $\sr_x$ in Experiment 3. In (a), the line represents the regression line; in (b), the line represents where predicted values match observed values.}
\label{fig:Ex3_DualGauss_X}
\end{figure}

\noindent As \autoref{fig:Ex3_DualGauss_X} shows, the estimation accuracy of $\sigma_x^2$ yielded $R^2 = .314$ (\autoref{table:Ex3 Model Regression}, \autoref{fig:Ex3_DualGauss_W_vs_Sigma}).
The estimation accuracy of $\sr_x$ yielded $R^2 = .750$ (\autoref{table:Ex3 Model Regression}, \autoref{fig:Ex3_DualGauss_SRx_vs_SRx}).
Similar to Experiment 1, a model considering only $W$ cannot fully capture the influence of the screen edges, and the observed $\sr_x$ exceeded the predicted value.

\begin{figure}[ht]
\centering
\begin{minipage}[b]{0.49\textwidth}
    \centering
    \includegraphics[width = 0.95\columnwidth]{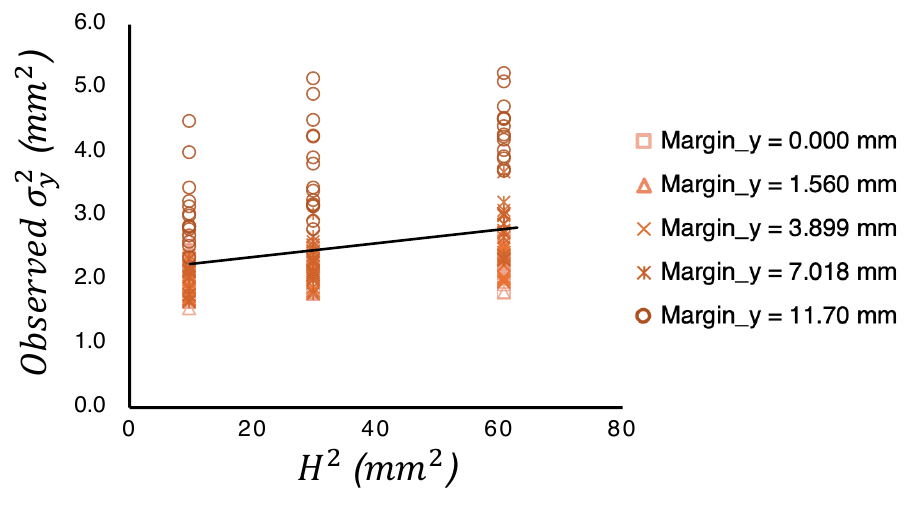}
    \subcaption{$H^2$ vs $\sigma_y^2$}
    \label{fig:Ex3_DualGauss_H_vs_Sigma}
\end{minipage}
\begin{minipage}[b]{0.49\textwidth}
    \centering
    \includegraphics[width = 0.95\columnwidth]{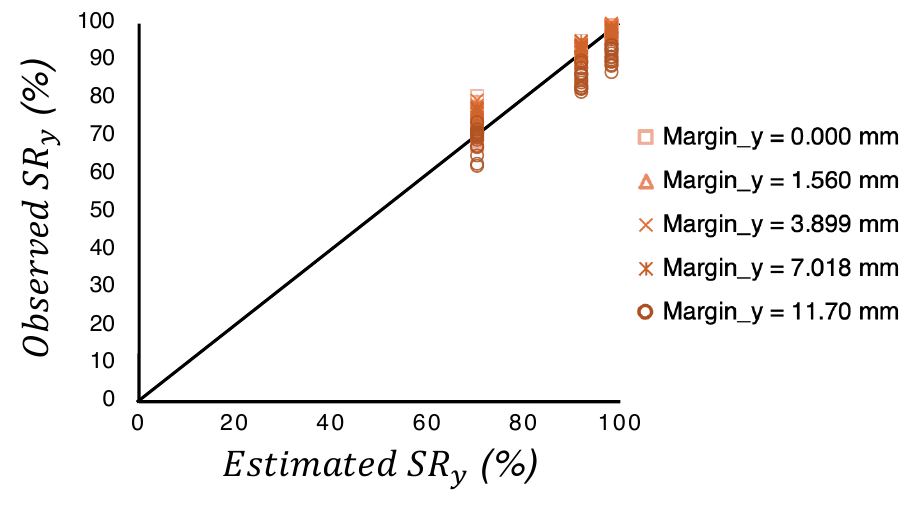}
    \subcaption{Predicted $SR_y$ vs. Observed $SR_y$}
    \label{fig:Ex3_DualGauss_SRy_vs_SRy}
\end{minipage}
\caption{Results regarding the prediction accuracy of the Dual Gaussian Distribution Model for $\sr_y$ in Experiment 3. In (a), the line represents the regression line; in (b), the line represents where predicted values match observed values.}
\label{fig:Ex3_DualGauss_Y}
\end{figure}

\autoref{fig:Ex3_DualGauss_Y} shows that the estimation accuracy of $\sigma_y^2$ was $R^2 = .089$ (\autoref{table:Ex3 Model Regression}, \autoref{fig:Ex3_DualGauss_H_vs_Sigma}) and that of $\sr_y$ was $R^2 = .865$ (\autoref{table:Ex3 Model Regression}, \autoref{fig:Ex3_DualGauss_SRy_vs_SRy}).
While this represents higher estimation accuracy compared to Experiment 2 ($R^2 = .699$), we observed the same trend where the observed $\sr_y$ exceeded the predicted value at $\margin_y = 0$.

\begin{figure}[ht]
    \centering
    \includegraphics[width = 0.4\columnwidth]{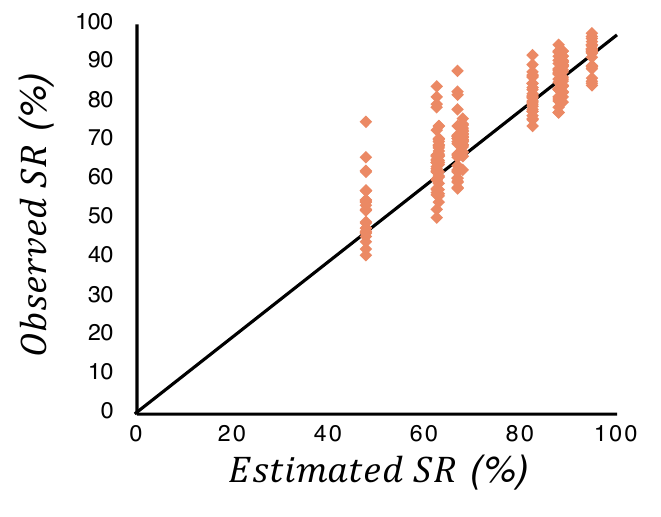}
    \caption{Prediction accuracy of $\sr$ for the Dual Gaussian Distribution Model in Experiment 3. The line represents where predicted values match observed values.}
    \label{fig:Ex3_DualGauss_SR}
\end{figure}

The estimated $\sr$ computed from $\sr_x$ and $\sr_y$ yielded $R^2 = .784$ (\autoref{table:Ex3 Model Regression}, \autoref{fig:Ex3_DualGauss_SR}). 
The $R^2$ and $\mae$ from the LOOCV were largely consistent with the results of the regression analysis using all data.
Given the recurring trend that observed values exceed predicted $\sr_x$ and $\sr_y$ at $\margin_x = 0$ and $\margin_y = 0$, there is potential for the proposed model to improve prediction accuracy.

\vspace{3mm}
\noindent\textit{Proposed Model: Skewed Dual Normal Distribution Model}
\begin{figure}[ht]
\centering
\begin{minipage}[b]{0.49\textwidth}
    \centering
    \includegraphics[width = 0.95\columnwidth]{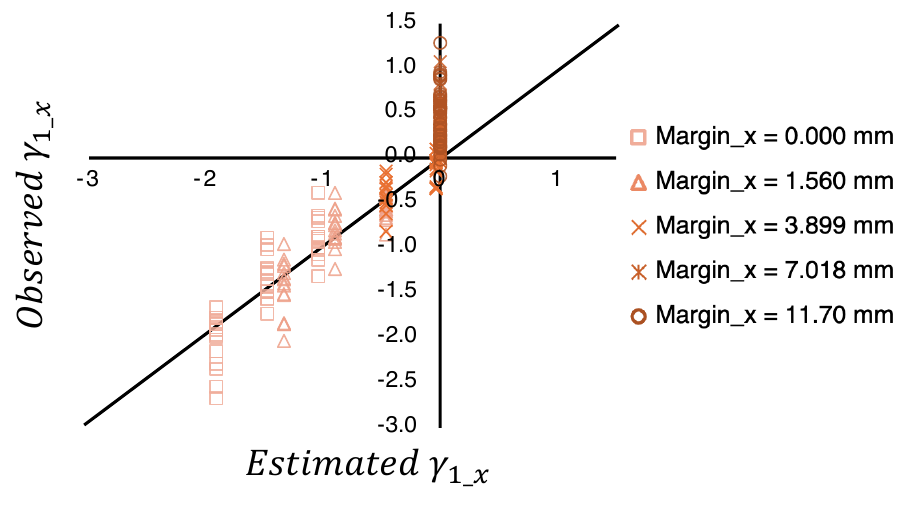}
    \subcaption{Predicted $\gamma_{1\_x}$ vs. Observed $\gamma_{1\_x}$}
    \label{fig:Ex3_DualSkew_Gamma_X}
\end{minipage}
\begin{minipage}[b]{0.49\textwidth}
    \centering
    \includegraphics[width = 0.95\columnwidth]{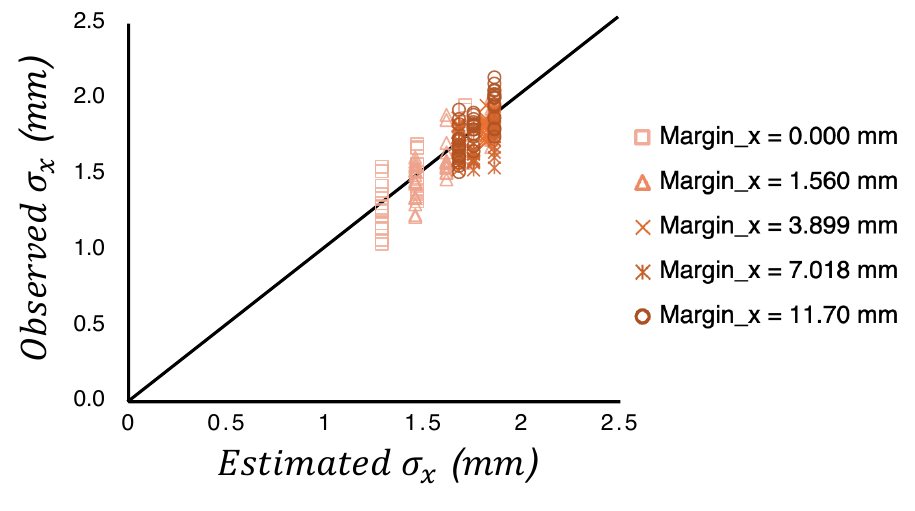}
    \subcaption{Predicted $\sigma_x$ vs. Observed $\sigma_x$}
    \label{fig:Ex3_DualSkew_Sigma_X}
\end{minipage}
\begin{minipage}[b]{0.49\textwidth}
    \centering
    \includegraphics[width = 0.95\columnwidth]{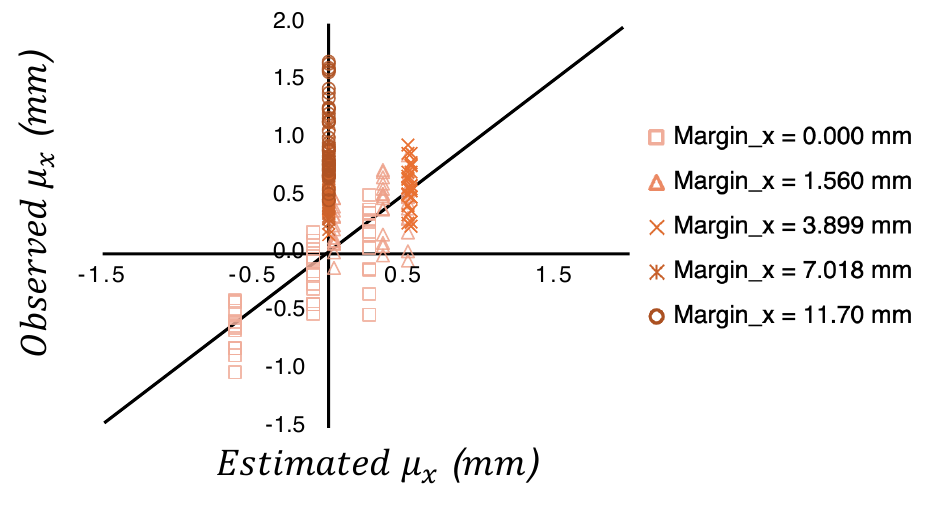}
    \subcaption{Predicted $\mu_x$ vs. Observed $\mu_x$}
    \label{fig:Ex3_DualSkew_Mu_X}
\end{minipage}
\begin{minipage}[b]{0.49\textwidth}
    \centering
    \includegraphics[width = 0.95\columnwidth]{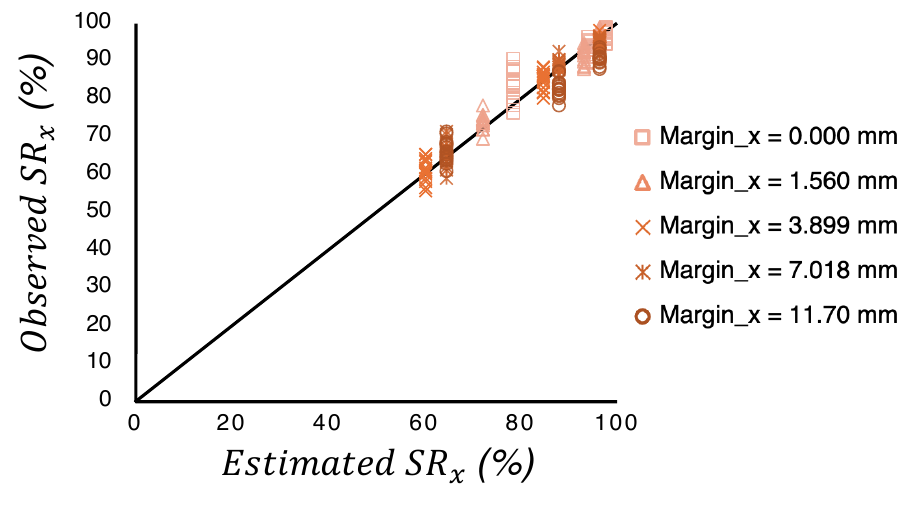}
    \subcaption{Predicted $SR_x$ vs. Observed $SR_x$}
    \label{fig:Ex3_DualSkew_SR_X}
\end{minipage}
\caption{Results regarding the prediction accuracy of the proposed model for $\sr_x$ in Experiment 3. The line represents the identity line where predicted values match observed values.}
\label{fig:Ex3_DualSkewX}
\end{figure}

\autoref{fig:Ex3_DualSkewX} shows the prediction accuracy for parameters along the $x$-axis.
The estimation accuracy of $\gammax$ yielded $R^2 = .806$ (\autoref{table:Ex3 Model Regression}, \autoref{fig:Ex3_DualSkew_Gamma_X}).
As in Experiment 1, $|\gamma_{1_x}|$ increased as the target approached the screen edge; however, the sign was negative because the screen edge was located on the right side (\autoref{fig:Ex3 GammaX Interaction X}, \autoref{fig:Ex3_DualSkew_Gamma_X}).
This consistently demonstrates that, even for targets in the top-right corner, the distribution skews such that the peak shifts toward the screen edge as the distance to the right edge decreases, aligning with the findings of Experiments 1 and 2.
Calculating $-c_x/d_x$ (the x-intercept of the regression line) yielded 6.70.
Similar to Experiment 1 ($-c_x/d_x = 6.40$), the model predicted that the tap-coordinate distribution would skew due to edge effects within 6.70 mm of the right screen edge and follow a normal distribution beyond that distance.

Next, verifying the estimation accuracy of $\sigma_x$ yielded $R^2 = .679$ (\autoref{table:Ex3 Model Regression}, \autoref{fig:Ex3_DualSkew_Sigma_X}), which was superior to the fit of the existing model that regresses $\sigma_x$ solely on $W$ ($R^2 = .314$).
We then performed a regression analysis for $\mu_x$ using \autoref{Formula:DualSkewNormalEstimateMuX} specifically for data where $\dedgex < -c_x/d_x$, yielding $R^2 = .730$ (\autoref{table:Ex3 Model Regression}, \autoref{fig:Ex3_DualSkew_Mu_X}). In the region $\dedgex \ge -c_x/d_x$, where $\mu_x$ was approximated as zero, positive $\mu_x$ values (approximately 0.0 to 1.5) were observed as deviations from the model, similar to Experiment 1 (\autoref{fig:Ex3_DualSkew_Mu_X}).

Finally, the prediction accuracy of $\sr_x$ yielded $R^2 = .917$ (\autoref{table:Ex3 Model Regression}, \autoref{fig:Ex3_DualSkew_SR_X}).
Thus, even when a target is near a corner and influenced by two screen edges, the variation in $\sr_x$ caused by proximity to the right edge can be largely modeled by $x$-axis factors. LOOCV results for $R^2$ and $\mae$ were comparable to the full-data regression, indicating no overfitting and suggesting robust predictive performance for unverified conditions.

\begin{figure}[ht]
\centering
\begin{minipage}[b]{0.49\textwidth}
    \centering
    \includegraphics[width = 0.95\columnwidth]{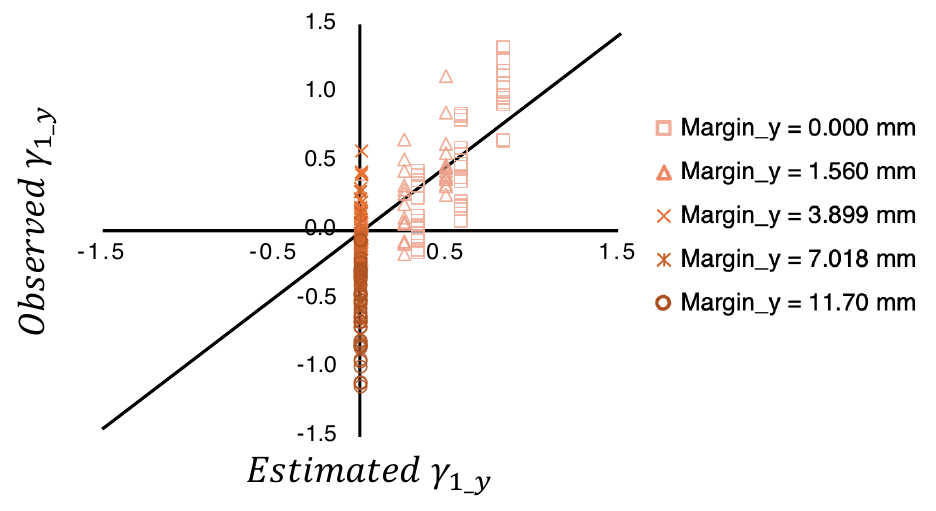}
    \subcaption{Predicted $\gamma_{1\_y}$ vs. Observed $\gamma_{1\_y}$}
    \label{fig:Ex3_DualSkew_Gamma_Y}
\end{minipage}
\begin{minipage}[b]{0.49\textwidth}
    \centering
    \includegraphics[width = 0.95\columnwidth]{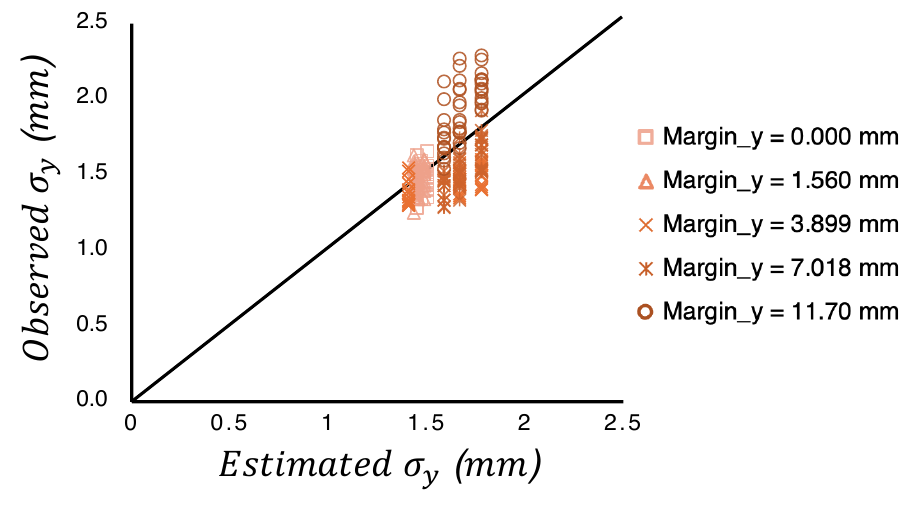}
    \subcaption{Predicted $\sigma_y$ vs. Observed $\sigma_y$}
    \label{fig:Ex3_DualSkew_Sigma_Y}
\end{minipage}
\begin{minipage}[b]{0.49\textwidth}
    \centering
    \includegraphics[width = 0.95\columnwidth]{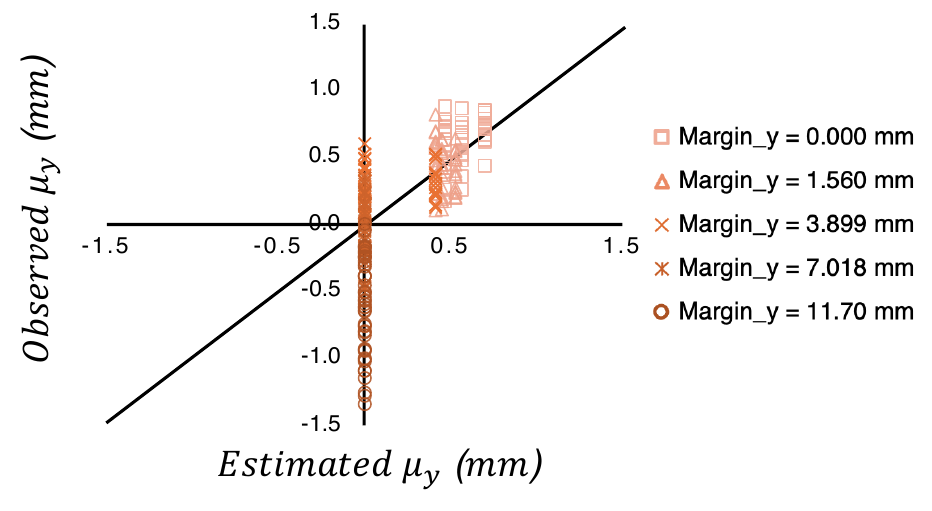}
    \subcaption{Predicted $\mu_y$ vs. Observed $\mu_y$}
    \label{fig:Ex3_DualSkew_Mu_Y}
\end{minipage}
\begin{minipage}[b]{0.49\textwidth}
    \centering
    \includegraphics[width = 0.95\columnwidth]{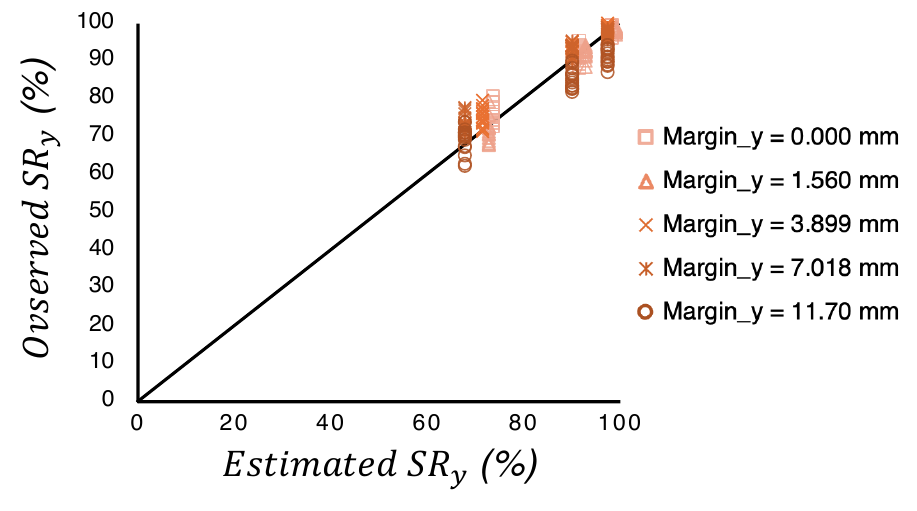}
    \subcaption{Predicted $SR_y$ vs. Observed $SR_y$}
    \label{fig:Ex3_DualSkew_SR_Y}
\end{minipage}
\caption{Results regarding the prediction accuracy of the proposed model for $\sr_y$ in Experiment 3. The line represents the identity line.}
\label{fig:Ex3_DualSkewY}
\end{figure}

\autoref{fig:Ex3_DualSkewY} shows the results for prediction accuracy of parameters and $\sr_Y$ along the $y$-axis.
The overall trends were the same as those for the $x$-axis.
The exceptions were that, contrary to the model's assumptions, many conditions exhibited skewness in the opposite direction (negative $\gamma_{1\_y}$, \autoref{fig:Ex3_DualSkew_Gamma_Y}).
The prediction accuracy for $\sr_y$ was $R^2 = .881$, which is nearly identical to the existing model ($R^2 = .865$).

\begin{figure}[ht]
    \centering
    \includegraphics[width = 0.4\columnwidth]{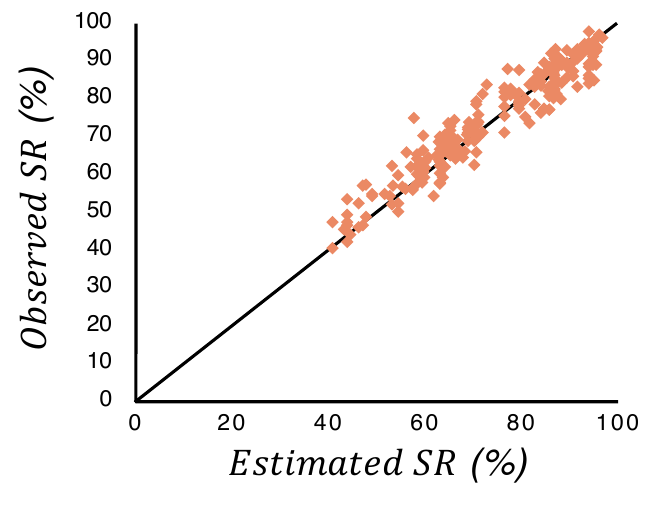}
    \caption{Prediction accuracy of $\sr$ for the proposed model in Experiment 3. The line represents the identity line.}
    \label{fig:Ex3_DualSkew_SR}
\end{figure}

The estimation accuracy of the overall $\sr$ computed from $\sr_x$ and $\sr_y$ yielded $R^2 = .888$ (\autoref{table:Ex3 Model Regression}, \autoref{fig:Ex3_DualSkew_SR}).
Compared to the existing model ($R^2 = .784$), ours showed improved prediction accuracy, suggesting its capability to account for edge effects.
Since the prediction accuracy for $\sr_y$ showed minimal improvement, the overall enhancement in $\sr$ prediction would stem primarily from the improved modeling of $\sr_x$.
LOOCV results confirmed the robustness of these findings.

\vspace{3mm}
\noindent\textit{Machine Learning Models}
\\
\noindent Random Forest with default and optimized, SVR with optimized, and MLP Neural Net with optimized achieved higher $R^2$ and lower $\mae$ than the proposed model (\autoref{table:Ex3 Model Regression}).
Furthermore, in the LOOCV analysis, the ML models demonstrated higher $R^2$ values than the proposed model.
The maximum difference in $R^2$ during LOOCV was $.065$ ($.948$ for optimized SVR versus $.883$ for the proposed model). 
Thus, the ML models were shown to possess higher $\sr$ prediction accuracy than the proposed model, even under unknown conditions.

\subsubsection{Analyzing Tap-Coordinate Distributions}
\begin{figure}[ht]
\centering
\begin{minipage}[b]{0.49\columnwidth}
    \centering
    \includegraphics[width=0.95\columnwidth]{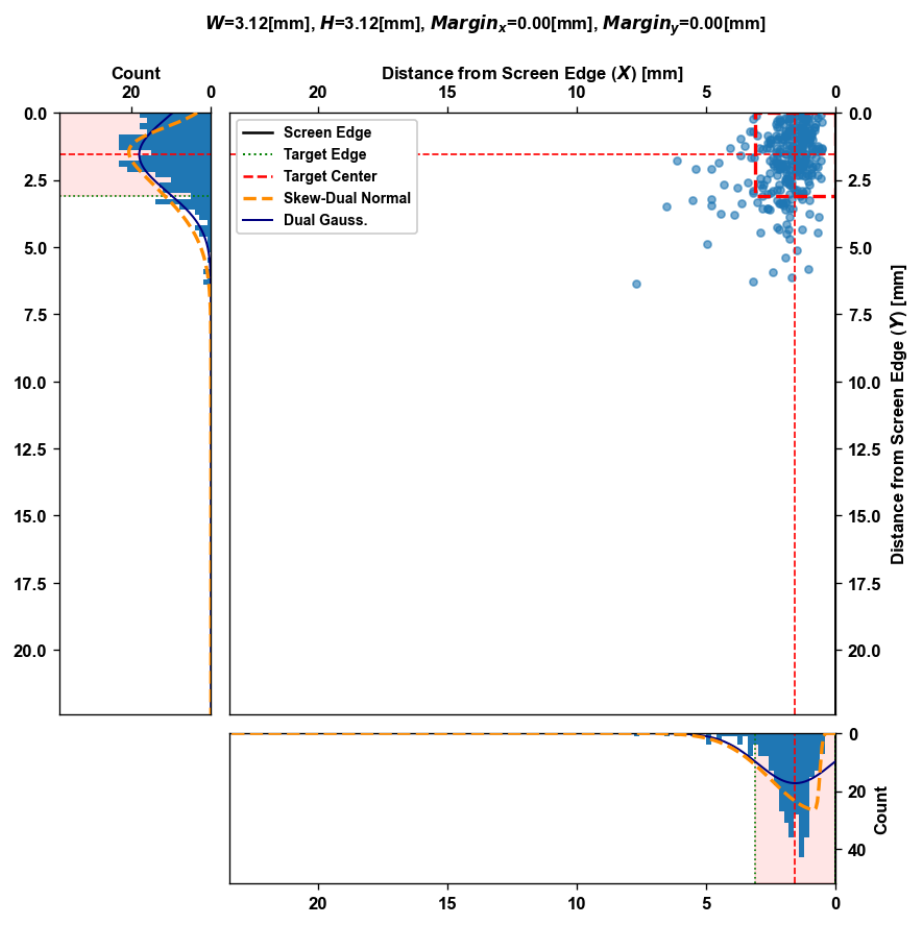}
    \subcaption{Small target touching the screen corner}
    \label{fig:Ex3 Distribution Both Margin 0 Small}
\end{minipage}
\begin{minipage}[b]{0.49\columnwidth}
    \centering
    \includegraphics[width=0.95\columnwidth]{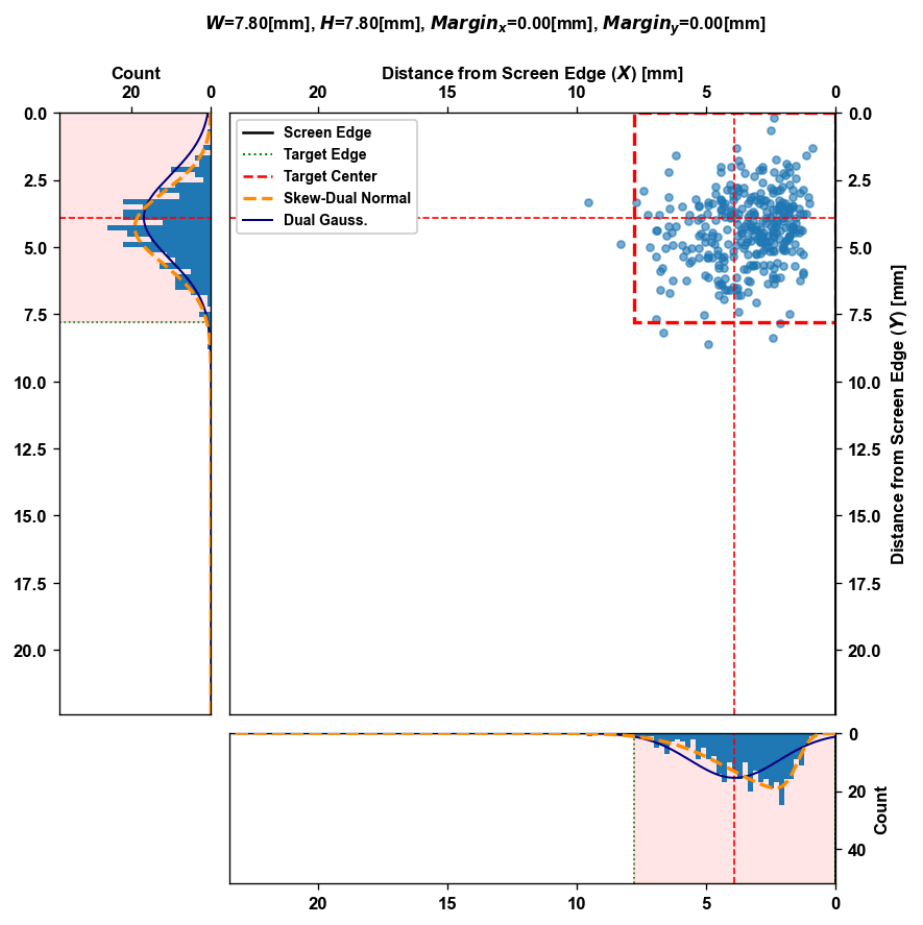}
    \subcaption{Large target touching the screen corner}
    \label{fig:Ex3 Distribution Both Margin 0 Big}
\end{minipage}
\begin{minipage}[b]{0.49\columnwidth}
    \centering
    \includegraphics[width=0.95\columnwidth]{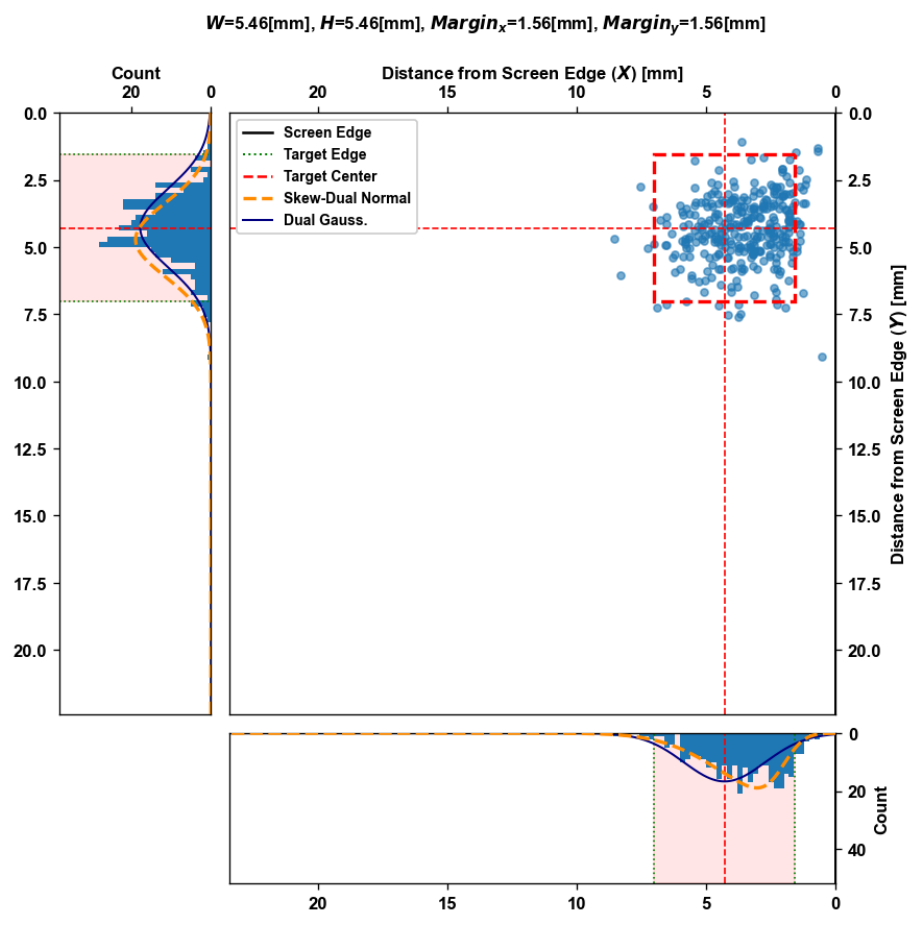}
    \subcaption{Target with small $\margin$ on both axes}
    \label{fig:Ex3 Distribution Both Small Margin}
\end{minipage}
\begin{minipage}[b]{0.49\columnwidth}
    \centering
    \includegraphics[width=0.95\columnwidth]{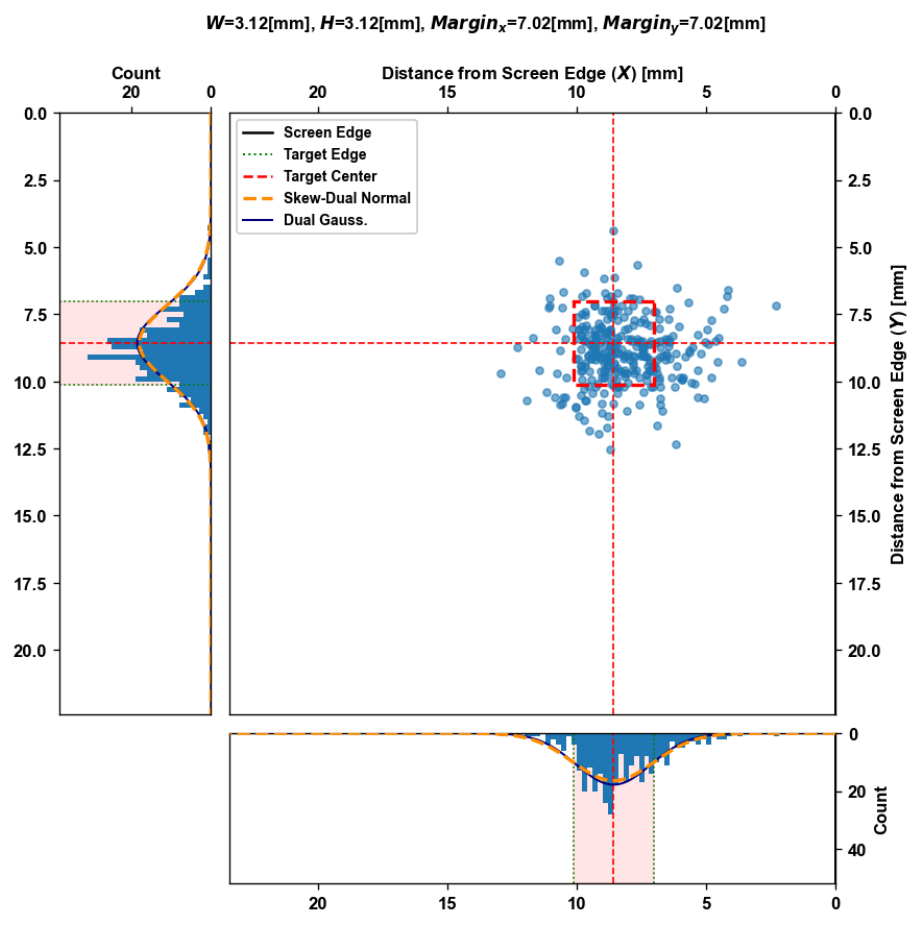}
    \subcaption{Target far from the screen corner}
    \label{fig:Ex3 Distribution Both Normal}
\end{minipage}
\caption{Tap-coordinate distributions in Experiment 3.}
\label{Fig:Ex3 TapDistribution}
\end{figure}

Even near the screen corners, where two edges exert influence, the proposed model generally captured the edge-induced skewness appropriately (see orange dashed lines in \autoref{fig:Ex3 Distribution Both Margin 0 Small}, \autoref{fig:Ex3 Distribution Both Margin 0 Big}). 
For targets sufficiently far from the corner, the distributions became normal; the black lines of the Dual Gaussian Distribution Model match the orange dashed lines in \autoref{fig:Ex3 Distribution Both Normal}. 
However, the skewness along the $x$- and $y$-axes was not symmetrical; the skewness along the $x$-axis persisted up to a certain $\margin_x$, but it rapidly disappeared along the $y$-axis as $H$ or $\margin_y$ increased (\autoref{fig:Ex3 Distribution Both Margin 0 Big}, \autoref{fig:Ex3 Distribution Both Small Margin}).

\begin{figure}[ht]
\centering
\begin{minipage}[b]{0.49\columnwidth}
    \centering
    \includegraphics[width=0.95\columnwidth]{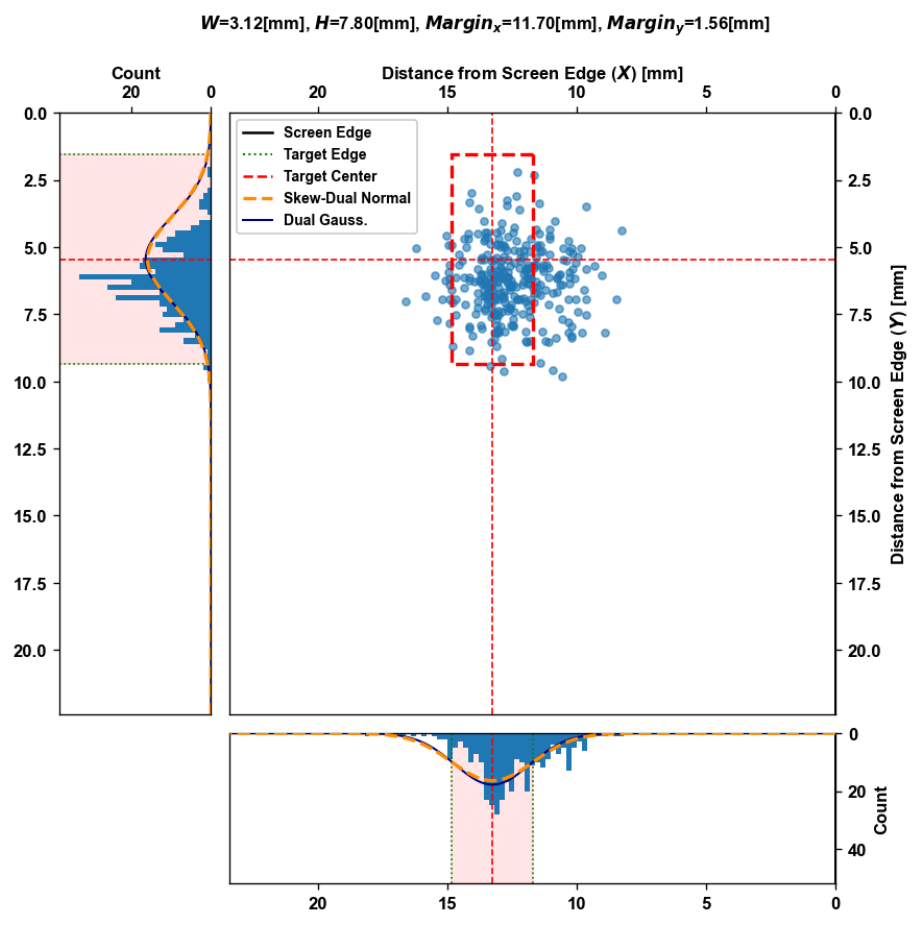}
    \subcaption{Inverse skewness}
    \label{fig:Ex3 Distribution Inverse Skew}
\end{minipage}
\begin{minipage}[b]{0.49\columnwidth}
    \centering
    \includegraphics[width=0.95\columnwidth]{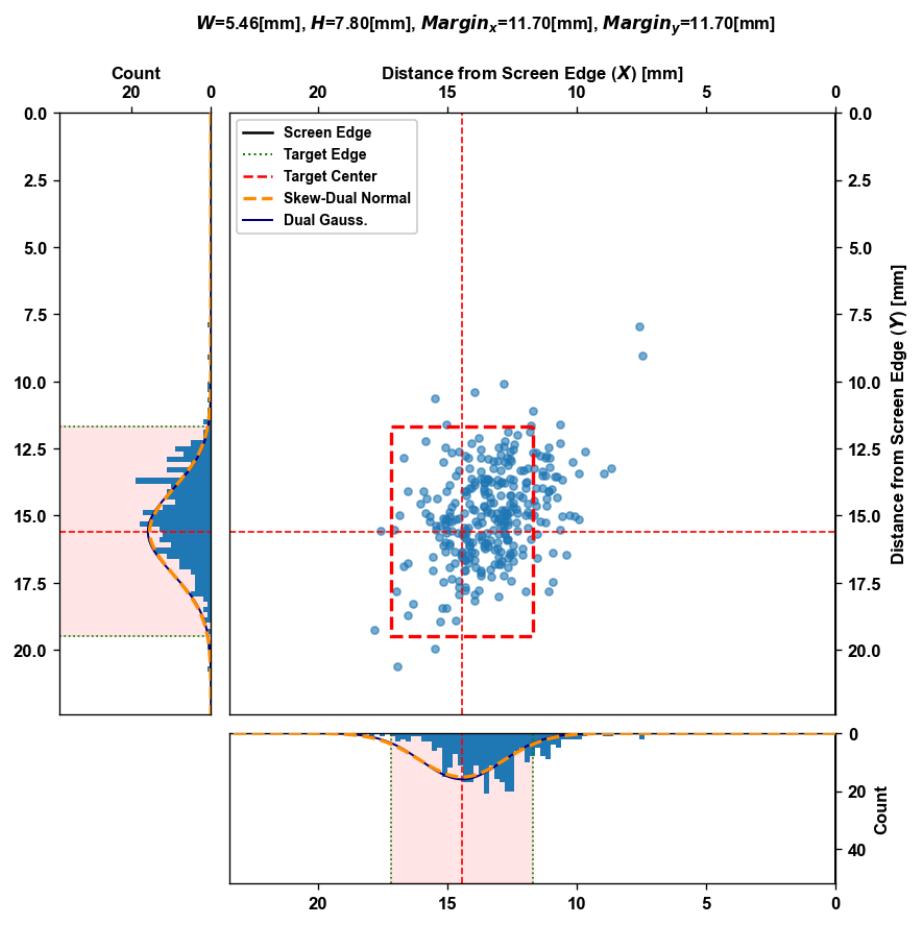}
    \subcaption{Upward and rightward shifts}
    \label{fig:Ex3 Distribution Right Shift Above Shift}
\end{minipage}
\caption{Tap-coordinate distributions deviating from the model in Experiment 3.}
\label{Fig:Ex3 TapDistribution Exception}
\end{figure}

Under several conditions, tap-coordinate distributions that deviated from the model's assumptions were observed (\autoref{Fig:Ex3 TapDistribution Exception}). For targets with large $H$, skewness along the $y$-axis occurred in the opposite direction of the assumed edge-ward peak shift (\autoref{fig:Ex3 Distribution Inverse Skew}). This trend was not observed in Experiments 1 and 2. Additionally, for targets far enough from the screen corner, both rightward and upward shifts of the entire distribution were observed (\autoref{fig:Ex3 Distribution Right Shift Above Shift}). This is consistent with the average shifts in distribution observed in Experiment 1 (\autoref{fig:Ex1 Distribution Exception}).

\subsubsection{Participant Questionnaire}
Four participants reported that they ``intentionally tapped the target and the bezel simultaneously when the target was touching the screen edge.'' Additionally, one participant reported, ``I tried to tap as close to the corner as possible for targets touching the corner.'' Other diverse operational strategies were also reported, such as ``tilted the smartphone,'' ``angled my finger,'' ``aimed for the center,'' or ``didn't think about anything.'' These reports followed trends similar to those in Experiments 1 and 2.
Furthermore, there were comments specific to 2D targets, such as ``It felt difficult when the aspect ratio of the target was extremely skewed.''
Notably, while six participants in Experiment 2 had reported that they ``aimed for the upper part of the target because the contact point was lower than the fingertip,'' no such responses to aim for a specific position were observed in Experiment 3.

\subsection{Discussion}
\subsubsection{Consistency with Experiment 1 and 2}
We showed that our model is generally applicable to $\sr$ prediction and tap-coordinate distributions for 2D targets. 
As in Experiments 1 and 2, the tap-coordinate distribution approached a skew-normal distribution as the target neared the screen edge (\autoref{fig:Ex3 LikelihoodRatio}), and the proposed model effectively captured this transition (\autoref{fig:Ex3_DualSkew_SR}). 
Furthermore, the strategy of simultaneously tapping the bezel and the target was also observed in Experiment 3 (\autoref{fig:Ex3 Distribution Both Margin 0 Small}). 
These findings suggest that the proposed model is applicable to 2D targets at screen corners, where two edges exert influence.

\subsubsection{Differences from Experiment 1 and 2}
\vspace{3mm}
\noindent\textit{Asymmetry of Skewness between Axes}\\
\noindent While Experiments 1 and 2 showed no clear difference between the $x$- and $y$-axes regarding edge-induced skewness, Experiment 3 revealed a distinct asymmetry between the axes.
In the $x$-axis direction of all experiments, tap-coordinate skewness vanished, and the distribution became normal when $\dedgex$ exceeded approximately 6 mm.
However, along the $y$-axis in Experiment 3, skewness tended to be relatively small (\autoref{fig:Ex3 GammaX Interaction X}, \autoref{fig:Ex3 GammaY Interaction Y}), and even showed inverse skewness in some cases (\autoref{Fig:Ex3 TapDistribution}, \autoref{fig:Ex3 Distribution Inverse Skew}, \autoref{fig:Ex3 LikelihoodRatio Y}).
A comparison between \autoref{fig:Ex3 GammaX Interaction X} and \autoref{fig:Ex3 GammaY Interaction Y} shows that while skewness occurred at $\margin_x=0$ regardless of $W$, it did not occur at the top edge when $H$ was large.
This indicates that for large targets touching the top edge, participants did not employ the model's assumed strategy of simultaneously tapping the bezel and the target to avoid errors in the opposite direction.

Notably, Experiment 2 showed significant edge-ward skewness even when $H$ was large (\autoref{fig:Ex2 Gamma1Y Interaction}).
This suggests that the observed asymmetry is a phenomenon unique to the top edge, potentially stemming from an interaction with gravity.
In our handheld operation study, tapping further up the $y$-axis required lifting the hand higher.
Considering that humans often adopt strategies that minimize physical effort \citep{Uno1989}, it is likely that participants favored tapping lower points.
Near the bottom edge, avoiding upward errors and minimizing physical effort both favor downward skewness, aligning the two motivations.
Near the top edge, however, these goals conflict.
Consequently, the mixture of minimizing errors by tapping higher and minimizing physical movement by tapping lower may have reduced skewness or caused inverse skewness.

The disappearance of skewness when $H$ was large suggests that, as the need for error minimization decreased, the minimization of physical effort took priority, potentially supporting this hypothesis.
Additionally, while six participants in Experiment 2 reported aiming higher because the contact point is lower than the fingertip \citep{Holz2010FingerShift,Holz2011UnderstandingTouch}, no such reports were made in Experiment 3, further suggesting that the strategy of tapping higher was not adopted and supporting this hypothesis.

\vspace{3mm}
\noindent\textit{Tap-Coordinate Distribution Shift}\\
\noindent Unlike Experiment 1 (shift toward the start position) and Experiment 2 (no shift), Experiment 3 exhibited a shift in the opposite direction from the start position.
Considering that Experiment 1, which used the right knee as the start position, showed a rightward shift (\autoref{fig:Ex1 Distribution Exception}), a shift toward the starting target (bottom-left) was expected in Experiment 3.
However, the observed shift was actually toward the top-right (\autoref{fig:Ex3 Distribution Right Shift Above Shift}).
This is likely due to the strategy reported in the open-ended questionnaire: ``I was conscious of moving my hand to the top-right immediately after pressing the starting target.''
Since participants knew the target always appear in the top-right, they may have adopted a strategy of moving their arm toward the top-right while simultaneously identifying the target.
Consequently, for targets far from the screen corner, a top-right overshoot occurred, which potentially caused the tap-coordinate distribution to shift toward the top-right.

\section{General Discussion}
\subsection{Influence of Screen Edges and Generalizability of the Model}
Through a series of experiments, we investigated touch-pointing behavior near screen edges and corners, focusing on tap success rates.
The results demonstrated that across all screen edges, the tap-coordinate distribution skewed such that the peak shifted toward the screen boundary.
This behavior was considered a rational operational strategy to avoid errors occurring in the direction opposite to the screen edge.
While these results conflict with prior studies suggesting that performance degrades near edges and that users tend to avoid them \citep{Usuba2023EdgeTarget,Henze2011LargeExperiment,Avrahami2015EdgeTouch}, this discrepancy was likely caused by differences in target and bezel shapes (\autoref{fig:FingerBezelCircleRectangle}).

However, it was also shown that the skewness caused by the edge decreased at the top of the screen, potentially due to a conflict between two rational strategies: error minimization and physical effort minimization.
Our model appropriately captured these edge-induced skews and predicted tap success rates in skewed distributions with higher accuracy than existing models.
Furthermore, the series of experiments suggested that the proposed model can be treated as a general model independent of tap-coordinate distribution direction ($x$-axis, $y$-axis) or target shape (1D, 2D).

\subsection{Practical Contribution}
As a practical contribution to UI design and HCI, this study offers three key insights: concrete design guidelines, applications for UI design support tools, and implications for adaptive interfaces.
First, regarding design guidelines, tap-coordinate distributions skew when the distance between the target center and the screen edge is within approximately 6 mm.
Furthermore, since targets with $\margin=0$ exhibited high tap success rates throughout the experiments, we suggest that targets near screen edges should be designed to touch the boundary completely.

Second, for UI design support tools, integration with existing tools such as Tappy \citep{usuba24arxivTappy,Yamanaka24arXivFigmaTappy} and Tap Analyzer \citep{LIFULL24tapAnalyzer} is a potential application.
Since these tools visualize tap success rates estimated by the Dual Gaussian Distribution Model, they assume a normal distribution even for targets near screen edges.
Integrating our model would provide more accurate predictions, thereby supporting UI designs that effectively utilize areas near screen boundaries.

Finally, as an implication for adaptive interfaces, the proposed model could be utilized in scrollable interfaces in which all UI elements have the potential to approach the screen edges.
Therefore, by using the proposed model to predict the tap success rate of each target within the current viewing area after scrolling, it may be possible to realize adaptive interfaces that automatically resize targets as they approach/leave the screen edges.

\subsection{Advantages of Analytical Models Relative to Machine Learning Models}
Throughout the series of experiments, the ML models demonstrated high predictive accuracy equivalent to or higher than that of the proposed model. 
This indicates that ML models are useful from the perspective of pure tap success rate prediction accuracy. 
However, from the viewpoint of understanding user behavior in HCI, analytical models like the proposed one are considered beneficial. 
Since ML models learn the relationship between inputs (target size and distance from the screen edge) and tap success rates as a black box, it is difficult to explain why errors occur. 
In contrast, the proposed model is constructed based on a geometric mechanism where proximity to screen edges affects tap success rates by altering the mean, variance, and skewness of the distribution. 
This ``explainability'' potentially functions effectively when making concrete design decisions, such as determining in which direction and by how much a UI element should be expanded to prevent mistaps.

\subsection{Limitations}
There are several limitations to this study.
First, we used only a single smartphone model (Google Pixel 6a) without a case.
Using a protective case could induce phenomena similar to a physical bezel (\autoref{fig:FingerBezel}) and create areas that are physically impossible to tap near the screen edges, potentially leading to different operational strategies.
Second, this study verified only index-finger interaction.
One-handed thumb interaction may exhibit different trends from the skewness observed here due to finger joint structure, occlusion, and reachability.
Third, off-screen errors were not observed.
Since this analysis focused only on tap success rates for generated touch events, it does not account for unobservable off-screen errors, such as cases where the bezel was touched but the screen was not.
Therefore, further analysis is needed, including extending the model to cased devices and thumb interaction, as well as investigating off-screen errors using motion-capturing systems.

\section{Conclusion}
In this study, we proposed and verified the Skewed Dual Normal Distribution Model, a new tap success rate prediction model that accounts for the skewness of tap-coordinate distributions due to nearby screen edges and corners of smartphones.
In Experiments 1 and 2, one-dimensional pointing tasks near the left and bottom edges of the screen demonstrated that, regardless of the tap-coordinate distribution direction ($x$-axis or $y$-axis), the distribution skews such that the peak shifts toward the boundary as the target approaches the screen edge.
In Experiment 3, a two-dimensional pointing task for the top-right corner suggested that the proposed model is also applicable to 2D operations influenced by two screen edges simultaneously.
These results indicate that our model predicts tap success rates with higher accuracy than the existing model, regardless of the edge position (top, bottom, left, or right) or task shape.
Furthermore, unlike ML models that showed equivalent or higher predictive accuracy across the experiments, the proposed model provides interpretable parameters, such as the range of influence exerted by the screen edge ($-c/d$), which are useful for UI layout design and theory-based UI development.
This study deepens the understanding of user behavior near screen edges and contributes to the advancement of theoretical UI-design tools by extending the Dual Gaussian Distribution Model to the entire screen.


\section*{Funding}
This work was not supported by any grant.

\section*{Disclosure Statement}
The second author is employed by LY Corporation.
The authors declare no other conflicts of interest.

\section*{Author Contributions}
Nobuhito Kasahara contributed to the conceptualization, model derivation, implementation of the experimental system, conducting experiments, data analysis, visualization, drafting the manuscript, and manuscript editing.
Shota Yamanaka contributed to the conceptualization, data analysis, and manuscript editing.
Homei Miyashita contributed to the conceptualization, manuscript editing, and supervision.
All authors critically revised the manuscript for important intellectual content, approved the final version for submission, and agreed to be accountable for all aspects of the work.

\section*{Use of Large Language Models}
Generative AI (ChatGPT 5.2 Pro) was used to improve the quality of writing, including translation, style, phrasing, and grammar, as well as to check the mathematical correctness of model derivations.
The original manuscript was written by the authors.

\bibliography{zinteractapasample}

\appendix

\section*{Biography}
\noindent\textbf{Nobuhito Kasahara} is a master's course student at Meiji University. His research interests focus on modeling user performance in target-pointing and path-steering tasks.

\noindent\textbf{Shota Yamanaka} is a senior chief researcher at LY Corporation Research, LY Corporation. He received his Ph.D. in engineering from Meiji University in 2016. His research interests include human-computer interaction, graphical user interfaces, and human performance modeling.

\noindent\textbf{Homei Miyashita} is a professor and chair of the department of Frontier Media Science, School of Interdisciplinary Mathematical Sciences, Meiji University. He specializes in human-computer interaction and has recently developed taste media that can record and reproduce taste.

\end{document}